\documentclass[12pt,preprint]{aastex}

\usepackage[margin=1.0in]{geometry}

\usepackage{amssymb}
\usepackage{booktabs}
\usepackage{epstopdf}
\usepackage{footnote}
\usepackage{graphicx}
\usepackage{longtable}
\usepackage{lmodern} 
\usepackage{lscape}
\usepackage{natbib}
\usepackage[caption=false]{subfig}

\begin{document}
\title{Studying Relation Between Star Formation and Molecular Clumps on Subparsec Scales in 30 Doradus}

\author{O. Nayak\altaffilmark{1}, M. Meixner\altaffilmark{2,1}, R. Indebetouw\altaffilmark{3,4}, G. De Marchi\altaffilmark{5}, A. Koekemoer\altaffilmark{2}, N. Panagia\altaffilmark{2}, E. Sabbi\altaffilmark{2}}

\altaffiltext{1}{Department of Physics $\&$ Astronomy, Johns Hopkins University, 3400 N. Charles St., Baltimore, MD 21218, USA}
\altaffiltext{2}{Space Telescope Science Institute, 3700 San Martin Drive, Baltimore, MD 21218, USA}
\altaffiltext{3}{Department of Astronomy, University of Virginia, P.O. Box 3818, Charlottesville, VA 22903, USA}
\altaffiltext{4}{National Radio Astronomy Observatory, 520 Edgemont Road Charlottesville, VA 22903, USA}
\altaffiltext{5}{European Space Agency, Space Science Department, Keplerlaan 1, 2200 AG Noordwijk, The Netherlands}

\begin{abstract}
\noindent
We present $\mathrm{^{12}CO}$ and $\mathrm{^{13}CO}$ molecular gas data observed by ALMA, massive early stage young stellar objects identified by applying color-magnitude cuts to \textit{Spitzer} and \textit{Herschel} photometry, and low-mass late stage young stellar objects identified via H$\mathrm{\alpha}$ excess. Using dendrograms, we derive properties for the molecular cloud structures. This is the first time a dendrogram analysis has been applied to extragalactic clouds. The majority of clumps have a virial parameter equal to unity or less. The size-linewidth relations of $\mathrm{^{12}CO}$ and $\mathrm{^{13}CO}$ show the clumps in this study have a larger linewidth for a given size (by factor of 3.8 and 2.5, respectively) in comparison to several, but not all, previous studies. The larger linewidths in 30 Doradus compared to typical Milky Way quiescent clumps are probably due to the highly energetic environmental conditions of 30 Doradus. The slope of the size-linewidth relations of $\mathrm{^{12}CO}$, 0.65 $\pm$ 0.04, and $\mathrm{^{13}CO}$, 0.97 $\pm$ 0.12, are on the higher end but consistent within 3$\mathrm{\sigma}$ of previous studies. Massive star formation occurs in clumps with high masses ($\textgreater 1.83 \times 10^{2}\;\mathrm{M_{\odot}}$), high linewidths (v $\textgreater 1.18\;\mathrm{km/s}$), and high mass densities ($\textgreater 6.67 \times 10^{2}\;\mathrm{M_{\odot}\;pc^{-2}}$). The majority of embedded, massive young stellar objects are associated with a clump. However the majority of more evolved, low-mass young stellar objects are not associated with a clump.
\end{abstract}


\section{Introduction}
\subsection{Star Formation in the Large Magellanic Cloud}
\noindent
Massive stars have a profound impact on galaxies. Their great luminosities and intense ionizing radiation change the thermodynamic state of the surrounding interstellar medium (ISM) \citep{blak87, krau13}, their strong stellar winds compress the ISM \citep{lada85, heck90, kudr00, hopk12}, and their explosive ends enrich it with heavy elements \citep{woos95, woos02, hege03}. Massive star formation is not a scaled up version of low-mass star formation. Formation of low-mass stars have been studied in detail within the Milky Way \citep[e.g.,][]{krou93, xue08, heid10, kafl14}. The intense radiation from high-mass stars has a significant effect on the surrounding medium as well as limits the accretion rate onto a high-mass protostar. Low-mass stars have very little gravitational effect on the surrounding gas or on other nearby stars. Massive stars play a more important role in defining the dynamics on the stellar cluster in which they reside: they can steal material from other stars (competitive accretion) \citep{tan14}. Massive stars have stellar winds and high ultraviolet radiation pressures that may trigger further star formation in nearby regions, which low-mass stars cannot do.
\\
\\
The Large Magellanic Cloud (LMC) provides an opportunity to study star formation at reduced metallicity of Z $\approx$ 0.5 $\mathrm{Z_{\odot}}$ \citep{west97}. The nearly face-on viewing angle provides a unique view of resolved stellar populations without the confusion and extinction of the Galactic plane. Extinction to young stellar objects (YSOs) is nearly all associated with the star-forming region itself. Also, unlike in the Galaxy where distances are ambiguous, the distance to the LMC is well-determined (50 kpc, \citet{feas99}), leading to robust luminosity determinations. The recent discoveries of thousands of YSO candidates in the LMC provides an opportunity to study massive star formation in both a statistical and detailed way \citep{meix06, whit08, seal09, chen10, carl12, meix13, seal14}. \citet{fuku15.1} study previously identified YSO candidates in the N159 region of the LMC, and related the known YSO candidates to $\mathrm{^{12}CO}$ and $\mathrm{^{13}CO}$ ALMA observations of molecular gas. One of the young, massive YSO candidates in N159 is suspected to have formed via filamentary collision. \citet{fuku15.2} look at two molecular clouds colliding in RCW 38, a young super star cluster in the Milky Way. The collision between two filaments in RCW 38 has lead to the formation of 20 O stars. Colliding filaments is one theory on how high-mass stars form. Another theory is based on competitive accretion. Star formation surveys show that the majority of low-mass, more evolved YSOs are found in clusters \citep{lada93}. \citet{bonn01} simulate 100 stars in a cluster and find stars located at the centers of clusters accrete more gas than those at the outskirts. Competitive accretion is highly non-uniform and depends on the initial mass of the star as well as the star's position in the cluster.
\\
\\
Previous observations, theories, and simulations all suggest the molecular gas structure is connected to the formation of high-mass stars \citep{gold11}. In this paper we will discuss the impact of molecular gas and dust on massive star formation and make comments on how our results relate to previously observed thresholds necessary for high-mass star formation to occur.
\subsection{Star Formation and Molecular Clumps in 30 Doradus}
\noindent
The 30 Doradus region, a.k.a. the Tarantula Nebula, provides a unique opportunity to study star formation over a range of masses and its feedback on the environment. 30 Doradus is one of the most active star formation regions in the Magellanic Clouds and the only Giant HII region in the Local group.  At the heart of 30 Doradus is R136, the super star cluster with extraordinarily high stellar densities of $\textgreater 1.5 \times 10^{4}-10^{7}\;\mathrm{M_{\odot}\;pc^{-3}}$ \citep{selm13}, containing the most massive stars known  \citep{crow10}. Combining data from the \textit{Spitzer} SAGE survey \citep{meix06}, the \textit{Herschel} Heritage survey \citep{meix13}, and the Hubble Tarantula Treasury Project (HTTP) \citep{sabb13, sabb16} we can study young and massive YSOs and low-mass and more evolved YSOs over the bulk of the mass function ($0.5 - 35\;\mathrm{M_{\odot}}$). Studying both low-mass and high-mass stars that are located in the same low-metallicity region can shed some light on the conditions necessary for star formation to occur.
\\
\\
The 30 Doradus region has been previously investigated in the tracers of the interstellar medium (ISM) that fuel the star formation. The dust content has been measured and quantified with the analysis of \textit{Spitzer} SAGE \citep{meix06} and \textit{Herschel} HERITAGE \citep{meix13} surveys by \citet{gord14}. $\mathrm{^{12}CO}$ (1-0) has been imaged as part of whole LMC surveys at $2.6^{\prime}$ with NANTEN \citep{fuku08} and at  $45^{\prime\prime}$ with MOPRA \citep{wong11}. $\mathrm{^{12}CO}$ (1-0), (2-1), and (3-2) have been observed in 30 Doradus by SEST \citep{joha98}. 
\\
\\
ALMA observations of $\mathrm{^{12}CO}$ (2-1), $\mathrm{^{13}CO}$ (2-1), and $\mathrm{C^{18}O}$ (2-1) clumps have been analyzed by \citet{inde13}. The $\mathrm{CO}$ observed in 30 Doradus is very clumpy, similar to that observed in the Milky Way. Analysis of the ALMA data shows a decrease in $\mathrm{^{12}CO}$ (2-1) emission relative to the total gas mass compared to massive star formation regions in the Milky Way, consistent with theory: there is less shielding in a lower-metallicity environment and $\mathrm{^{12}CO}$ (2-1) is not well shielded in comparison to $\mathrm{H_{2}}$ gas. High density gas tracers such as $\mathrm{HCN}$ (1-0) and $\mathrm{HCO^{+}}$ (1-0) show clumps in 30 Doradus have similar, but slightly larger linewidths than other regions in the LMC \citep{ande13}.  The physical conditions of the gas surrounding the R136 cluster have previously been constrained with ionized gas models \citep{inde09, lope11, pell11}. \citet{inde09} find photodissociated regions (PDRs) dominate over collisional excitation of shocks and that the effects of local hot stars play a more important role in shaping the gas chemistry than large scale trends with distance from the R136 cluster. The hot gas in 30 Doradus may be leaking out of the pores of the HII shells, since the X-ray gas pressure is measured to be much weaker than the radiation pressure \citep{lope11}. $\mathrm{H_{2}}$ gas correlates well with $\mathrm{Br\gamma}$ and $\mathrm{CO}$ in the region, implying that the $\mathrm{H_{2}}$ gas comes from PDRs \citep{yeh15}. Further work on PDR modeling has been done by \citet{chev16} who find 90$\%$ of [CII] emissions originates from PDRs and 70$\%$ of the far infrared luminosity is associated with the ionized gas component. \citet{chev16} conclude that the gas in 30 Doradus is very porous and dominated by photoionization.
\\
\\
Our goal in this paper is to conduct a thorough investigation of early stage massive YSOs and later stage low-mass YSOs, and their relation to the clump structure of $\mathrm{^{12}CO}$ (2-1) and $\mathrm{^{13}CO}$ (2-1) molecular gas. This comprehensive study of high-mass and low-mass star formation with multi wavelength photometric and spectroscopic observations of molecular gas and dust will help shed light on the physical conditions necessary for high-mass star formation to take place. In Section 2 we describe the observations we use in more detail. In Section 3 we describe the dendrogram program we use to study the hierarchical structure of the clumps. In Section 4 we describe the YSOs in the study in more detail. In Section 5 we discuss the trends we see in molecular gas and how the how star formation is affected by the distribution of molecular gas. Lastly we present our conclusions in Section 6.

\section{Observations}
\subsection{\textit{Spitzer} Surveying the Agents of Galaxy Evolution (SAGE) and \textit{Herschel} HERschel Inventory of the Agents of Galaxy Evolution (HERITAGE) Surveys}
\noindent
The SAGE survey imaged the LMC using IRAC (3.6 $\mu$m, 4.5 $\mu$m, 5.8 $\mu$m, 8.0 $\mu$m) and MIPS (24 $\mu$m, 70 $\mu$m, 160 $\mu$m) instruments on the \textit{Spitzer} Space Telescope. One of the primary goals of the SAGE survey was to catalog YSOs in the LMC and study the current star formation rate in the LMC. The \textit{Herschel} Heritage survey used the PACS and SPIRE instruments to image the LMC at 100 $\mu$m, 160 $\mu$m, 250 $\mu$m, 350 $\mu$m, 500 $\mu$m wavelengths in order to study the very cold dust surrounding the deeply embedded YSOs. Several YSO candidates we use in this work have previously been studied by \citet{whit08, grue09, carl12}; and \citet{seal14} who used the \textit{Spitzer} SAGE \citep{meix06} and \textit{Herschel} Heritage \citep{meix13} surveys.
\subsection{Hubble Tarantula Treasury Project (HTTP)}
\noindent
HTTP is a survey of stellar populations in 30 Doradus that reaches into the sub-solar mass regime ($\textless 0.5\;\mathrm{M_{\odot}}$) \citep{sabb13, sabb16}. The survey includes optical (F555W and F658N with ACS and WFC), infrared (F110W and F160W with WFC3 and IR), and ultraviolet (F257W and F336W with WFC3 and UVIS). In addition, archival monochromatic survey in the F775W filter was used (realized using ACS/WFC and WFC3/UVIS in parallel). HTTP covers a projected area of 14' x 12' in the sky. We focus on the late-stage and more evolved YSOs \citep{sabb16} that are located within the 30 Doradus ALMA footprint \citep{inde13}. 
\subsection{ALMA $\mathrm{^{12}CO}$ (2-1) and $\mathrm{^{13}CO}$ (2-1) Observation}
\noindent
ALMA allows us to resolve giant molecular clouds (GMCs) at the distance of the LMC with comparable resolution as HST, as well as provides us with high spectral resolution data. The frequency axis of the data is converted to velocity and results in a position-position-velocity (PPV) cube. The ALMA Cycle 0 footprint of 30 Doradus is a small region north of R136 and has been mapped with $\mathrm{^{12}CO}$ (2-1), $\mathrm{^{13}CO}$ (2-1), $\mathrm{C^{18}O}$, and $\mathrm{H30\alpha}$ \citep{inde13}. \citet{inde13} analyze the PPV data cubes and calculate mass, velocity, and size of the clumps. Figure ~\ref{fig:d1} shows the HST F160W in greyscale, ALMA+APEX $\mathrm{^{13}CO}$ (2-1) in contour, and the location of the low-mass YSOs. Figure ~\ref{fig:d1} shows the HST F160W in greyscale, ALMA+APEX $\mathrm{^{13}CO}$ (2-1) in contour, and the location of point sources from the SAGE catalog that are in the ALMA footprint. We use the ALMA Cycle 0 combined with APEX single dish observations of $\mathrm{^{12}CO}$ (2-1) and $\mathrm{^{13}CO}$ (2-1) made by \citet{inde13} in our analysis.

\section{Dendrogram Analysis}
\subsection{Astrophysical Motivation for Using Dendrogram Analysis}
\noindent
The molecular cloud structure determines where star formation occurs and the mass distribution of stars in a cluster. Molecular cloud sizes range several orders of magnitudes: high density parsec size clumps are nested within larger low density clumps that span several tens of parsecs \citep{lada92}. We use dendrograms as a way to study the hierarchical structure of the $\mathrm{^{12}CO}$ (2-1) and $\mathrm{^{13}CO}$ (2-1) emission from molecular clouds. We define clumps to mean any entity that is bound by an iso-intensity surface. Dendrogram algorithms look for the largest scale sizes first and then for smaller clumps embedded within the larger ones. This is different than previous studies which look at individual clumps and not nested structures. The dendrogram of a PPV data cube is a way to keep track of the iso-intensity surfaces over a range of size scales. If the intensities of a structure are not contiguous, then this structure is split into separate entities. The low-density gas is represented at the bottom of the hierarchical structure in a dendrogram, the starting branch of the structure. This branch connects to other branches and leaves that represent smaller and denser clumps embedded in the low-density media. The leaves and branches correspond to the different volumes in the data cube bounded by a given isosurface level. 
\\
\\
Dendrogram analysis allows us to relate the low density ISM to the high density clumps in which star formation takes place. This novel approach is different than previous studies where a GMC would be separated into different clumps, however the structure as a whole would not be taken into account. Previous studies using dendrograms find that using this approach can lead to selecting GMCs in an automated way using only physically motivated criteria \citep{roso08}. Clumps in very active star forming regions like the central molecular zone (CMZ) follow the same size-linewidth relation slope as clumps in other Milky Way clouds but are offset higher by a certain factor \citep{shet12}. Clumps with star formation tend to be more massive (for a given radius) than clumps without star formation \citep{kauf10}. With the use of dendrograms we can relate the parent clump (the starting branch of the dendrogram) to the child clumps (the leaves of the dendrogram) and study the contiguous nature that is inherent in the ISM.
\subsection{Method}
\noindent
The \textit{astrodendro} program uses what is known as the clipping method to calculate clump properties from a 2D image or a 3D data cube. The clipping method only accounts for emission above a contour level to be associated with an object, as oppose to extrapolating down to a zero intensity isosurface. The user has to input the minimum value to be considered in the data set, the threshold value that determines if a leaf will be a single entity or not, and the minimum number of pixels for a leaf to be considered a single entity. The larger branch breaks down into leaves depending on the three input criteria. Figure ~\ref{fig:d2a} shows the dendrogram structure of $\mathrm{^{13}CO}$ (2-1) column density, Figure ~\ref{fig:d2b} shows how the dendrogram structure relates to the molecular clouds. Figure ~\ref{fig:d3a} shows the $\mathrm{^{12}CO}$ (2-1) brightness temperature, Figure ~\ref{fig:d3b} shows the corresponding emission map. The individual dendrogram structures of $\mathrm{^{13}CO}$ (2-1) and $\mathrm{^{12}CO}$ (2-1) are different because we are looking at two different quantities (column density and brightness temperature) and because of optical depth effects. The dendrogram structures highlighted in Figures ~\ref{fig:d2b} and ~\ref{fig:d3b} were chosen in order to illustrate how branches and leaves translate to an image map. The \textit{astrodendro} program outputs the total integrated luminosity of an isosurface (a.k.a. clump), the linewidth ($\mathrm{v_{RMS}}$) of the clump, and the area of the clump, the effective radius of the clump, and the orientation angle of the clump. These properties can be used to calculate relevant astrophysical properties such as the column density or mass.
\subsection{Results from Dendrogram Analysis of $\mathrm{^{12}CO}$ (2-1) and $\mathrm{^{13}CO}$ (2-1) Emission in 30 Doradus}
\noindent
The most dominant molecular species in the ISM is $\mathrm{H_{2}}$. Unfortunately $\mathrm{H_{2}}$ does not have any dipole moment, all of the low energy transitions are quadrupole transitions with small transition probabilities and high excitation energies. This means that the $\mathrm{H_{2}}$ is excited at high temperatures or in the vicinity of strong ultraviolet radiation. The most abundant molecule in the ISM is invisible to direct observation since the majority of ISM conditions cannot excite it. The second most abundant molecule in the ISM is CO and is frequently used to estimate the $\mathrm{H_{2}}$ gas mass via a conversion factor. \citet{lero11} find the $\mathrm{^{12}CO-to-H_{2}}$ conversion factor (or the X-factor) in the LMC to be $3-9\;\mathrm{M_{\odot}\;pc^{-2}\;(K\;km\;s^{-1})^{-1}}$ in the local group which includes: M21, M33, the LMC, NGC 6822, and the SMC. The $\mathrm{M_{mol}}$/F($\mathrm{^{12}CO}\;1-0$) X-factor of 30 Doradus is higher than the average X-factor in the LMC: \citet{inde13} compare the calculated molecular mass to the $\mathrm{^{12}CO}$ (2-1) intensity and find that the X-factor is $12 \pm 4\;\mathrm{M_{\odot}\;pc^{-2}\;(K\;km\;s^{-1})^{-1}}$.  We calculate the brightness temperature integrated over the velocity for $\mathrm{^{12}CO}$ (2-1) and use a X-factor of $12 \pm 4\;\mathrm{M_{\odot}\;pc^{-2}\;(K\;km\;s^{-1})^{-1}}$. The total mass of all dendrogram branches in Figure ~\ref{fig:d3a} is $1.3 \pm 0.4 \times 10^{5}\;\mathrm{M_{\odot}}$. Table 1 lists the mass of the $\mathrm{^{12}CO}$ (2-1) clumps. 
\\
\\
Table 2 lists column density and mass values we obtain on the $\mathrm{^{13}CO}$ (2-1) clumps. The equations we used to convert the $\mathrm{^{13}CO}$ (2-1) column density to the $\mathrm{H_{2}}$ mass can be found in Appendix A. We use a $\mathrm{H_{2}/^{13}CO}$ abundance ratio of $5 \times 10^{6}$ to convert the $\mathrm{^{13}CO}$ column density to the $\mathrm{H_{2}}$ column density, the same abundance factor that was used by \citet{inde13}. The total $\mathrm{H_{2}}$ mass we obtain by summing up the masses of each dendrogram trunk in Figure ~\ref{fig:d2a} is $6.7 \times 10^{4}\;\mathrm{M_{\odot}}$. The mass estimated from $\mathrm{^{13}CO}$ (2-1) and $\mathrm{^{12}CO}$ (2-1) in this work are consistent with each other as well as those estimated by \citet{inde13} using dust continuum emission ($6.0 \pm 1.0 \times 10^{4}\;\mathrm{M_{\odot}}$) and LTE analysis of $\mathrm{^{13}CO}$ (2-1) ($6.8 \times 10^{4}\;\mathrm{M_{\odot}}$).

\section{Massive YSOs and Low-Mass YSOs}
\noindent
The molecular clump sizes range from 0.2 - 1.4 $\mathrm{pc}$. To trace star formation at this scale we need an inventory of the forming stars. We use recent surveys (HTTP, \textit{Spitzer} SAGE, \textit{Herschel} HERITAGE) to make a catalog of all the low-mass and massive YSOs. We define the stage of the YSO candidates in our sample by the following: Stage 0/I objects have $\mathrm{\frac{{\dot{M}_{env}}}{M_{star}}} \textgreater 10^{-6}\;\mathrm{yr^{-1}}$, Stage II objects are those that have $\mathrm{\frac{{\dot{M}_{env}}}{M_{star}}} \textless 10^{-6}\;\mathrm{yr^{-1}}$ and $\mathrm{\frac{M_{disk}}{M_{star}}} \textgreater 10^{-6}$, and Stage III objects have $\mathrm{\frac{{\dot{M}_{env}}}{M_{star}}} \textless 10^{-6}\;\mathrm{yr^{-1}}$ and $\mathrm{\frac{M_{disk}}{M_{star}}} \textless 10^{-6}$ \citep{robi06, robi07, chen10}.
\subsection{Stage III Low-Mass YSOs}
\noindent
Stage III low-mass YSOs are usually identified by their location in an optical color-magnitude diagram, which leads to two problems: 1) contamination by older field stars and the effects of differential extinction that may lead to an overestimate of the actual number of candidates 2) older YSOs that are close to the main sequence (MS) cannot be identified accurately \citep{dema11.1}. An alternate method of looking for Stage III low-mass YSOs is by H$\mathrm{\alpha}$ excess from accretion shocks \citep{dema11.1}. Strong and fast variability of the H$\mathrm{\alpha}$ line intensity has been detected spectroscopically \citep[e.g.,][]{fern95, reip96, smit99, alen01, alen05, sous16} that can move a star above and below our 10 Angstrom equivalent width threshold in a few days and sometimes in a few hours. Comparing observations of the SN 1987A field taken at 3 different epochs \citep{roma98, pana00} revealed that the number of stars that have strong H$\mathrm{\alpha}$ excess is essentially the same at all epochs, but only about about 1/3 of them are that strong at all epochs. This suggests a duty cycle of about 1/3. An ongoing study of the PMS stars in NGC 346, based on the original work of \citet{dema11.2, dema11.3} revealed that the number of H$\mathrm{\alpha}$ strong PMS stars in the `classical' region of the color magnitude diagram (CMD) (i.e. stars brighter and redder than the MS) are between 1/3 and 2/3 of all the stars in the same region (De Marchi $\&$ Panagia, private communication). Note that in the case of NGC 346 the reddening is not so high ($\mathrm{A_{v}} \approx 0.5$) and, therefore, there is no serious problem of contamination in that region of the CMD. All those elements taken together suggest that the duty cycle of relatively young PMS stars (those still above and redder than the MS) should be of the order of 1/3.
\\
\\
We have used the HTTP catalog \citep{sabb16} to select the low-mass Stage III YSO candidates with H$\mathrm{\alpha}$ excess within the ALMA footprint using the stringent selection criteria as defined by \citet{dema11.1}.  In particular, all selected stars must have uncertainties lower than 0.1 mag in each of the V, I and H$\mathrm{\alpha}$ bands, simultaneously.  Furthermore, they must display  V-H$\mathrm{\alpha}$ excesses higher than the 4$\mathrm{\sigma}$ level  as compared to stars of the same V-I color. There are 87 stars that meet the above criteria and are located within the ALMA footprint.
\\
\\
Figures ~\ref{fig:d4a} and ~\ref{fig:d4b} shows the F160W versus F110W - F160W color-magnitude diagram (CMD) with isochrones and mass evolutionary tracks overplotted. We plot isochrones and mass evolutionary tracks taken from the Padova code \citep{bres12, chen15} on these figures. For our purposes we use Milky Way extinction of $\mathrm{A_{v}}=0.06$ and metallicity of $\mathrm{Z}=0.008$ to redden the Padova isochrones to their location in the LMC. In addition we use the local extinction of 30 Doradus as calculated by \citet{dema16} to de-redden the low-mass YSOs. The majority of stars fall between $10^{5} - 10^{7}\;\mathrm{yrs}$ isochrones and $1 - 3\;\mathrm{M_{\odot}}$ mass evolutionary tracks. Table 3 lists the approximate masses of these low-mass YSOs. These masses were calculated by the best fit mass evolutionary track from F160W versus F110W - F160W for each source. When we check these mass estimates with the best fit mass evolutionary track to F775W versus F555W - F775W for each source, we find that 75$\%$ of the masses do not match. The masses from the best fit evolutionary track to F775W versus F555W - F775W are always lower than those from F160W versus F110W - F160W. We choose the near infrared CMD mass evolution tracks to determine the mass because these wavelengths are not as contaminated from extinction as optical bands. All masses of low-mass YSOs have an uncertainty of $\pm 0.25\;\mathrm{M_{\odot}}$. This is due to the the $0.5\;\mathrm{M_{\odot}}$ spacing between the mass evolutionary track (Figure ~\ref{fig:d4b}).
\\
\\
We are not complete in our catalog of low-mass YSOs due to the H$\mathrm{\alpha}$ duty cycle - selection of low-mass YSOs by H$\mathrm{\alpha}$ excess means at any given time we can observe 1/3 of the stars. Selecting the low-mass YSOs by H$\mathrm{\alpha}$ excess means that we are targeting Stage III objects. Figure ~\ref{fig:d7} shows a histogram of all massive and low-mass YSO star masses. The majority of low-mass YSOs in our analysis are $3\;\mathrm{M_{\odot}}$. We do not select any Stage I or Stage II low-mass YSO candidates. This is the result of our bias towards selecting low-mass Stage III YSOs by H$\mathrm{\alpha}$ excess. 
\subsection{Stage I Massive YSOs}  
\subsubsection{Overview}
\noindent
We include massive YSO stage I candidates found by \citet{grue09} and  \citet{seal09, seal14} in our study. There are no massive stage I YSO candidates from \citet{whit08} that are found in the ALMA CO footprint. \citet{grue09, whit08} implements color cuts of different kinds, \citet{seal09} was a spectroscopic follow-up of \citet{grue09} sources, and \citet{walb13} study 10 Spitzer selected YSOs in 30 Doradus in more detail in order to provide mass estimates of these objects. \citet{seal09} use \textit{Spitzer} IRS spectra and group YSO candidates in six categories. The S Group objects have silicate absorption features including the 10 $\mu$m absorption feature and sometimes including the 18 $\mu$m absorption feature, the SE Group objects have both silicate absorption features (similar to S Group) and strong fine-structure emission lines, the P group objects have strong PAH features, the PE group objects have strong PAH and fine-structure line emission, the E group contain sources that only have very strong fine-structure emission lines, and the F group all other sources whose spectra looks similar to that of a YSO but they do not fit any of the above criteria.
\\
\\
We look at all point source objects in the SAGE catalog that are in the ALMA footprint of 30 Doradus to see if any possible massive YSO candidates were missed by previous galaxy wide studies because of stringent cuts. There are 6 YSO candidates that have previously been identified: J84.703995-69.079110, J84.699755-69.069803, J84.688990-69.084695, J84.695173-69.084857, J84.726173-69.082254, and J84.720292-69.077084. And there are 9 point sources in the ALMA footprint that have not been studied before: J84.695932-69.083807, J84.688372-69.078168, J84.669113-69.081638, J84.674734-69.077374, J84.709403-69.075682, J84.688168-69.071013, J84.694286-69.074499, J84.676469-69.082774, and J84.671132-69.077168. We fit these sources to the \citet{robi06, robi07} spectral energy distribution (SED) fitter. We develop a code which we use in conjunction with the \citet{robi06} SED fitter. This code outputs the maximum likelihood of the YSO parameters such as mass, luminosity, accretion rate, and associated 1$\mathrm{\sigma}$ errors. Tables 4 and 5 list the near infrared and far infrared photometry we used. For candidates with no PACS or SPIRE photometry we use the upper limit in each far infrared band as given in Table 6 of \citet{meix13}. We assume the 6 previously identified YSO candidates are real candidates. For the 9 new point sources in the Spitzer footprint that have not been previously studied we require the reduced $\mathrm{\chi^{2}}$ of the best fit YSO SED to be less than 10 and the reduced $\mathrm{\chi^{2}}$ of the best fit stellar photosphere to be greater than 10. If an object is a good match to a YSO SED and not a good match to a reddened stellar photosphere SED, then we say it is a YSO candidate. With this definition we identify 4 new YSO candidates: J84.695932-69.083807, J84.688372-69.078168, J84.669113-69.081638, and J84.674734-69.077374. There are 5 \textit{Spitzer} point sources that do not meet our $\mathrm{\chi^{2}}$ criteria: J84.709403-69.075682, J84.688168-69.071013, J84.694286-69.074499, J84.676469-69.082774 and J84.671132-69.077168. We look at the $\mathrm{^{12}CO}$ (2-1) map and find that the emission at the location of these 5 sources is less than $3\mathrm{\sigma}$. Figures ~\ref{fig:d5} shows SEDs of all YSO candidates. Figure ~\ref{fig:d6} shows spectra for 3 massive YSO candidates.
\\
\\
\citet{carl12} also used multi-color cuts, but included more of a grey zone with $\mathrm{\alpha}$ and $\mathrm{\beta}$ cuts to include bluer sources at a lower reliability in comparison to \citet{grue09}. The \citet{carl12} cuts are based on \textit{Spitzer} IRAC (3.6 $\mu$m, 4.5 $\mu$m, 5.8 $\mu$m, 8.0 $\mu$m) and MIPS (24 $\mu$m) color-magnitude diagrams (CMDs). The $\mathrm{\alpha}$ cut criteria can be applied to galaxy-wide surveys and has a low contamination rate. Applying $\mathrm{\alpha}$ cuts selects the most massive and luminous YSOs. The $\mathrm{\beta}$ cut criteria should be applied to star-forming regions with less contamination from background sources, where it is more likely the cuts will select a YSO. Figure  ~\ref{fig:d8} shows [3.6] versus [3.5]-[5.8] CMD. YSO candidates numbered 1, 3, 4, and 7 have been previously studied \citep{grue09, seal09, seal14}. These four YSO candidates meet the \citet{carl12} $\mathrm{\alpha}$ cut criteria (all located to the right of the orange line). YSO candidates 2, 5, 8, and 10 have not been previously identified. YSO candidate 8 meets the \citet{carl12} $\mathrm{\alpha}$ cut criteria. YSO candidates 2, 5, and 10 meet the \citet{carl12} $\mathrm{\beta}$ cut criteria. The YSO numbering refers to ranking from most massive to least massive YSO, as given in Table 6.
\\
\\
The 6 previously identified YSO candidates and the 4 new YSO candidates in this work are all Stage I objects as given by the peak of the likelihood distribution of $\mathrm{\frac{\dot{M_{env}}}{M_{star}}}$. Our goal is to find out how massive star formation relates to molecular clumps. Even though we are most likely missing many later stage low-mass YSOs, our catalog of massive Stage I YSO candidates (objects $\textgreater 8\;\mathrm{M_{\odot}}$ in Figure ~\ref{fig:d7}) is complete. 
\\
\\
In addition to Stage, we can characterize the SED by Type. \citet{chen10} conduct an empirical analysis by looking at the shape of the SEDs to classify it by Type: Type I YSOs have SEDs with a steady rise in the near infrared and a high peak at 8-24 $\mu$m, Type II YSOs have low peak in the optical and a peak in the near infrared as well, Type III have SEDs peaking in the optical with some peak in the near or mid infrared wavelengths. There is correlation between Type I and Stage I YSOs, because these objects are deeply embedding within a molecular cloud. However there is little to no correlation between the any other Type and Stage classification. Below we list each of the YSO candidates in this study and describe the SED morphology in more detail.
\subsubsection{Details of Each YSO Candidate in Order of Most Massive to Least Massive}
\noindent
J84.703995-69.079110: This massive YSO candidate in our work is located in the central, most massive clump as can be seen in Figure ~\ref{fig:d1}. The spectra for this source is shown in the top panel of Figure ~\ref{fig:d6}. This YSO candidate is in the PE group \citep{seal09}. There is a silicon absorption feature visible, as well as strong PAH and fine-structure line emission. There is a steep rise in the SED as shown in Figure ~\ref{fig:d5}, and a peak in the mid infrared. This object is a Type I YSO. This object also meets one of the \citet{carl12} $\mathrm{\alpha}$ cut criteria. \citet{walb13} find there are two water masers associated with this object and the spectral type of the object is O2.
\\
\\
J84.695932-69.083807: This second most massive YSO candidate is located in clump 13 ($\mathrm{^{13}CO}$ (2-1) clump as listed in Table 1), as well as clumps 14 and 25. This object also meets one of the \citet{carl12} $\mathrm{\beta}$ cut criteria. There is a shallow rise from the near infrared to the far infrared wavelengths in the SED, therefore we say this is a Type II YSO. This object along with J84.695173-69.084857 is a double source and labeled as IRSN 122 and IRSN 126 by \citet{rubi08}.
\\
\\
J84.699755-69.069803: The SED of the YSO candidate shows a steep rise in the IRAC bands, therefore we say this is a Type I YSO. This object also meets one of the \citet{carl12} $\mathrm{\alpha}$ cut criteria. \citet{walb13} find this object to be a bright source surrounded by a cluster of faint sources, and the entire cluster is surrounded by a thick annulus of emission in the \textit{Spitzer}/MIPS 24$\mathrm{\mu m}$ image.
\\
\\
J84.688990-69.084695: \citet{seal09} categorize this object to be in the PE group. The spectra is shown in the central panel of Figure ~\ref{fig:d6}. There is no silicon absorption feature, however there are strong PAH and fine-structure line emission features. A lack of silicon absorption may mean this candidate is more evolved that the other two YSO candidates that have spectra in our work(J84.703995-69.079110 and J84.720292-69.077084). This YSO candidate is located inside clump 8 ($\mathrm{^{13}CO}$ (2-1) clump as listed in Table 1). There is a double peak in the SED of this object, with one peak in the near infrared and the other peak in the far infrared. Therefore we classify this objects as a Type II YSO candidate. his object also meets one of the \citet{carl12} $\mathrm{\alpha}$ cut criteria.
\\
\\
J84.695173-69.084857: This YSO candidate is located inside clump 15 ($\mathrm{^{13}CO}$ (2-1) clump as listed in Table 1). This object meets \citet{carl12} $\mathrm{\alpha}$ and $\mathrm{\beta}$ cut criteria. There is a shallow rise from the near infrared to the far infrared wavelengths in the SED, therefore we say this is a Type II YSO. This object and J84.695932-69.083807 are part of a double source as studied by \citet{walb13}.
\\
\\
J84.726173-69.082254: The SED of the YSO candidate shows a steep rise in the IRAC bands, therefore we say this is a Type I YSO.
\\
\\
J84.720292-69.077084: This object is in the PE group \citep{seal09}. The spectra shown in the bottom panel of Figure ~\ref{fig:d6} shown silicon absorption, PAH emission, and fine-structure line emission. This YSO candidate is located inside clump 9 ($\mathrm{^{13}CO}$ (2-1) clump as listed in Table 1). There is a steep rise in the SED as shown in Figure ~\ref{fig:d5}, and a peak in the far infrared. This object is a Type I YSO. This objects meets \citet{carl12} $\mathrm{\alpha}$ cut criteria. \citet{walb13} find a water maser near this object.
\\
\\
J84.688372-69.078168: This YSO candidate is located within $\mathrm{^{13}CO}$ (2-1) clumps 29 and 30. This object also meets one of the \citet{carl12} $\mathrm{\alpha}$ cut criteria. This YSO is Type I YSO due to the prominent peak of the SED in mid infrared wavelengths. There is a steep rise in the SED as shown in Figure ~\ref{fig:d5}, and a peak in the far infrared. This object is a Type I YSO.
\\
\\
J84.669113-69.081638:This YSO candidate is located within $\mathrm{^{13}CO}$ (2-1) clump 46 and meets one of the \citet{carl12} $\mathrm{\alpha}$ cut criteria. This object is a Type III YSO candidate due to the similar strength of the near infrared and far infrared peaks in the SED.
\\
\\
J84.674734-69.077374: This object meets the \citet{carl12} $\mathrm{\alpha}$ and $\mathrm{\beta}$ cut criteria. This object is a Type III YSO candidate due to the similar strength of the near infrared and far infrared peak in the SED.

\section{Results and Discussion}
\subsection{Size, Linewidth, and Mass Surface Density}
\noindent
The `universality' of Larson's law has been verified by other observations \citep{lada85, bola08, heye09} and simulations \citep{fede12, gold11}. \citet{shet12} study the central molecular zone (CMZ) and analyze $\mathrm{N_{2}H^{+}}$, $\mathrm{HCN}$, $\mathrm{H^{13}CN}$, and $\mathrm{HCO^{+}}$ using dendrograms. They find a slope of 0.67, 0.46, 0.78, 0.64 and a coefficient of 2.6, 3.8, 2.6, 2.1 for $\mathrm{N_{2}H^{+}}$, $\mathrm{HCN}$, $\mathrm{H^{13}CN}$, and $\mathrm{HCO^{+}}$ respectively. There is a similarity in the slope of the size-linewidth relation that is independent of the local environmental factors and independent of the particular molecular gas that is studied. \citet{bola08} find the slope to be 0.60 in extra-galactic clouds, \citet{heye09} find the slope to be 0.50 in Milky Way clouds, \citet{shet12} find the slope to range between 0.46 and 0.78 in the CMZ for several high density tracers. This agreement in the slope of the size-linewidth relation is theorized to be because of the universality of turbulence \citep{heye04}. 
\\
\\
Figure ~\ref{fig:d9} shows the corresponding size-linewidth relation of the $\mathrm{^{12}CO (2-1)}$ brightness temperature clumps. In order to define a radius we use the geometric mean of the major axis radius and minor axis radius of the best-fit ellipse to the clump. The best fit line to the $\mathrm{^{12}CO}$ (2-1) clumps in this study is given by the equation $\mathrm{\sigma = (1.66 \pm 0.06) r^{(0.65 \pm 0.04)}}$. We find the slope of the $\mathrm{^{12}CO}$ (2-1) size-linewidth relation to be similar to extragalactic clouds studied by \citet{bola08}, but the clumps in this 30 Doradus study have a systematic offset to larger linewidths by a factor of 3.8.
\\
\\
Figure ~\ref{fig:d10} shows the size-linewidth of $\mathrm{^{13}CO (2-1)}$ clumps analyzed using dendrograms and $\mathrm{^{13}CO (2-1)}$ clumps analyzed by \citet{inde13} using cprops \citet{roso06}. The astrodendro algorithm calculated smaller linewidths for a given clump size. This can be seen in Figure ~\ref{fig:d10} and is further explained in Appendix B: the astrodendro clumps have smaller linewidth for a given size. The best-fit line for the $\mathrm{^{13}CO (2-1)}$ clumps from this study is given by the solid black line ($\mathrm{\sigma = (1.58 \pm 0.18) r^{(0.97 \pm 0.12)}}$). The best-fit line for the $\mathrm{^{13}CO (2-1)}$ clumps from \citep{inde13} is given by the solid cyan line ($\mathrm{\sigma = (2.39 \pm 0.33) r^{(0.91 \pm 0.15)}}$). The cyan points in Figure ~\ref{fig:d10} use the same regions in position-position-velocity (PPV) space as the clumps defined using cprops and analyzed in \citep{inde13}, but a different method to calculate size and linewidth, to be more consistent with the astrodentro calculation. In this paper we use the same definition of size for both astrodendro clumps and cprops clumps. The size is calculated from the weighted second moment in two spatial directions. The radius is 1.91 times the geometric mean of those two spatial second moments \citep{solo87}. The linewidth is calculated using the weighted velocity second moment of the pixels assigned to the cprops clump or astrodendro structure. More details of the sizes and linewidths from cprops and astrodendro can be found in Appendix C. The analysis in this work is in agreement with \citet{inde13}: the $\mathrm{^{13}CO (2-1)}$ clumps derived from astrodendro lie in the same parameter space as the clumps analyzed by cprops \citep{inde13}.
\\
\\
One comparable analysis to our work in that of the Perseus cloud in the Milky Way by \citet{shet12}. The dashed line in Figure ~\ref{fig:d10} shows the results from \citet{shet12} who analyzed the Perseus cloud, an ordinary Milky Way molecular cloud, using dendrograms. The best-fit equation to the dash line is given by $\mathrm{\sigma = 0.62 r^{0.54}}$. The slope of the line that best fits the $\mathrm{^{13}CO (2-1)}$ clumps in this study is within $3\mathrm{\sigma}$ of the slope of the line that best fits the Perseus clump analysis. The linewidths for a given clumps size in 30 Doradus are larger by a factor of 2.5 than those in Perseus. Linewidths of clumps in high pressure, high density, high star-forming environments are on average higher than those in quiescent Milky Way clumps. This is similar to studies by \citet{shet12} who find $\mathrm{N_{2}H^{+}}$ and $\mathrm{HCN}$ clumps in the CMZ  are offset by a factor of 5 in comparison to $\mathrm{^{13}CO}$ clumps in Perseus. The results of this work and the results found by \citet{inde13} show 30 Doradus clumps to have larger linewidths in comparison to ordinary Milky Way clouds.  We use the same data as \citet{inde13} in this work. However we do a more thorough analysis: we re-calculate the `size' and `linewidth' values output from cprops in order to match the same definition by astrodendro, we fit a line through the cprops size-linewidth relation, and compare our results quantitatively to other studies. The results of this paper are different and supersede the results of Indebetouw et al. (2013).
\\
\\
Similar offset to higher linewidths has also been seen in infrared dark cloud studies in the Milky Way, as is pointed out by \citet{inde13}. There are star forming infrared dark clumps in our Galaxy with similar size, mass, linewidth, and mass surface density as those in an extreme environment like 30 Doradus. \citet{pere13} study infrared dark cloud SDC335.579-0.272, located 3.25 kpc away from the Sun. Their ALMA observations show filamentary collisions with two massive star-forming cores at the intersection of the collision. \citet{bont10} study six massive and dense cores located in Cygnus X. Cygnus X is located 1.7 kpc away and contains 40 known massive dense cores (representative of the earliest phase of massive star formation). \citet{gibs09} use the MSX database to probe the physical conditions of several dozens of infrared dark clouds. We plot the size-linewidth values of infrared dark cloud studies by \citet{gibs09, bont10, pere13} in Figure ~\ref{fig:d10}. The Milky Way infrared dark clumps are in the same size-linewidth parameter space as 30 Doradus $\mathrm{^{13}CO (2-1)}$ clumps. Clumps in 30 Doradus have a larger linewidths than typical Milky Way clumps, however they have similar linewidth as infrared dark clouds that are very likely forming protostars.
\\
\\
We investigate if the size-linewidth relation is dependent on the individual dendrogram structures. The size-linewidth relation of $\mathrm{^{13}CO}$ (2-1) column density clumps is $\mathrm{\sigma = (1.58 \pm 0.18) r^{(0.97 \pm 0.12)}}$ (Figure ~\ref{fig:d11}). The red squares are clumps in the red dendrogram structure in Figure ~\ref{fig:d3a} ($\mathrm{\sigma = (1.91 \pm 0.31) r^{(1.14 \pm 0.19)}}$), the green stars are those that are in the green structure in Figure ~\ref{fig:d3a} ($\mathrm{\sigma = (1.12 \pm 0.11) r^{(0.58 \pm 0.11)}}$), and the cyan circles are clumps in the cyan structure in Figure ~\ref{fig:d3a} ($\mathrm{\sigma = (1.25 \pm 0.69) r^{(0.60 \pm 0.48)}}$). For each structure traced in different colors, we fit a slope and intercept. We compare the intercepts with each other and all the intercept values are 1-2$\mathrm{\sigma}$ away from each other. We then compare the slopes of the line fit to the different structures and we find the slopes are also 1-2$\mathrm{\sigma}$ away from each other. We then investigate if the size-linewidth is dependent upon if a clump is forming a star or not. The size-linewidth relation of $\mathrm{^{13}CO}$ (2-1) clumps that are associated with YSOs ($\mathrm{\sigma = (1.58 \pm 0.18) r^{(0.90 \pm 0.17)}}$) and those that are not ($\mathrm{\sigma = (1.86 \pm 0.26) r^{(0.91 \pm 0.14)}}$) in Figure ~\ref{fig:d12}. The size-linewidth of $\mathrm{^{13}CO}$ (2-1) clumps is the same for clumps with YSOs and those without. Individual star forming clumps have different mass, temperature, pressure conditions. Clumps with star formation taking place within them, and those with no observable star formation also should have different conditions. We thought local variations due to properties of the cloud or star formation happening would result in a different size-linewidth relation. However these differences in the local environment are negligible in comparison to the overall conditions in the regions of 30 Doradus we study in this work.
\\
\\
Larson's first law suggests that the linewidth of a clump is only dependent on the size, $\mathrm{\sigma \propto r^{0.5}}$. We look into whether the universality of Larson's first law applied to the clumps in 30 Doradus. We assume the clumps are self-gravitating and say the observed cloud mass is the virial mass:
\begin{equation}
M_{obs}=\frac{5 \sigma^{2} r}{G}.
\end{equation}
We can substitute the molecular gas surface density ($\mathrm{\Sigma=\frac{M_{obs}}{\pi r^{2}}}$) and solve for the linewidth:
\begin{equation}
\sigma=\left(\frac{\pi G}{5}\right)^{0.5} \Sigma^{0.5} r^{0.5}.
\end{equation}
The linewidth of a clump depends on the spatial extent as well as the mass surface density. Figure ~\ref{fig:d13} shows the size-line width for $\mathrm{^{13}CO}$ (2-1) colorized by mass surface density: black diamonds for clumps with mass density less than 1000 $\mathrm{M_{\odot}\;pc^{-2}}$, red squares are clumps with mass density between 1000-2000 $\mathrm{M_{\odot}\;pc^{-2}}$, and cyan circles are clumps with mass density above 2000 $\mathrm{M_{\odot}\;pc^{-2}}$. We find clumps with larger sizes and linewidths tend to have larger mass surface densities.
\\
\\
We follow a similar approach to \citet{heye09} and look at the dependence of the size and linewidth on the mass surface density. \citet{heye09} find that Larson's scaling relations are not universal. If $\mathrm{\Sigma}$ in Equation 2 is constant, then you can recover Larson's first law ($\mathrm{\sigma=\nu\;r^{0.5}}$, where $\mathrm{\nu}$ is a constant). This means that $\mathrm{\frac{\sigma}{r^{0.5}}}$ = $(\mathrm{\frac{\pi G \Sigma}{5}})^{0.5}$ = $\mathrm{\nu}$, which is a constant. If Larson's first law is to hold, then we should see a flat horizontal line when we plot $\mathrm{\nu}$ versus $\mathrm{\Sigma}$. Figure ~\ref{fig:d14} shows that $\mathrm{\nu}$ is not constant with respect to $\mathrm{\Sigma}$, which is contradictory to Larson's first scaling law. We find that $\mathrm{\nu \propto \Sigma ^{0.45}}$. This is similar to the study done by \citet{heye09}. They too find a dependence of $\mathrm{\frac{\sigma}{r^{0.5}}}$ on $\mathrm{\Sigma}$ (Figure 7 in their paper). The slope of the line is not reported in their paper. We take the linewidth, size, and mass values they report and calculate $\mathrm{\nu}$ and $\mathrm{\Sigma}$. \citet{heye09} results show that $\mathrm{\nu \propto \Sigma ^{0.31}}$. Our findings of scaling relations between size, linewidth, and mass surface density lead us to the same conclusion as \citet{heye09}: Larson's velocity scaling relationships are not universal.
\subsection{Energy Balance}
\noindent
We can further analyze the dynamics of the clumps by comparing the kinetic energy and the gravitational energy. We define the virial parameter, $\alpha$, to equal to $\mathrm{\frac{5 \sigma^{2}\;r}{G\;M}}$ \citep{roso08} \citep{mcke92} - the ratio of the kinetic potential energy to the gravitational potential energy. When $\alpha \textless 1$, the object is self-gravitating. Figure ~\ref{fig:d15} shows the virial parameter as a function of clump mass on a log-log plot. The smaller mass clumps embedded within the larger mass clumps have a larger kinetic energy compare to the larger mass. We find larger clumps to have a larger gravitational potential energy than the smaller clumps: 72$\%$ of the $\mathrm{^{13}CO}$ (2-1) clumps have a virial parameter equal to 1 or less. Larger clumps in this study tend to have multiple YSO candidates, whereas smaller clumps tend to have one or no YSO candidates associated with them. 
\\
\\
We compare our results with Milky Way studies in order to see how star-forming clumps in two different galaxies differ from each other. We plot virial parameter versus clump mass of infrared dark clouds studies by \citet{gibs09, bont10, pere13} in Figure ~\ref{fig:d15}. These Milky Way clumps found via analysis of dark clouds could contain a few massive YSOs or hundreds of low-mass YSOs \citep{bont10}. Mass constraints on these stars have not been placed yet. ALMA observations of Dark cloud SDC335.579-0.272 show that it is one of the most massive, compact protostellar cores ever observed in the Galaxy and could potentially form an OB cluster similar to the Trapezium cluster in Orion \citep{pere13}. The virial parameter versus mass value of this cluster is located close to the gravitationally bound massive star forming clumps in Figure ~\ref{fig:d15} (the green star located at the bottom right). Orion is the nearest region of massive star formation, and therefore Orion and massive star formation sites similar to Orion provide the best comparison to 30 Doradus. Figure ~\ref{fig:d15} shows star formation can take place in a wide range of conditions within infrared dark clouds: there are several low-mass gravitationally-unbound clumps that are forming stars and gravitationally-bound clumps 1000s of solar masses in size that are forming stars. However the virial conditions of dark cloud SDC335.579-0.272, a potential massive star formation location in the Milky Way, seems similar to massive star forming clumps in 30 Doradus.
\\
\\
There are a total of 15 star forming clumps, out of which 14 have a virial parameter equal to 1 or less. The majority (75$\%$) of clumps with masses greater than log(clump mass)=3.5 are associated with a YSO, but only 17$\%$ of clumps with masses less than log(clump mass)=3.5 are associated with a newly forming star. It is possible that the $\mathrm{^{13}CO}$ (2-1) line is not fully thermalized, therefore leading to an underestimate of the true mass. If the mass we report is underestimated, then the virial parameter we calculate would be overestimated. The points in Figure ~\ref{fig:d15} would shift down and to the right in this hypothetical scenario if we could calculate the true mass. We would still find 14, or perhaps even all 15, of the star forming clumps to be gravitationally bound ($\alpha \textless 1$). We would also still find the majority of high-mass clumps are associated with star formation. We do not have the whole CO ladder to calculate the true mass, however the possibility of $\mathrm{^{13}CO}$ (2-1) not being fully thermalized does not affect our conclusions about the dynamical state of the clumps.
\subsection{Which Clumps Have Newly Forming Stars and Which Do Not}
\noindent
We use the effective radius of the column density clumps to determine if there are any YSOs associated with the clumps. We find 7/87 more evolved low-mass YSOs and 7/10 massive young YSO candidates are associated with clumps. The 7 massive YSOs that are found within a clump are: [1] J84.703995-69.079110 , [2] J84.695932-69.083807, [4] J84.688990-69.084695, [5] J84.695173-69.084857, [7] J84.720292-69.077084, [8] J84.688372-69.078168, [9] J84.669113-69.081638 (the numbers are the same as those in Table  6). The reason for a higher percentage of massive YSOs to be associated with clumps than low-mass YSOs is consistent with massive YSOs being at an earlier evolutionary stage. The molecular cloud surrounding stars quickly dissipate due to the strong UV radiation from the central stars. Massive YSOs are much younger than low-mass YSOs in this study and therefore still can sometimes be found in their parental cloud. We are missing low-mass embedded YSOs. \citet{tach01} study the Lupus star-forming region with $\mathrm{^{12}CO}$ NANTEN data and find 40$\%$ of the more evolved low-mass YSOs in their study are associated with clumps less than $\textless 20\;\mathrm{M_{\odot}}$, which suggests that more evolved low-mass YSOs form in small clouds and then the parental cloud rapidly dissipates. We see more evolved low-mass YSOs (stars less than 3 $\mathrm{M_{\odot}}$ in Figure \ref{fig:d17a}) projected against, and thus perhaps forming in clouds a few hundred solar masses to several tens of thousands solar masses. It is not clear if the lack of many more evolved low-mass YSOs associated with clumps is due to them moving away from the parental clump over time or if the parental clump of the young star has dissipated over time. Another reason of finding a smaller fraction of more evolved low-mass YSOs associated with clumps is because low-mass YSOs are fainter and harder to detect due to the sensitivity limits of the surveys. 
\\
\\
Figure ~\ref{fig:d16} shows the difference in clump mass and $\mathrm{H_{2}}$ mass surface density distributions between clumps that do contain massive or low-mass YSOs versus clumps that do not. The distributions of clumps with massive or low-mass YSOs and those without are very different. The mass distribution of clumps without massive or low-mass YSOs is quasi normal, and the mass surface density distribution of clumps without massive or low-mass YSOs is uniform. Both the mass distribution and the mass surface density distribution of clumps with massive or low-mass YSOs are mostly uniform, but do have a noticeable peak in the center. The mass distribution and mass surface density distribution of clumps with star formation span a higher range in values than those without star formation: for example Figure ~\ref{fig:d16} shows clumps without any stars have masses up to log(clump mass) = 3.75, but clumps with YSOs have masses up to log(clump mass) = 4.50. The average mass, column density, and mass surface density of clumps without any YSOs are 957 $\mathrm{M_{\odot}}$, $6.1 \times 10^{22}\;\mathrm{cm^{-2}}$, and 957 $\mathrm{M_{\odot}\;pc^{-2}}$. The distribution of clumps with massive or low-mass YSOs is centered at x6 higher mass, x2 higher column density, and x2 higher mass surface density. Figure ~\ref{fig:d17a} shows the mass of YSO versus the clump mass. Structures are designed such that the clumps may be concentric (one clump within another clump), therefore there are stars that are counted more than once. If a YSO is associated with a small clump, then it is also associated with the larger clump the smaller clump is embedded in. Only 20$\%$ of clumps smaller than 1778 $\mathrm{M_{\odot}}$ ($10^{3.25}\;\mathrm{M_{\odot}}$) have a YSO above 10 $\mathrm{M_{\odot}}$, whereas  66$\%$ of clumps larger than 1778 $\mathrm{M_{\odot}}$ have a YSO above 10 $\mathrm{M_{\odot}}$. Figure ~\ref{fig:d17b} shows that 46$\%$ of clumps with $\mathrm{H_{2}}$ mass surface density less than 2512 $\mathrm{M_{\odot}\;pc^{-2}}$ ($10^{3.40}\;\mathrm{M_{\odot}\;pc^{-2}}$) have a massive YSO, but 71$\%$ of of clumps with mass surface density more than 2512 $\mathrm{M_{\odot}\;pc^{-2}}$ have a massive YSO. Larger mass clumps are more likely to form larger mass stars than smaller mass clumps. 
\\
\\
The lowest surface density clump with a massive YSO forming in it provides us with a lower limit on the threshold above which high-mass star formation can occur: massive star formation appears to occur in clumps with mass $\textgreater 1.83 \times 10^{2}\;\mathrm{M_{\odot}}$, surface density $\textgreater 6.67 \times 10^{2}\;\mathrm{M_{\odot}\;pc^{-2}}$, and turbulence $\textgreater 1.18\;\mathrm{km/s}$. It is possible there are clumps with lower masses and lower surface densities forming low-mass YSOs. Future studies that find lower mass YSOs may push this mass surface density value lower. However, 30 Doradus is a very extreme star forming environment and more turbulent than typical GMCs in the Milky Way. The specific region we are analyzing in this study is 10 pc North of the R136 supercluster and is most likely influenced by the radiation field of R136. If there is more turbulence, then we would need higher densities and mass surface densities for star formation to occur. The mass surface density threshold in this study is 5 to 6 times higher than mass thresholds necessary to form stars as observed by \citet{heid10, lada10, kenn12}. \citet{heid10} use extinction maps to determine the gas surface density of Galactic star forming regions and find a threshold of 129 $\mathrm{M_{\odot}\;pc^{-2}}$. \citet{lada10} find similar values for the gas surface density: 116 $\mathrm{M_{\odot}\;pc^{-2}}$. The thresholds determined from Galactic studies applies to low-mass star formation. The surface density threshold of  $\textgreater 6.67 \times 10^{2}\;\mathrm{M_{\odot}\;pc^{-2}}$ we place in the ALMA footprint of 30 Doradus is for massive star formation.
\\
\\
\citet{dunh11} study high density gas tracer $\mathrm{NH_{3}}$ and measure an average mass density threshold of 176 $\mathrm{M_{\odot}\;pc^{-2}}$ for star forming clumps in the Milky Way. \citet{rath14} study nearby clouds and compare them to clouds in the CMZ. Observations of solar neighborhood clouds suggest a column density threshold of $1.4 \times 10^{22}\;\mathrm{cm^{-2}}$ \citep{lada10}. However the universality of this value is questioned since theoretical models predict the threshold for star formation is dependent on the density and the Mach number \citep{krum05}. Even though the CMZ has a much higher column density than $1.4 \times 10^{22}\;\mathrm{cm^{-2}}$, it is producing orders of magnitude less stars than predicted. \citet{rath14} find that the density threshold for star formation locally is $10^{4}\;\mathrm{cm^{-3}}$, however the CMZ with much higher turbulence has a density threshold of $10^{8}\;\mathrm{cm^{-3}}$. Star formation is dependent on the local environment, with regions of high turbulence having a higher threshold to overcome in order for the star formation process to occur.

\section{Conclusion}
\noindent
We look at YSOs that are within the ALMA CO footprint of 30 Doradus to conduct a comprehensive analysis of the molecular gas and the stars that form within them. We analyze the CO clumps using dendrograms. The $\mathrm{^{13}CO}$ (2-1) molecular gas clumps analyzed using dendrograms have sizes that range from 0.23 - 1.17 pc, masses that range from 53 - 19300 $\mathrm{M_{\odot}}$, and linewidths that range from 0.27 - 2.81 km/s. There are several conclusions we come to with our analysis of CO clumps and YSOs as listed below.
\\
\\
1. We find $\mathrm{^{13}CO}$ (2-1) clumps to have a larger linewidth for a given size than previous studies done with the Perseus cloud \citep{shet12}. A larger linewidth is not dependent on the size scale, and not dependent on if star formation is taking place in a clump or not. Local environmental factors (high star formation rate, high densities, and  high pressures) in 30 Doradus can explain why we find clumps to have a larger linewidth for a given size. Our result is similar to studies by \citet{shet12} who find clumps in the CMZ have a slope of 0.5 in the size-linewidth relation, but are offset higher compared to the Perseus cloud due to the local environmental conditions. We find the slope of the size-linewidth relation of $\mathrm{^{12}CO}$ (2-1) and $\mathrm{^{13}CO}$ (2-1) molecular gas to be consistent within 3$\mathrm{\sigma}$ of previous studies \citep{bola08, heye09, shet12}.
\\
\\
2. Higher mass clumps have a tendency to have a lower viral parameter and contain multiple YSOs in comparison to lower mass clumps.
\\
\\
3. We find a total of 10 massive YSOs (4 new YSO candidates) with masses between 8.5-24.0 $\mathrm{M_{\odot}}$, and 87 more evolved low-mass YSOs with masses between 1-3 $\mathrm{M_{\odot}}$ in the region of interest. 
\\
\\
4. There is a difference in distribution of the mass and $\mathrm{H_{2}}$ column density of clumps that do not contain any newly forming stars and those that do. The mass and column density of clumps without stars falls off quickly and tend to be lower on average than clumps with stars. The clumps that has the lowest surface density, but still hosts a massive YSO candidate provides us with lower limits necessary for star formation. It is necessary for clumps to have high masses ($\textgreater 1.83 \times 10^{2}\;\mathrm{M_{\odot}}$), high linewidths (v $\textgreater 1.18\;\mathrm{km/s}$), and high mass densities ($\textgreater 6.67 \times 10^{2}\;\mathrm{M_{\odot}\;pc^{-2}}$) in order for massive star formation to occur. This threshold was found by looking at the least massive $\mathrm{^{13}CO}$ clump that is forming a massive YSO.
\\
\\
5. A higher fraction of young and massive YSO candidates are found within clumps (7/10) than the more evolved low-mass YSOs (7/87). This is because the more evolved low-mass YSOs in this study are older, and therefore have moved away from their parental clump. Our inventory of young forming stars is not sensitive to the very embedded low-mass YSOs.
\\
\\
6. We look at the dependence of the size-linewidth relation to the mass surface density. There is a dependence of $\mathrm{\frac{\sigma}{r^{0.5}}}$ on $\mathrm{\Sigma}$. This is contradictory to Larson's scaling relationships. Assuming the clumps are is gravitational equilibrium and making relevant substitutions (see Section 5.1), we derive $\sigma=\left(\frac{\pi G}{5}\right)^{0.5} \Sigma^{0.5} r^{0.5}$. Larson's third scaling relation states that $\Sigma$ is approximately constant for all clouds. Re-arranging the equation we derived and taking into account Larson's third scaling relations leads to the conclusion that the slope of $\mathrm{\frac{\sigma}{r^{0.5}}}$ versus $\mathrm{\Sigma}$ should be 0. However we do not find the slope to be 0 (Figure 14). It is possible that Larson's scaling relationships are not universal.
\\
\\
7. The virial parameter of massive clumps in 30 Doradus forming multiple YSOs is similar to SDC335.579-0.272, an infrared dark cloud in the Milky Way that is most likely forming massive stars.

\section*{Acknowledgements}
\noindent
This research made use of astrodendro, a Python package to compute dendrograms of Astronomical data (http://www.dendrograms.org). This research also made use of Astropy (http://www.astropy.org). Meixner, Indebetouw, and Nayak were supported by NSF grant 1312902. Sabbi and Panagia were supported by STScI grant GO-12939. We are grateful for discussions with the following astronomers while writing this paper:  Dr. Marta Sewilo, Dr. Olivia Jones, Dr. Bram Ochsendorf, Kirill Tchernyshov.

\section*{Appendix}
\appendix

\section{Calculating $\mathrm{H_{2}}$ mass using $\mathrm{N_{13CO}}$}
\noindent
We assume that $\mathrm{^{13}CO}$ is optically thin. The optical depth of $\mathrm{^{13}CO}$ is given by:
\begin{equation}
\mathrm{\tau_{^{13}CO(2-1)}} = \mathrm{-ln[1-\frac{T_{B,^{13}CO(2-1)}}{10.6}([e^{10.6/T_{ex,^{12}CO(2-1)}}-1]^{-1}-0.02)^{-1}]},
\end{equation}
where the excitation temperature is given by:
\begin{equation}
\mathrm{T_{ex,^{12}CO(2-1)}}=\mathrm{\frac{10.6\;K}{ln(1+\frac{10.6\;K}{T_{B,^{12}CO(2-1)} + 0.21\;K})}}.
\end{equation}
The $\mathrm{^{12}CO}$ excitation temperature is only dependent on the $\mathrm{^{12}CO}$ brightness temperature because we assume $\mathrm{^{12}CO}$ to be optically thick. The $\mathrm{^{12}CO}$ is likely to be optically thick if $\mathrm{^{13}CO}$ is present because a larger column of molecular has is needed to see $\mathrm{^{13}CO}$, as oppose to $\mathrm{^{12}CO}$. The $\mathrm{^{13}CO}$ map is the mask and we do not impose an artificial mask on the excitation temperature. There are places where the optical depth (equation A1) does not make sense, and therefore gives a column density value that is not believable, which has been masked out. For example, there are pixels for which the optical depth is negative or infinity because of the natural log and an exponent of -1 in equation A1. This is because of how you clean the ALMA data. These pixels with negative values of values that equal infinity are not real or believable, but rather an artifact of the data cube quality. Further assumptions of the local thermodynamic equilibrium (or LTE) analysis include: $\mathrm{^{13}CO}$ has the same excitation temperature as $\mathrm{^{12}CO}$, CO is thermalized (level population is described by the Boltzmann equation), and the abundance ratio between $\mathrm{^{12}CO}$ and $\mathrm{^{13}CO}$ is constant. The total $\mathrm{^{13}CO}$ column density is then given by the equation:
\begin{equation}
\mathrm{N_{^{13}CO,tot}}=\mathrm{\frac{1.5 \times 10^{14}\;T_{ex}\;e^{5.3/T_{ex}}\;\int \tau dv}{1-e^{-10.6/T_{ex}}}}.
\end{equation}
We assume $\mathrm{\frac{H_{2}}{^{13}CO}}=5 \times 10^{2}$ \citep{inde13} to convert the $\mathrm{^{13}CO}$ column density to $\mathrm{H_{2}}$ column density. The total $\mathrm{H_{2}}$ mass is then derived by:
\begin{equation}
\mathrm{M}=\mathrm{m_{H_{2}}\;\int N_{H_{2}} dA}.
\end{equation}

\section{Comparison Between Cprops and Astrodendro}
\noindent
Due to the different algorithm used by \citet{inde13} (cprops), it is not possible to have a 1:1 relation between the clumps identified in this work and those published by Indebetouw and collaborators. Therefore we pick three clumps we identify by eye that are similar in the output of both programs to compare to each other. We compare the clump properties listed in Table 2 in this work, to the clump properties from \citet{inde13}. We used the same definition of sizes and linewidths when comparing clump properties by \citet{inde13} and the results from this work. We got the size and linewidth values of $\mathrm{^{13}CO}$ clumps output from cprops from our personal correspondence with Remy Indebetouw. Definition of sizes and linewidths is further described in Appendix C. The two algorithms (cprops and \textit{astrodendro}) work in different ways and therefore not possible to select a clump output from one program, and match it to the same exact clump output from the other program. And we cannot do this for all the leaves and branches output from the dendrogram program. We match the $\mathrm{^{13}CO}$ clumps output from cprops to the same clumps identified by astrodendro, i.e. we search for clumps output from the two different algorithms and find those that are within 1 clump radii of each other. There are a total of 8 such clumps we can match from the two different algorithms and their properties are listed below. The purpose of this comparison is to see if we can find clump mass and size properties output from the two programs that are not different by orders of magnitude.
\\
\\
For the first clump listed in the table below we find that the astrodendro picks out a size that is twice as high as the size calculated from cprops, however the linewidth calculated by astrodendro is about 70$\%$ of the linewidth calculated by cprops. The next four clumps listed in the table below are comparable in size output from cprops and astrodendro, however the linewidths calculated by astrodendro are lower by 60-80$\%$. Clumps selected by astrodendro that are larger or comparable in size to clumps selected by cprops tend to have lower linewidths. The next two in the table below are $\mathrm{^{13}CO}$ clumps selected by astrodendro that are smaller in size than the same clumps selected by cprops. The size derived by astrodendro for the two clumps is smaller by a fact of 60-70$\%$, however the linewidths are smaller by a factor of 40-50$\%$. Astrodendro systematically calculates smaller linewidths, which explains the offset of the \citet{inde13} clumps from the clumps we derived in this paper. If we assume the clumps selected with astrodendro have lower linewidths than the clumps selected with cprops by a factor of 60-80$\%$ (as in the case for comparable size clumps), we can calculate this shift in the cprops clumps shown in cyan in Figure 10. The equation of the best fit line going through the cprops clumps in Figure 10 is given by $\mathrm{\sigma = (2.39 \pm 0.33) r^{(0.91 \pm 0.15)}}$. Shifting this line down by 60-80$\%$ changes the intercept of 2.39 to be between 1.69 to 1.99. The equation of the best fit line going through the astrodendro clumps in Figure 10 is given by $\mathrm{\sigma = (1.58 \pm 0.18) r^{(0.97 \pm 0.12)}}$. After taking into account the offset to lower linewidths inherent to astrodendro, the best fit-line going through the $\mathrm{^{13}CO}$ cprops clumps and the best-fit line going through the $\mathrm{^{13}CO}$ astrodendro clumps are in agreement within 1$\sigma$.

\setcounter{table}{0} 
\renewcommand{\thetable}{B.\arabic{table}}
\tabletypesize{\tiny}
\begin{deluxetable}{cccccccccc}
\tablecaption{Comparing Cprops and \textit{Astrodendro} Outputs}
\tablehead{\colhead{Size [$\mathrm{pc}$]} & \colhead{$\sigma$ [$\mathrm{km/s}$]} & \colhead{$\mathrm{^{12}CO}$} & \colhead{$\mathrm{^{12}CO}$} & \colhead{$\mathrm{^{12}CO}$} & \colhead{Mass Derived} & \colhead{$\mathrm{^{13}CO}$} & \colhead{$\mathrm{^{13}CO}$} & \colhead{$\mathrm{^{13}CO}$} & \colhead{Mass Derived}\\ 
 \colhead{From Cprops} & \colhead{From Cprops} &  \colhead{ID} & \colhead{ Size [$\mathrm{pc}$]} & \colhead{$\sigma$ [$\mathrm{km/s}$]} & \colhead{From $\mathrm{^{12}CO}$ [$\mathrm{M_{\odot}}$]} & \colhead{ID} & \colhead{Size [$\mathrm{pc}$]} & \colhead{$\sigma$ [$\mathrm{km/s}$]} & \colhead{From $\mathrm{^{13}CO}$ [$\mathrm{M_{\odot}}$]}}
\startdata
0.35 & 0.57 & 26 & 0.34 & 0.68 & 259 & 0 & 0.73 & 0.39 & 297 \\
0.28 & 0.87 & 53 & 0.17 & 0.94 & 62 & 16 & 0.32 & 0.55 & 112\\
0.32 & 0.94 & 69 & 0.32 & 0.95 & 359 & 22 & 0.32 & 0.73 & 612\\
0.36 & 0.78 & 28 & 0.31 & 0.84 & 271 & 4 & 0.35 & 0.48 & 281 \\
0.48 & 1.15 & 94 & 0.31 & 0.98 & 851 & 41 & 0.42 & 0.82 & 2050\\
0.44 & 1.37 & 67 & 0.20 & 0.65 & 206 & 15 & 0.29 & 0.58 & 385\\
0.39 & 0.96 & 80 & 0.41 & 1.0 & 681 & 27 & 0.28 & 0.48 & 232\\
0.51 & 1.36 & 98 & 0.19 & 0.47 & 125 & 39 & 0.34 & 1.07 & 934\\
\enddata
\tablecomments{Column 1: Clump size derived from cprops. Column 2: The linewidth derived from cprops. Column 3: $\mathrm{^{12}CO}$ clump ID as listed in Table 3. Column 4: $\mathrm{^{12}CO}$ clump radius as listed in Table 3. Column 5: The linewidth of $\mathrm{^{12}CO}$ clumps in this study. Column 6: $\mathrm{H_{2}}$ mass derived using $\mathrm{^{12}CO}$. Column 7: $\mathrm{^{13}CO}$ clump ID as listed in Table 4. Column 8: $\mathrm{^{13}CO}$ clump radius. Column 9: The linewidth of $\mathrm{^{13}CO}$ clumps in this study. Column 10: $\mathrm{H_{2}}$ mass derived using $\mathrm{^{13}CO}$.}
\end{deluxetable}

\section{Defining Sizes and Linewidths}
\noindent
There are many definitions of clump `sizes' and many ways to calculate a `linewidth' of a data cube. These definitions vary significantly in the literature. In \citet{inde13}, the size was measured by fitting a 2-dimensional ellipse to the half-power contour of each clump.  The ellipse was then deconvolved in two dimensions with the beamsize, and the radius quoted in \citet{inde13} was 1.91 times the geometric mean of that deconvolved ellipse semimajor and semiminor axes. At time of writing \citet{inde13}, the ellipse-fitting measure of size was compared to spatial moments and to fitting a two dimensional Gaussian to the spatial distribution of emission, and found to be in agreement within error, but the most robust of the three methods. In this paper, for both the \citet{inde13} clumps and those defined by astrodendro, the size is calculated from the weighted second moment in two spatial directions, specifically the direction of greatest spatial extent and perpendicular to that.  The radius is then 1.91 times the geometric mean of those two spatial second moments. In particular it should be noted that the current numbers thus include convolution by the $\sim$ 0.5pc beam.
\\
\\
In \citet{inde13}, the linewidth was calculated for each clump by extracting the `column' through the cube, i.e. for each pixel in the clump, the pixels covering the full velocity range of the cube.  A Gaussian was then fitted to the spectrum of that `column', and the width, deconvolved by the spectral resolution of 0.35km/s, was reported. The fitted spectrum thus includes velocity line wings which are not explicitly assigned to the clump by the decomposition.  At the time of writing \citet{inde13}, each spectrum was carefully examined, and verified that effects such as two clumps along the same line of sight artificially broadening the linewidth did not occur.  That Gaussian fitting linewidth calculation was compared to the weighted second moments in velocity of only the clump-assigned pixels, and of the entire `column'  through the cube. Results were within uncertainties, but fitting a Gaussian to the `column' was more robust. For the current comparison, we use the weighted velocity second moment of the pixels assigned to the cprops clump or astrodendro structure. The \citet{inde13} method could over-estimate the linewidth, if significant emission is along the line of sight, whereas the current method likely underestimates the linewidth of a structure, since PPV and PPP space do not precisely correspond.  As with the sizes, the current reported linewidths include the effects of finite spectral resolution, but that is less than a 2\% effect for the linewidths under consideration. The slight difference between the cprops and astrodendro 30 Doradus clumps seen in Figure ~\ref{fig:d10} can be attributed to the difference between the two algorithms, how they pick out the structure, and the input thresholds used.

\bibliographystyle{apj}
\bibliography{master}

\begin{thebibliography}{83}
\expandafter\ifx\csname natexlab\endcsname\relax\def\natexlab#1{#1}\fi

\bibitem[{{Alencar} {et~al.}(2005){Alencar}, {Basri}, {Hartmann}, \&
  {Calvet}}]{alen05}
{Alencar}, S.~H.~P., {Basri}, G., {Hartmann}, L., \& {Calvet}, N. 2005, \aap,
  440, 595

\bibitem[{{Alencar} {et~al.}(2001){Alencar}, {Johns-Krull}, \&
  {Basri}}]{alen01}
{Alencar}, S.~H.~P., {Johns-Krull}, C.~M., \& {Basri}, G. 2001, \aj, 122, 3335

\bibitem[{{Anderson} {et~al.}(2013){Anderson}, {Meier}, {Ott}, {Hughes}, \&
  {Wong}}]{ande13}
{Anderson}, C.~N., {Meier}, D.~S., {Ott}, J., {Hughes}, A., \& {Wong}, T. 2013,
  in IAU Symposium, Vol. 292, IAU Symposium, ed. T.~{Wong} \& J.~{Ott}, 95--95

\bibitem[{{Blake} {et~al.}(1987){Blake}, {Sutton}, {Masson}, \&
  {Phillips}}]{blak87}
{Blake}, G.~A., {Sutton}, E.~C., {Masson}, C.~R., \& {Phillips}, T.~G. 1987,
  \apj, 315, 621

\bibitem[{{Bolatto} {et~al.}(2008){Bolatto}, {Leroy}, {Rosolowsky}, {Walter},
  \& {Blitz}}]{bola08}
{Bolatto}, A.~D., {Leroy}, A.~K., {Rosolowsky}, E., {Walter}, F., \& {Blitz},
  L. 2008, \apj, 686, 948

\bibitem[{{Bonnell} {et~al.}(2001){Bonnell}, {Bate}, {Clarke}, \&
  {Pringle}}]{bonn01}
{Bonnell}, I.~A., {Bate}, M.~R., {Clarke}, C.~J., \& {Pringle}, J.~E. 2001,
  \mnras, 323, 785

\bibitem[{{Bontemps} {et~al.}(2010){Bontemps}, {Motte}, {Csengeri}, \&
  {Schneider}}]{bont10}
{Bontemps}, S., {Motte}, F., {Csengeri}, T., \& {Schneider}, N. 2010, \aap,
  524, A18

\bibitem[{{Bressan} {et~al.}(2012){Bressan}, {Marigo}, {Girardi}, {Salasnich},
  {Dal Cero}, {Rubele}, \& {Nanni}}]{bres12}
{Bressan}, A., {Marigo}, P., {Girardi}, L., {et~al.} 2012, \mnras, 427, 127

\bibitem[{{Carlson} {et~al.}(2012){Carlson}, {Sewilo}, {Meixner}, {Romita}, \&
  {Lawton}}]{carl12}
{Carlson}, L.~R., {Sewilo}, M., {Meixner}, M., {Romita}, K.~A., \& {Lawton}, B.
  2012, VizieR Online Data Catalog, 354, 29066

\bibitem[{{Chen} {et~al.}(2010){Chen}, {Indebetouw}, {Chu}, {Gruendl},
  {Testor}, {Heitsch}, {Seale}, {Meixner}, \& {Sewilo}}]{chen10}
{Chen}, C.-H.~R., {Indebetouw}, R., {Chu}, Y.-H., {et~al.} 2010, \apj, 721,
  1206

\bibitem[{{Chen} {et~al.}(2015){Chen}, {Bressan}, {Girardi}, {Marigo}, {Kong},
  \& {Lanza}}]{chen15}
{Chen}, Y., {Bressan}, A., {Girardi}, L., {et~al.} 2015, \mnras, 452, 1068

\bibitem[{{Chevance} {et~al.}(2016){Chevance}, {Madden}, {Lebouteiller},
  {Godard}, {Cormier}, {Galliano}, {Hony}, {Indebetouw}, {Le Bourlot}, {Lee},
  {Le Petit}, {Pellegrini}, {Roueff}, \& {Wu}}]{chev16}
{Chevance}, M., {Madden}, S.~C., {Lebouteiller}, V., {et~al.} 2016, ArXiv
  e-prints

\bibitem[{{Crowther} {et~al.}(2010){Crowther}, {Schnurr}, {Hirschi}, {Yusof},
  {Parker}, {Goodwin}, \& {Kassim}}]{crow10}
{Crowther}, P.~A., {Schnurr}, O., {Hirschi}, R., {et~al.} 2010, \mnras, 408,
  731

\bibitem[{{De Marchi} {et~al.}(2011{\natexlab{a}}){De Marchi}, {Panagia},
  {Romaniello}, {Sabbi}, {Sirianni}, {Prada Moroni}, \&
  {Degl'Innocenti}}]{dema11.3}
{De Marchi}, G., {Panagia}, N., {Romaniello}, M., {et~al.} 2011{\natexlab{a}},
  \apj, 740, 11

\bibitem[{{De Marchi} {et~al.}(2011{\natexlab{b}}){De Marchi}, {Panagia}, \&
  {Sabbi}}]{dema11.2}
{De Marchi}, G., {Panagia}, N., \& {Sabbi}, E. 2011{\natexlab{b}}, \apj, 740,
  10

\bibitem[{{De Marchi} {et~al.}(2011{\natexlab{c}}){De Marchi}, {Paresce},
  {Panagia}, {Beccari}, {Spezzi}, {Sirianni}, {Andersen}, {Mutchler}, {Balick},
  {Dopita}, {Frogel}, {Whitmore}, {Bond}, {Calzetti}, {Carollo}, {Disney},
  {Hall}, {Holtzman}, {Kimble}, {McCarthy}, {O'Connell}, {Saha}, {Silk},
  {Trauger}, {Walker}, {Windhorst}, \& {Young}}]{dema11.1}
{De Marchi}, G., {Paresce}, F., {Panagia}, N., {et~al.} 2011{\natexlab{c}},
  \apj, 739, 27

\bibitem[{{De Marchi} {et~al.}(2016){De Marchi}, {Panagia}, {Sabbi}, {Lennon},
  {Anderson}, {van der Marel}, {Cignoni}, {Grebel}, {Larsen}, {Zaritsky},
  {Zeidler}, {Gouliermis}, \& {Aloisi}}]{dema16}
{De Marchi}, G., {Panagia}, N., {Sabbi}, E., {et~al.} 2016, \mnras, 455, 4373

\bibitem[{{Dunham} {et~al.}(2011){Dunham}, {Rosolowsky}, {Evans}, {Cyganowski},
  \& {Urquhart}}]{dunh11}
{Dunham}, M.~K., {Rosolowsky}, E., {Evans}, II, N.~J., {Cyganowski}, C., \&
  {Urquhart}, J.~S. 2011, \apj, 741, 110

\bibitem[{{Feast}(1999)}]{feas99}
{Feast}, M. 1999, in IAU Symposium, Vol. 190, New Views of the Magellanic
  Clouds, ed. Y.-H. {Chu}, N.~{Suntzeff}, J.~{Hesser}, \& D.~{Bohlender}, 542

\bibitem[{{Federrath} \& {Klessen}(2012)}]{fede12}
{Federrath}, C., \& {Klessen}, R.~S. 2012, \apj, 761, 156

\bibitem[{{Fernandez} {et~al.}(1995){Fernandez}, {Ortiz}, {Eiroa}, \&
  {Miranda}}]{fern95}
{Fernandez}, M., {Ortiz}, E., {Eiroa}, C., \& {Miranda}, L.~F. 1995, \aaps,
  114, 439

\bibitem[{{Fukui} {et~al.}(2008){Fukui}, {Kawamura}, {Minamidani}, {Mizuno},
  {Kanai}, {Mizuno}, {Onishi}, {Yonekura}, {Mizuno}, {Ogawa}, \&
  {Rubio}}]{fuku08}
{Fukui}, Y., {Kawamura}, A., {Minamidani}, T., {et~al.} 2008, \apjs, 178, 56

\bibitem[{{Fukui} {et~al.}(2015{\natexlab{a}}){Fukui}, {Harada}, {Tokuda},
  {Morioka}, {Onishi}, {Torii}, {Ohama}, {Nayak}, {Meixner}, {Sewilo},
  {Indebetouw}, {Kawamura}, {Saigo}, {Yamamoto}, {Tachihara}, {Minamidani},
  {Inoue}, {Madden}, {Galametz}, {Lebouteiller}, {Mizuno}, \&
  {Chen}}]{fuku15.1}
{Fukui}, Y., {Harada}, R., {Tokuda}, K., {et~al.} 2015{\natexlab{a}}, ArXiv
  e-prints

\bibitem[{{Fukui} {et~al.}(2015{\natexlab{b}}){Fukui}, {Torii}, {Ohama},
  {Hasegawa}, {Hattori}, {Sano}, {Ohashi}, {Fujii}, {Kuwahara}, {Mizuno},
  {Dawson}, {Yamamoto}, {Tachihara}, {Okuda}, {Onishi}, \& {Mizuno}}]{fuku15.2}
{Fukui}, Y., {Torii}, K., {Ohama}, A., {et~al.} 2015{\natexlab{b}}, ArXiv
  e-prints

\bibitem[{{Gibson} {et~al.}(2009){Gibson}, {Plume}, {Bergin}, {Ragan}, \&
  {Evans}}]{gibs09}
{Gibson}, D., {Plume}, R., {Bergin}, E., {Ragan}, S., \& {Evans}, N. 2009,
  \apj, 705, 123

\bibitem[{{Goldbaum} {et~al.}(2011){Goldbaum}, {Krumholz}, {Matzner}, \&
  {McKee}}]{gold11}
{Goldbaum}, N.~J., {Krumholz}, M.~R., {Matzner}, C.~D., \& {McKee}, C.~F. 2011,
  \apj, 738, 101

\bibitem[{{Gordon} {et~al.}(2014){Gordon}, {Roman-Duval}, {Bot}, {Meixner},
  {Babler}, {Bernard}, {Bolatto}, {Boyer}, {Clayton}, {Engelbracht}, {Fukui},
  {Galametz}, {Galliano}, {Hony}, {Hughes}, {Indebetouw}, {Israel}, {Jameson},
  {Kawamura}, {Lebouteiller}, {Li}, {Madden}, {Matsuura}, {Misselt}, {Montiel},
  {Okumura}, {Onishi}, {Panuzzo}, {Paradis}, {Rubio}, {Sandstrom}, {Sauvage},
  {Seale}, {Sewi{\l}o}, {Tchernyshyov}, \& {Skibba}}]{gord14}
{Gordon}, K.~D., {Roman-Duval}, J., {Bot}, C., {et~al.} 2014, \apj, 797, 85

\bibitem[{{Gruendl} \& {Chu}(2009)}]{grue09}
{Gruendl}, R.~A., \& {Chu}, Y.-H. 2009, \apjs, 184, 172

\bibitem[{{Heckman} {et~al.}(1990){Heckman}, {Armus}, \& {Miley}}]{heck90}
{Heckman}, T.~M., {Armus}, L., \& {Miley}, G.~K. 1990, \apjs, 74, 833

\bibitem[{{Heger} {et~al.}(2003){Heger}, {Fryer}, {Woosley}, {Langer}, \&
  {Hartmann}}]{hege03}
{Heger}, A., {Fryer}, C.~L., {Woosley}, S.~E., {Langer}, N., \& {Hartmann},
  D.~H. 2003, \apj, 591, 288

\bibitem[{{Heiderman} {et~al.}(2010){Heiderman}, {Evans}, {Allen}, {Huard}, \&
  {Heyer}}]{heid10}
{Heiderman}, A., {Evans}, II, N.~J., {Allen}, L.~E., {Huard}, T., \& {Heyer},
  M. 2010, \apj, 723, 1019

\bibitem[{{Heyer} {et~al.}(2009){Heyer}, {Krawczyk}, {Duval}, \&
  {Jackson}}]{heye09}
{Heyer}, M., {Krawczyk}, C., {Duval}, J., \& {Jackson}, J.~M. 2009, \apj, 699,
  1092

\bibitem[{{Heyer} \& {Brunt}(2004)}]{heye04}
{Heyer}, M.~H., \& {Brunt}, C.~M. 2004, \apjl, 615, L45

\bibitem[{{Hopkins} {et~al.}(2012){Hopkins}, {Quataert}, \& {Murray}}]{hopk12}
{Hopkins}, P.~F., {Quataert}, E., \& {Murray}, N. 2012, \mnras, 421, 3522

\bibitem[{{Indebetouw} {et~al.}(2009){Indebetouw}, {de Messi{\`e}res},
  {Madden}, {Engelbracht}, {Smith}, {Meixner}, {Brandl}, {Smith}, {Boulanger},
  {Galliano}, {Gordon}, {Hora}, {Sewilo}, {Tielens}, {Werner}, \&
  {Wolfire}}]{inde09}
{Indebetouw}, R., {de Messi{\`e}res}, G.~E., {Madden}, S., {et~al.} 2009, \apj,
  694, 84

\bibitem[{{Indebetouw} {et~al.}(2013){Indebetouw}, {Brogan}, {Chen}, {Leroy},
  {Johnson}, {Muller}, {Madden}, {Cormier}, {Galliano}, {Hughes}, {Hunter},
  {Kawamura}, {Kepley}, {Lebouteiller}, {Meixner}, {Oliveira}, {Onishi}, \&
  {Vasyunina}}]{inde13}
{Indebetouw}, R., {Brogan}, C., {Chen}, C.-H.~R., {et~al.} 2013, \apj, 774, 73

\bibitem[{{Johansson} {et~al.}(1998){Johansson}, {Greve}, {Booth}, {Boulanger},
  {Garay}, {de Graauw}, {Israel}, {Kutner}, {Lequeux}, {Murphy}, {Nyman}, \&
  {Rubio}}]{joha98}
{Johansson}, L.~E.~B., {Greve}, A., {Booth}, R.~S., {et~al.} 1998, \aap, 331,
  857

\bibitem[{{Kafle} {et~al.}(2014){Kafle}, {Sharma}, {Lewis}, \&
  {Bland-Hawthorn}}]{kafl14}
{Kafle}, P.~R., {Sharma}, S., {Lewis}, G.~F., \& {Bland-Hawthorn}, J. 2014,
  \apj, 794, 59

\bibitem[{{Kauffmann} {et~al.}(2010){Kauffmann}, {Pillai}, {Shetty}, {Myers},
  \& {Goodman}}]{kauf10}
{Kauffmann}, J., {Pillai}, T., {Shetty}, R., {Myers}, P.~C., \& {Goodman},
  A.~A. 2010, \apj, 716, 433

\bibitem[{{Kennicutt} \& {Evans}(2012)}]{kenn12}
{Kennicutt}, R.~C., \& {Evans}, N.~J. 2012, \araa, 50, 531

\bibitem[{{Krause} {et~al.}(2013){Krause}, {Fierlinger}, {Diehl}, {Burkert},
  {Voss}, \& {Ziegler}}]{krau13}
{Krause}, M., {Fierlinger}, K., {Diehl}, R., {et~al.} 2013, \aap, 550, A49

\bibitem[{{Kroupa} {et~al.}(1993){Kroupa}, {Tout}, \& {Gilmore}}]{krou93}
{Kroupa}, P., {Tout}, C.~A., \& {Gilmore}, G. 1993, \mnras, 262, 545

\bibitem[{{Krumholz} \& {McKee}(2005)}]{krum05}
{Krumholz}, M.~R., \& {McKee}, C.~F. 2005, \apj, 630, 250

\bibitem[{{Kudritzki} \& {Puls}(2000)}]{kudr00}
{Kudritzki}, R.-P., \& {Puls}, J. 2000, \araa, 38, 613

\bibitem[{{Lada}(1985)}]{lada85}
{Lada}, C.~J. 1985, \araa, 23, 267

\bibitem[{{Lada} {et~al.}(2010){Lada}, {Lombardi}, \& {Alves}}]{lada10}
{Lada}, C.~J., {Lombardi}, M., \& {Alves}, J.~F. 2010, \apj, 724, 687

\bibitem[{{Lada}(1992)}]{lada92}
{Lada}, E.~A. 1992, \apjl, 393, L25

\bibitem[{{Lada} {et~al.}(1993){Lada}, {Strom}, \& {Myers}}]{lada93}
{Lada}, E.~A., {Strom}, K.~M., \& {Myers}, P.~C. 1993, in Protostars and
  Planets III, ed. E.~H. {Levy} \& J.~I. {Lunine}, 245--277

\bibitem[{{Leroy} {et~al.}(2011){Leroy}, {Bolatto}, {Gordon}, {Sandstrom},
  {Gratier}, {Rosolowsky}, {Engelbracht}, {Mizuno}, {Corbelli}, {Fukui}, \&
  {Kawamura}}]{lero11}
{Leroy}, A.~K., {Bolatto}, A., {Gordon}, K., {et~al.} 2011, \apj, 737, 12

\bibitem[{{Lopez} {et~al.}(2011){Lopez}, {Krumholz}, {Bolatto}, {Prochaska}, \&
  {Ramirez-Ruiz}}]{lope11}
{Lopez}, L.~A., {Krumholz}, M.~R., {Bolatto}, A.~D., {Prochaska}, J.~X., \&
  {Ramirez-Ruiz}, E. 2011, \apj, 731, 91

\bibitem[{{McKee} \& {Zweibel}(1992)}]{mcke92}
{McKee}, C.~F., \& {Zweibel}, E.~G. 1992, \apj, 399, 551

\bibitem[{{Meixner} {et~al.}(2006){Meixner}, {Gordon}, {Indebetouw}, {Hora},
  {Whitney}, {Blum}, {Reach}, {Bernard}, {Meade}, {Babler}, {Engelbracht},
  {For}, {Misselt}, {Vijh}, {Leitherer}, {Cohen}, {Churchwell}, {Boulanger},
  {Frogel}, {Fukui}, {Gallagher}, {Gorjian}, {Harris}, {Kelly}, {Kawamura},
  {Kim}, {Latter}, {Madden}, {Markwick-Kemper}, {Mizuno}, {Mizuno}, {Mould},
  {Nota}, {Oey}, {Olsen}, {Onishi}, {Paladini}, {Panagia}, {Perez-Gonzalez},
  {Shibai}, {Sato}, {Smith}, {Staveley-Smith}, {Tielens}, {Ueta}, {van Dyk},
  {Volk}, {Werner}, \& {Zaritsky}}]{meix06}
{Meixner}, M., {Gordon}, K.~D., {Indebetouw}, R., {et~al.} 2006, \aj, 132, 2268

\bibitem[{{Meixner} {et~al.}(2013){Meixner}, {Panuzzo}, {Roman-Duval},
  {Engelbracht}, {Babler}, {Seale}, {Hony}, {Montiel}, {Sauvage}, {Gordon},
  {Misselt}, {Okumura}, {Chanial}, {Beck}, {Bernard}, {Bolatto}, {Bot},
  {Boyer}, {Carlson}, {Clayton}, {Chen}, {Cormier}, {Fukui}, {Galametz},
  {Galliano}, {Hora}, {Hughes}, {Indebetouw}, {Israel}, {Kawamura}, {Kemper},
  {Kim}, {Kwon}, {Lebouteiller}, {Li}, {Long}, {Madden}, {Matsuura}, {Muller},
  {Oliveira}, {Onishi}, {Otsuka}, {Paradis}, {Poglitsch}, {Reach},
  {Robitaille}, {Rubio}, {Sargent}, {Sewi{\l}o}, {Skibba}, {Smith},
  {Srinivasan}, {Tielens}, {van Loon}, \& {Whitney}}]{meix13}
{Meixner}, M., {Panuzzo}, P., {Roman-Duval}, J., {et~al.} 2013, \aj, 146, 62

\bibitem[{{Panagia} {et~al.}(2000){Panagia}, {Romaniello}, {Scuderi}, \&
  {Kirshner}}]{pana00}
{Panagia}, N., {Romaniello}, M., {Scuderi}, S., \& {Kirshner}, R.~P. 2000,
  \apj, 539, 197

\bibitem[{{Pellegrini} {et~al.}(2011){Pellegrini}, {Baldwin}, \&
  {Ferland}}]{pell11}
{Pellegrini}, E.~W., {Baldwin}, J.~A., \& {Ferland}, G.~J. 2011, \apj, 738, 34

\bibitem[{{Peretto} {et~al.}(2013){Peretto}, {Fuller}, {Duarte-Cabral},
  {Avison}, {Hennebelle}, {Pineda}, {Andr{\'e}}, {Bontemps}, {Motte},
  {Schneider}, \& {Molinari}}]{pere13}
{Peretto}, N., {Fuller}, G.~A., {Duarte-Cabral}, A., {et~al.} 2013, \aap, 555,
  A112

\bibitem[{{Rathborne} {et~al.}(2014){Rathborne}, {Longmore}, {Jackson},
  {Kruijssen}, {Alves}, {Bally}, {Bastian}, {Contreras}, {Foster}, {Garay},
  {Testi}, \& {Walsh}}]{rath14}
{Rathborne}, J.~M., {Longmore}, S.~N., {Jackson}, J.~M., {et~al.} 2014, \apjl,
  795, L25

\bibitem[{{Reipurth} {et~al.}(1996){Reipurth}, {Pedrosa}, \& {Lago}}]{reip96}
{Reipurth}, B., {Pedrosa}, A., \& {Lago}, M.~T.~V.~T. 1996, \aaps, 120, 229

\bibitem[{{Robitaille} {et~al.}(2007){Robitaille}, {Whitney}, {Indebetouw}, \&
  {Wood}}]{robi07}
{Robitaille}, T.~P., {Whitney}, B.~A., {Indebetouw}, R., \& {Wood}, K. 2007,
  \apjs, 169, 328

\bibitem[{{Robitaille} {et~al.}(2006){Robitaille}, {Whitney}, {Indebetouw},
  {Wood}, \& {Denzmore}}]{robi06}
{Robitaille}, T.~P., {Whitney}, B.~A., {Indebetouw}, R., {Wood}, K., \&
  {Denzmore}, P. 2006, \apjs, 167, 256

\bibitem[{{Romaniello}(1998)}]{roma98}
{Romaniello}, M. 1998, PhD thesis, , Scuola Normale Superiore di Pisa, (1998)

\bibitem[{{Rosolowsky} \& {Leroy}(2006)}]{roso06}
{Rosolowsky}, E., \& {Leroy}, A. 2006, \pasp, 118, 590

\bibitem[{{Rosolowsky} {et~al.}(2008){Rosolowsky}, {Pineda}, {Kauffmann}, \&
  {Goodman}}]{roso08}
{Rosolowsky}, E.~W., {Pineda}, J.~E., {Kauffmann}, J., \& {Goodman}, A.~A.
  2008, \apj, 679, 1338

\bibitem[{{Rubio} {et~al.}(1998){Rubio}, {Barb{\'a}}, {Walborn}, {Probst},
  {Garc{\'{\i}}a}, \& {Roth}}]{rubi08}
{Rubio}, M., {Barb{\'a}}, R.~H., {Walborn}, N.~R., {et~al.} 1998, \aj, 116,
  1708

\bibitem[{{Sabbi} {et~al.}(2013){Sabbi}, {Anderson}, {Lennon}, {van der Marel},
  {Aloisi}, {Boyer}, {Cignoni}, {de Marchi}, {de Mink}, {Evans}, {Gallagher},
  {Gordon}, {Gouliermis}, {Grebel}, {Koekemoer}, {Larsen}, {Panagia}, {Ryon},
  {Smith}, {Tosi}, \& {Zaritsky}}]{sabb13}
{Sabbi}, E., {Anderson}, J., {Lennon}, D.~J., {et~al.} 2013, \aj, 146, 53

\bibitem[{{Sabbi} {et~al.}(2016){Sabbi}, {Lennon}, {Anderson}, {Cignoni}, {van
  der Marel}, {Zaritsky}, {De Marchi}, {Panagia}, {Gouliermis}, {Grebel},
  {Gallagher}, {Smith}, {Sana}, {Aloisi}, {Tosi}, {Evans}, {Arab}, {Boyer}, {de
  Mink}, {Gordon}, {Koekemoer}, {Larsen}, {Ryon}, \& {Zeidler}}]{sabb16}
{Sabbi}, E., {Lennon}, D.~J., {Anderson}, J., {et~al.} 2016, \apjs, 222, 11

\bibitem[{{Seale} {et~al.}(2009){Seale}, {Looney}, {Chu}, {Gruendl}, {Brandl},
  {Chen}, {Brandner}, \& {Blake}}]{seal09}
{Seale}, J.~P., {Looney}, L.~W., {Chu}, Y.-H., {et~al.} 2009, \apj, 699, 150

\bibitem[{{Seale} {et~al.}(2014){Seale}, {Meixner}, {Sewi{\l}o}, {Babler},
  {Engelbracht}, {Gordon}, {Hony}, {Misselt}, {Montiel}, {Okumura}, {Panuzzo},
  {Roman-Duval}, {Sauvage}, {Boyer}, {Chen}, {Indebetouw}, {Matsuura},
  {Oliveira}, {Srinivasan}, {van Loon}, {Whitney}, \& {Woods}}]{seal14}
{Seale}, J.~P., {Meixner}, M., {Sewi{\l}o}, M., {et~al.} 2014, \aj, 148, 124

\bibitem[{{Selman} \& {Melnick}(2013)}]{selm13}
{Selman}, F.~J., \& {Melnick}, J. 2013, \aap, 552, A94

\bibitem[{{Shetty} {et~al.}(2012){Shetty}, {Beaumont}, {Burton}, {Kelly}, \&
  {Klessen}}]{shet12}
{Shetty}, R., {Beaumont}, C.~N., {Burton}, M.~G., {Kelly}, B.~C., \& {Klessen},
  R.~S. 2012, \mnras, 425, 720

\bibitem[{{Smith} {et~al.}(1999){Smith}, {Lewis}, {Bonnell}, {Bunclark}, \&
  {Emerson}}]{smit99}
{Smith}, K.~W., {Lewis}, G.~F., {Bonnell}, I.~A., {Bunclark}, P.~S., \&
  {Emerson}, J.~P. 1999, \mnras, 304, 367

\bibitem[{{Solomon} {et~al.}(1987){Solomon}, {Rivolo}, {Barrett}, \&
  {Yahil}}]{solo87}
{Solomon}, P.~M., {Rivolo}, A.~R., {Barrett}, J., \& {Yahil}, A. 1987, \apj,
  319, 730

\bibitem[{{Sousa} {et~al.}(2016){Sousa}, {Alencar}, {Bouvier}, {Stauffer},
  {Venuti}, {Hillenbrand}, {Cody}, {Teixeira}, {Guimar{\~a}es}, {McGinnis},
  {Rebull}, {Flaccomio}, {F{\"u}r{\'e}sz}, {Micela}, \& {Gameiro}}]{sous16}
{Sousa}, A.~P., {Alencar}, S.~H.~P., {Bouvier}, J., {et~al.} 2016, \aap, 586,
  A47

\bibitem[{{Tachihara} {et~al.}(2001){Tachihara}, {Toyoda}, {Onishi}, {Mizuno},
  {Fukui}, \& {Neuh{\"a}user}}]{tach01}
{Tachihara}, K., {Toyoda}, S., {Onishi}, T., {et~al.} 2001, \pasj, 53, 1081

\bibitem[{{Tan} {et~al.}(2014){Tan}, {Beltr{\'a}n}, {Caselli}, {Fontani},
  {Fuente}, {Krumholz}, {McKee}, \& {Stolte}}]{tan14}
{Tan}, J.~C., {Beltr{\'a}n}, M.~T., {Caselli}, P., {et~al.} 2014, Protostars
  and Planets VI, 149

\bibitem[{{Walborn} {et~al.}(2013){Walborn}, {Barb{\'a}}, \&
  {Sewi{\l}o}}]{walb13}
{Walborn}, N.~R., {Barb{\'a}}, R.~H., \& {Sewi{\l}o}, M.~M. 2013, \aj, 145, 98

\bibitem[{{Westerlund}(1997)}]{west97}
{Westerlund}, B.~E. 1997, {The Magellanic Clouds}

\bibitem[{{Whitney} {et~al.}(2008){Whitney}, {Sewilo}, {Indebetouw},
  {Robitaille}, {Meixner}, {Gordon}, {Meade}, {Babler}, {Harris}, {Hora},
  {Bracker}, {Povich}, {Churchwell}, {Engelbracht}, {For}, {Block}, {Misselt},
  {Vijh}, {Leitherer}, {Kawamura}, {Blum}, {Cohen}, {Fukui}, {Mizuno},
  {Mizuno}, {Srinivasan}, {Tielens}, {Volk}, {Bernard}, {Boulanger}, {Frogel},
  {Gallagher}, {Gorjian}, {Kelly}, {Latter}, {Madden}, {Kemper}, {Mould},
  {Nota}, {Oey}, {Olsen}, {Onishi}, {Paladini}, {Panagia}, {Perez-Gonzalez},
  {Reach}, {Shibai}, {Sato}, {Smith}, {Staveley-Smith}, {Ueta}, {Van Dyk},
  {Werner}, {Wolff}, \& {Zaritsky}}]{whit08}
{Whitney}, B.~A., {Sewilo}, M., {Indebetouw}, R., {et~al.} 2008, \aj, 136, 18

\bibitem[{{Wong} {et~al.}(2011){Wong}, {Hughes}, {Ott}, {Muller}, {Pineda},
  {Bernard}, {Chu}, {Fukui}, {Gruendl}, {Henkel}, {Kawamura}, {Klein},
  {Looney}, {Maddison}, {Mizuno}, {Paradis}, {Seale}, \& {Welty}}]{wong11}
{Wong}, T., {Hughes}, A., {Ott}, J., {et~al.} 2011, \apjs, 197, 16

\bibitem[{{Woosley} {et~al.}(2002){Woosley}, {Heger}, \& {Weaver}}]{woos02}
{Woosley}, S.~E., {Heger}, A., \& {Weaver}, T.~A. 2002, Reviews of Modern
  Physics, 74, 1015

\bibitem[{{Woosley} \& {Weaver}(1995)}]{woos95}
{Woosley}, S.~E., \& {Weaver}, T.~A. 1995, \apjs, 101, 181

\bibitem[{{Xue} {et~al.}(2008){Xue}, {Rix}, {Zhao}, {Re Fiorentin}, {Naab},
  {Steinmetz}, {van den Bosch}, {Beers}, {Lee}, {Bell}, {Rockosi}, {Yanny},
  {Newberg}, {Wilhelm}, {Kang}, {Smith}, \& {Schneider}}]{xue08}
{Xue}, X.~X., {Rix}, H.~W., {Zhao}, G., {et~al.} 2008, \apj, 684, 1143

\bibitem[{{Yeh} {et~al.}(2015){Yeh}, {Seaquist}, {Matzner}, \&
  {Pellegrini}}]{yeh15}
{Yeh}, S.~C.~C., {Seaquist}, E.~R., {Matzner}, C.~D., \& {Pellegrini}, E.~W.
  2015, \apj, 807, 117

\end{thebibliography}

\noindent
\setcounter{figure}{0} 
\renewcommand{\thefigure}{\arabic{figure}}
\begin{figure*}
\centering
\includegraphics[scale=0.65,clip,trim=0cm 0cm 0cm 0cm,angle=0]{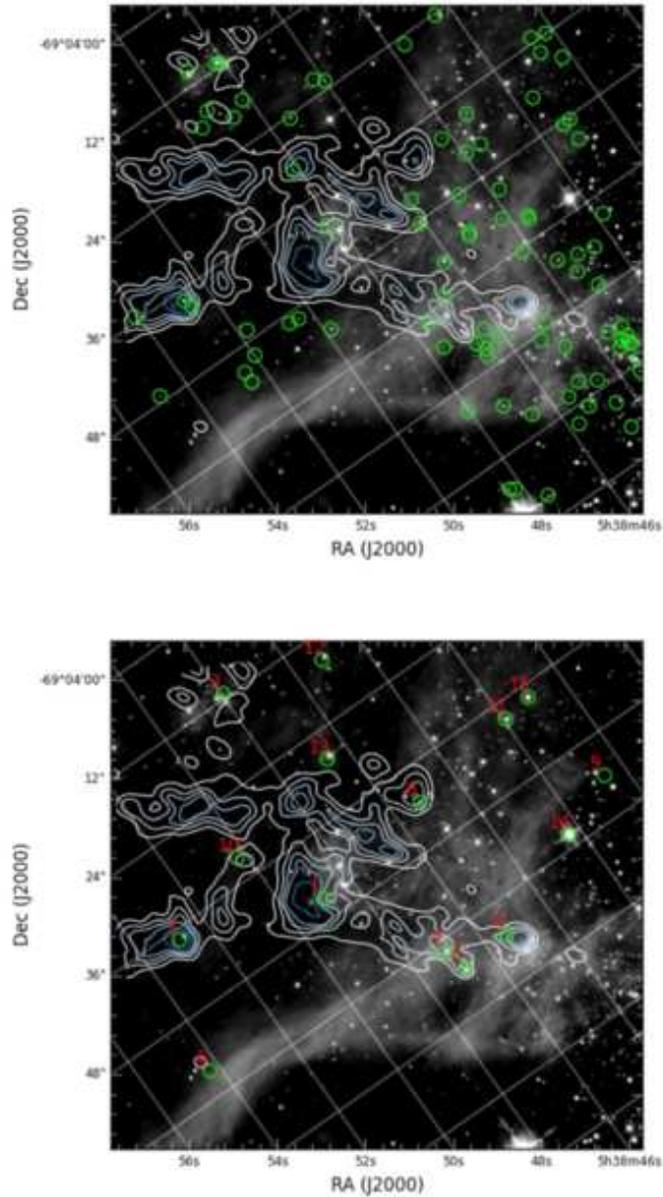}
\caption{Top: HST F160W in greyscale, ALMA+APEX $\mathrm{^{12}CO}$ (2-1) in contour (contour levels signify 10$\%$, 20$\%$, 30$\%$, 40$\%$, 60$\%$, and 80$\%$ of peak integrated flux), and the location of more evolved low-mass YSOs (as discussed in section 4.1) in green circles . North is to the upper-left in this figure. Bottom: HST F160W in greyscale, ALMA+APEX $\mathrm{^{12}CO}$ (2-1) in contour (contour levels signify 10$\%$, 20$\%$, 30$\%$, 40$\%$, 60$\%$, and 80$\%$ of peak integrated flux), and the location of all point sources in the SAGE catalog in green circles. Numbers 1-10 are massive YSO candidates (as discussed in section 4.2), and number 11-15 are SAGE point sources in the footprint we do not consider to be YSOs. The number near the YSO candidates indicates rank in mass (1 is the most massive YSO candidate and 10 in the least massive YSO candidate).} 
\label{fig:d1}
\end{figure*}

\noindent
\begin{figure*}
\setcounter{figure}{1} 
\renewcommand{\thefigure}{\arabic{figure} a}
\centering
\includegraphics[scale=0.65,clip,trim=0cm 0cm 0cm 0cm,angle=0]{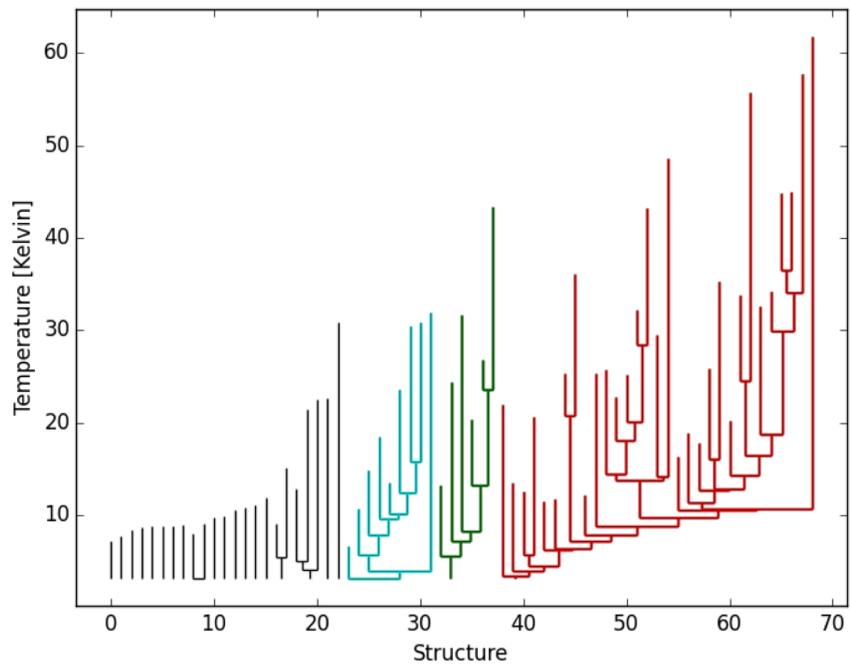}
\caption{Dendrogram hierarchy of $\mathrm{^{12}CO}$ (2-1) brightness temperature. The trunks represent the lowest lever of emission. The trunks are connected to leaves, which represent higher emission clumps.} 
\label{fig:d2a}
\end{figure*}

\noindent
\begin{figure}
\setcounter{figure}{1} 
\renewcommand{\thefigure}{\arabic{figure} b}
\centering
\includegraphics[scale=0.95,clip,trim=0cm 0cm 0cm 0cm,angle=0]{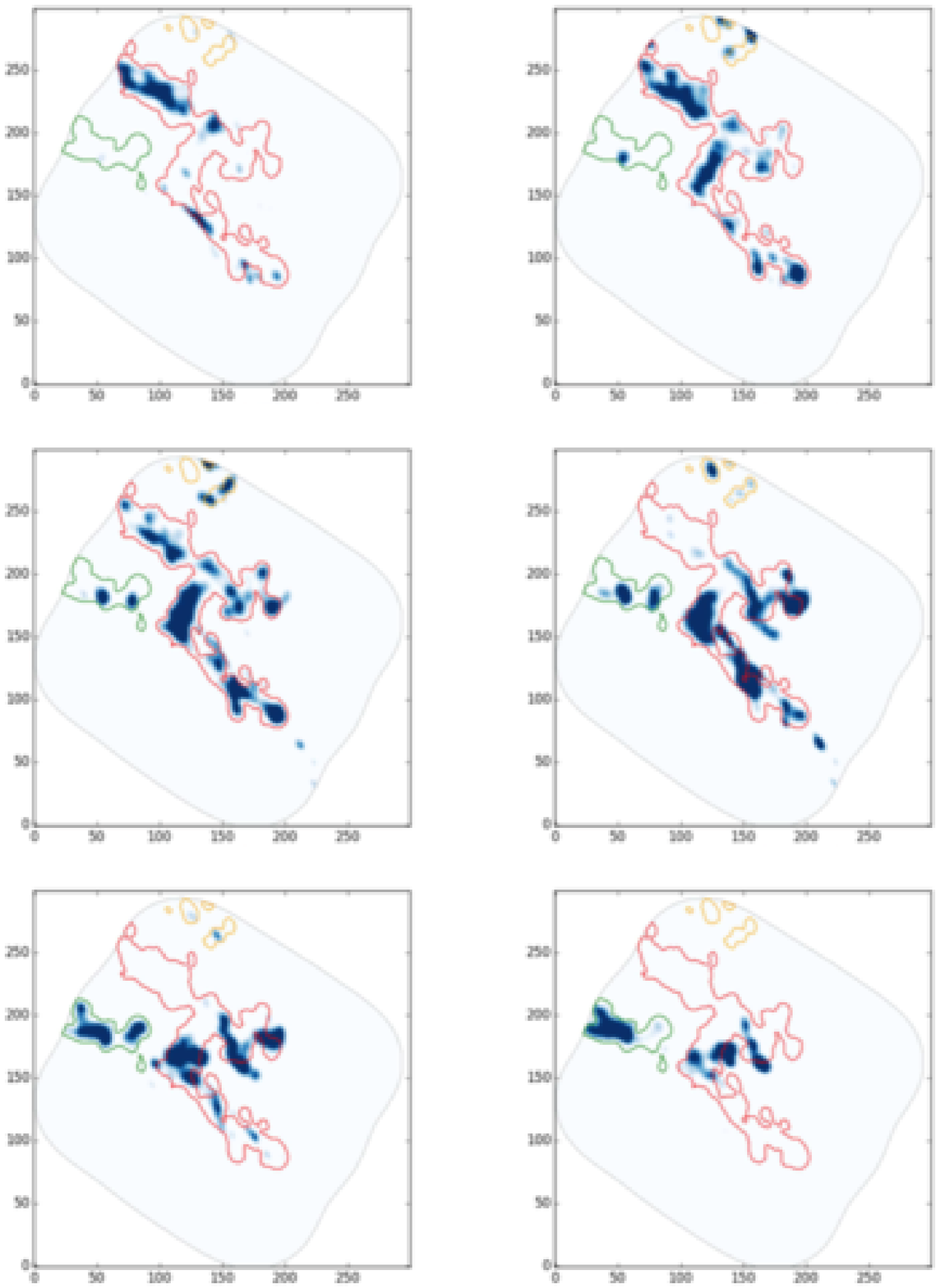}
\caption{Slices of the $\mathrm{^{12}CO}$ (2-1) brightness temperature velocity cube is shown in blue color. From left-to-right, top-to-bottom the channel velocities are 245.17 km/s, 247.17 km/s, 249.17 km/s, 251.17 km/s, 253.17 km/s, and 255.17 km/s. The red, green, and orange molecular structures highlighted correspond to the red, green, and cyan dendrogram structure in Figure ~\ref{fig:d2a}.} 
\label{fig:d2b}
\end{figure}

\noindent
\setcounter{figure}{2} 
\renewcommand{\thefigure}{\arabic{figure} a}
\begin{figure*}
\centering
\includegraphics[scale=0.65,clip,trim=0cm 0cm 0cm 0cm,angle=0]{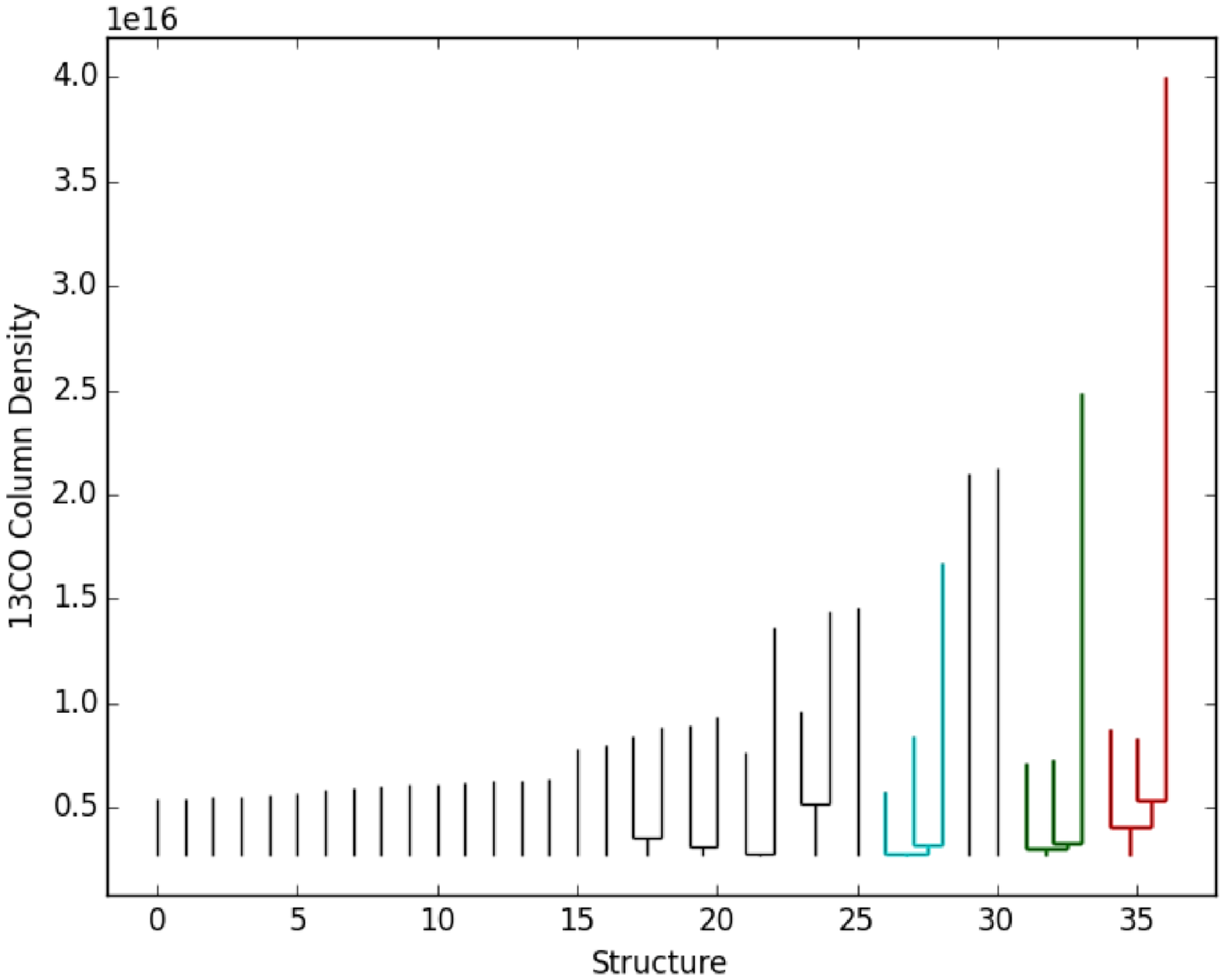}
\caption{Dendrogram hierarchy of $\mathrm{^{13}CO}$ (2-1) column density.} 
\label{fig:d3a}
\end{figure*}

\noindent
\begin{figure}
\setcounter{figure}{2} 
\renewcommand{\thefigure}{\arabic{figure} b}
\centering
\includegraphics[scale=0.95,clip,trim=0cm 0cm 0cm 0cm,angle=0]{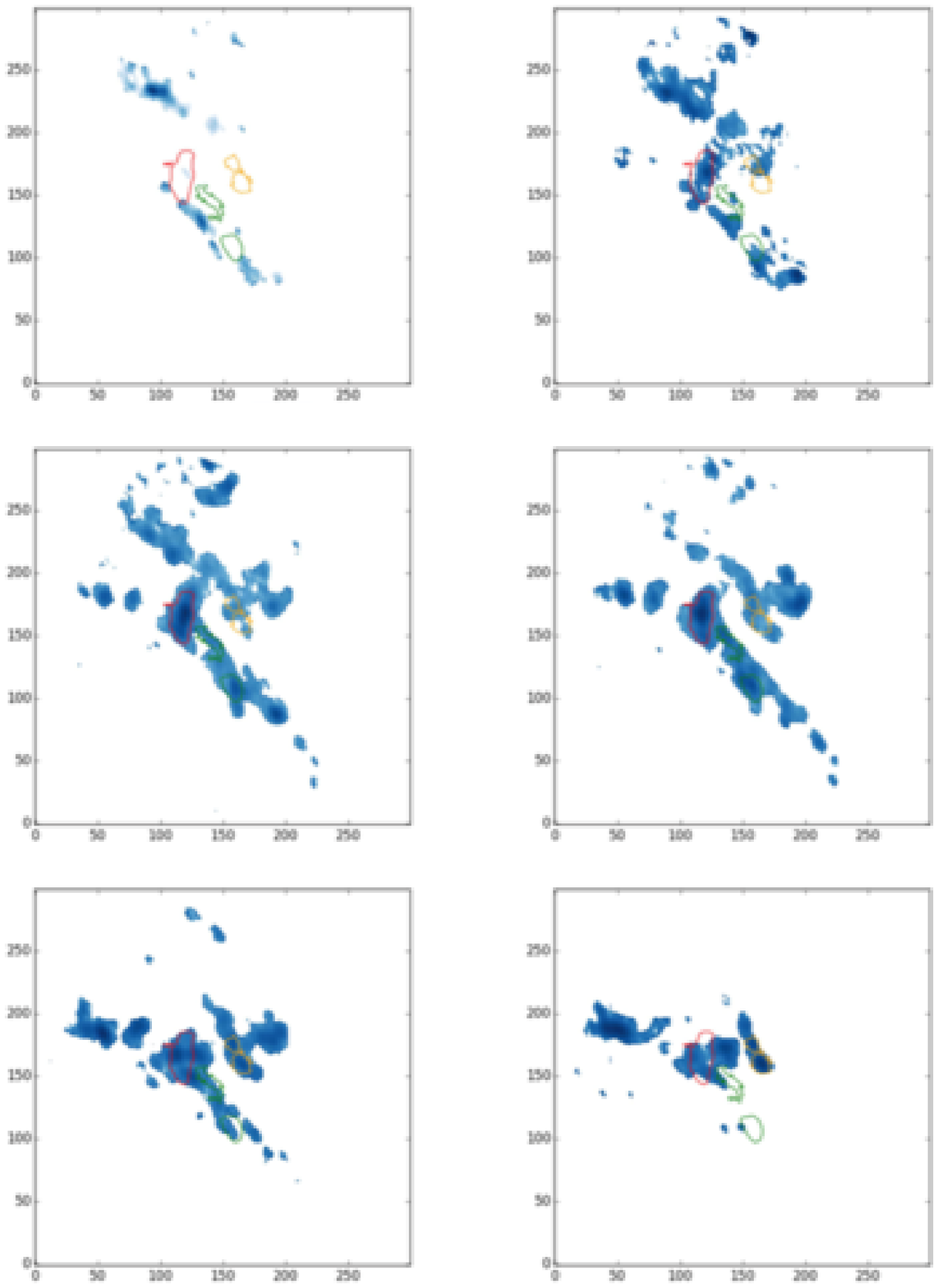}
\caption{Slices of the $\mathrm{^{13}CO}$ (2-1) column density velocity cube is shown in blue color. From left-to-right, top-to-bottom the channel velocities are 245.17 km/s, 247.17 km/s, 249.17 km/s, 251.17 km/s, 253.17 km/s, and 255.17 km/s. The red, green, and orange molecular structures highlighted correspond to the red, green, and cyan dendrogram structure in Figure ~\ref{fig:d3a}.} 
\label{fig:d3b}
\end{figure}

\noindent
\begin{figure*}
\setcounter{figure}{3} 
\renewcommand{\thefigure}{\arabic{figure} a}
\centering
\includegraphics[scale=0.65,clip,trim=0cm 0cm 0cm 0cm,angle=0]{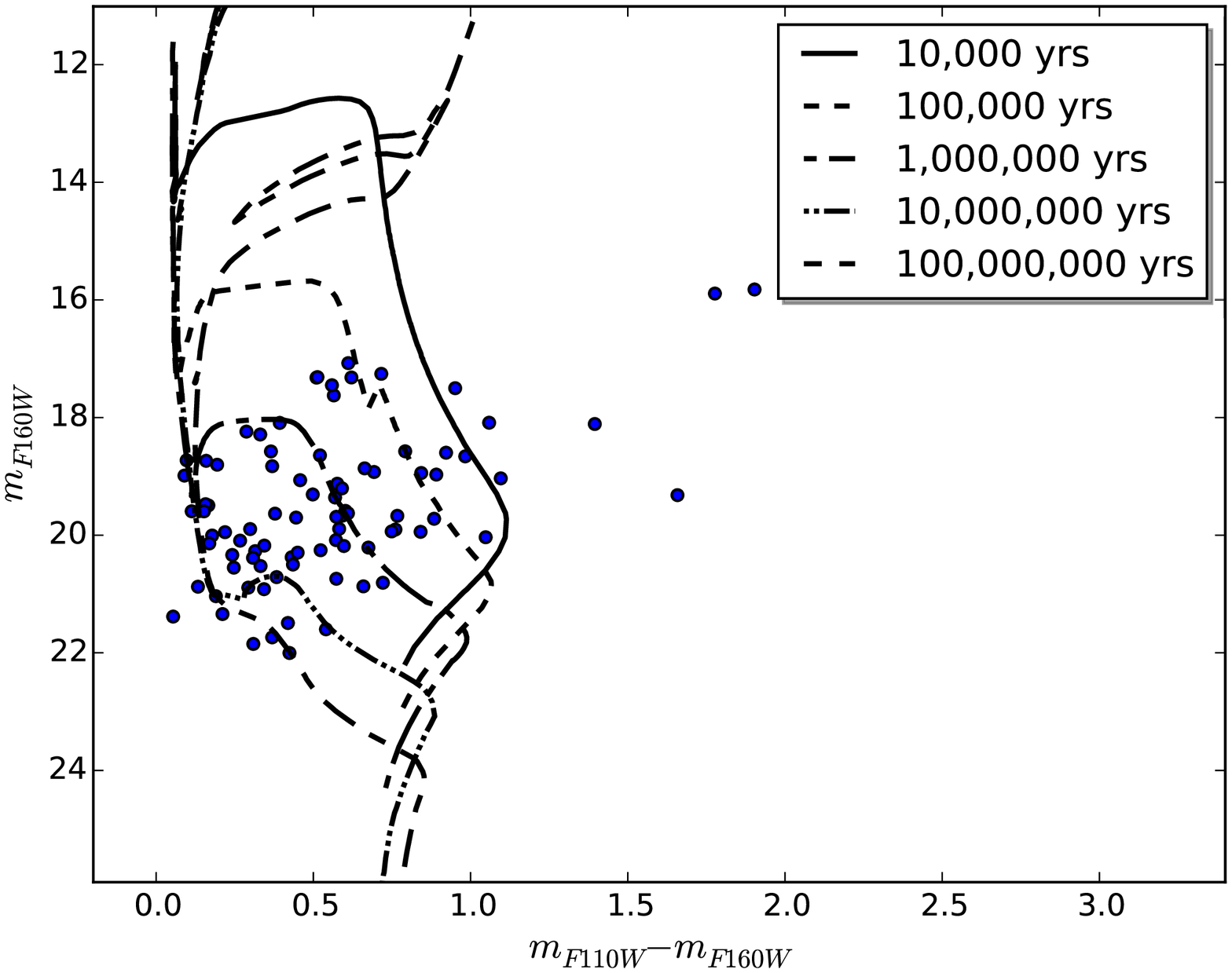}
\caption{F160W versus F110W - F160W CMD with isochrones. The isochrones are from the Padova code \citep{bres12, chen15}. The points are more evolved low-mass YSOs that have been selected via H$\mathrm{\alpha}$ excess and have been de-reddened \citep{dema16}, as discussed in section 4.1.} 
\label{fig:d4a}
\end{figure*}

\noindent
\begin{figure*}
\setcounter{figure}{3} 
\renewcommand{\thefigure}{\arabic{figure} b}
\centering
\includegraphics[scale=0.65,clip,trim=0cm 0cm 0cm 0cm,angle=0]{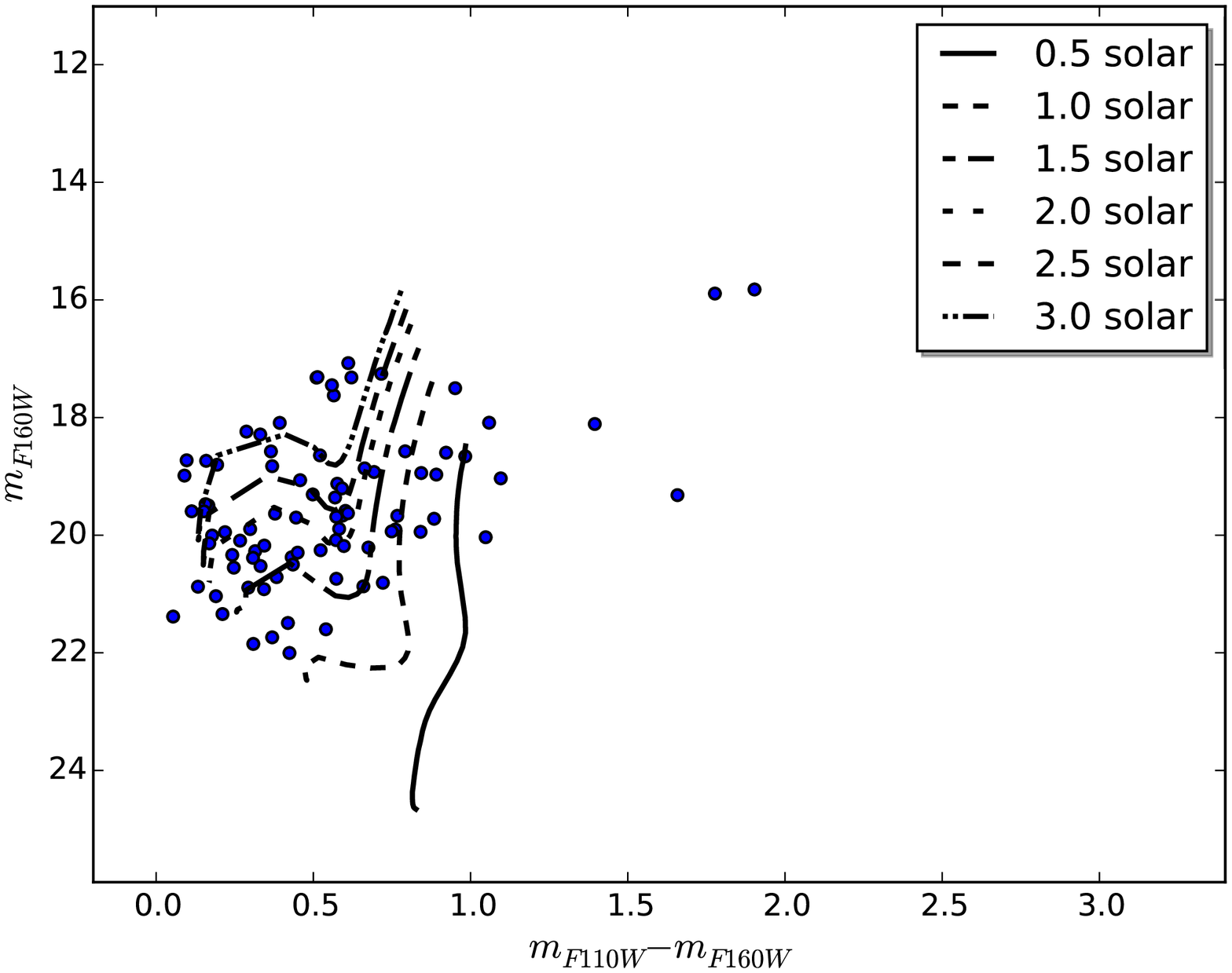}
\caption{This is the same as Figure ~\ref{fig:d4a}, however we plot F160W versus F110W - F160W CMD.} 
\label{fig:d4b}
\end{figure*}

\noindent
\begin{figure}
\setcounter{figure}{4} 
\renewcommand{\thefigure}{\arabic{figure}}
\centering
\includegraphics[scale=0.40,clip,trim=0cm 0cm 0cm 0cm,angle=0]{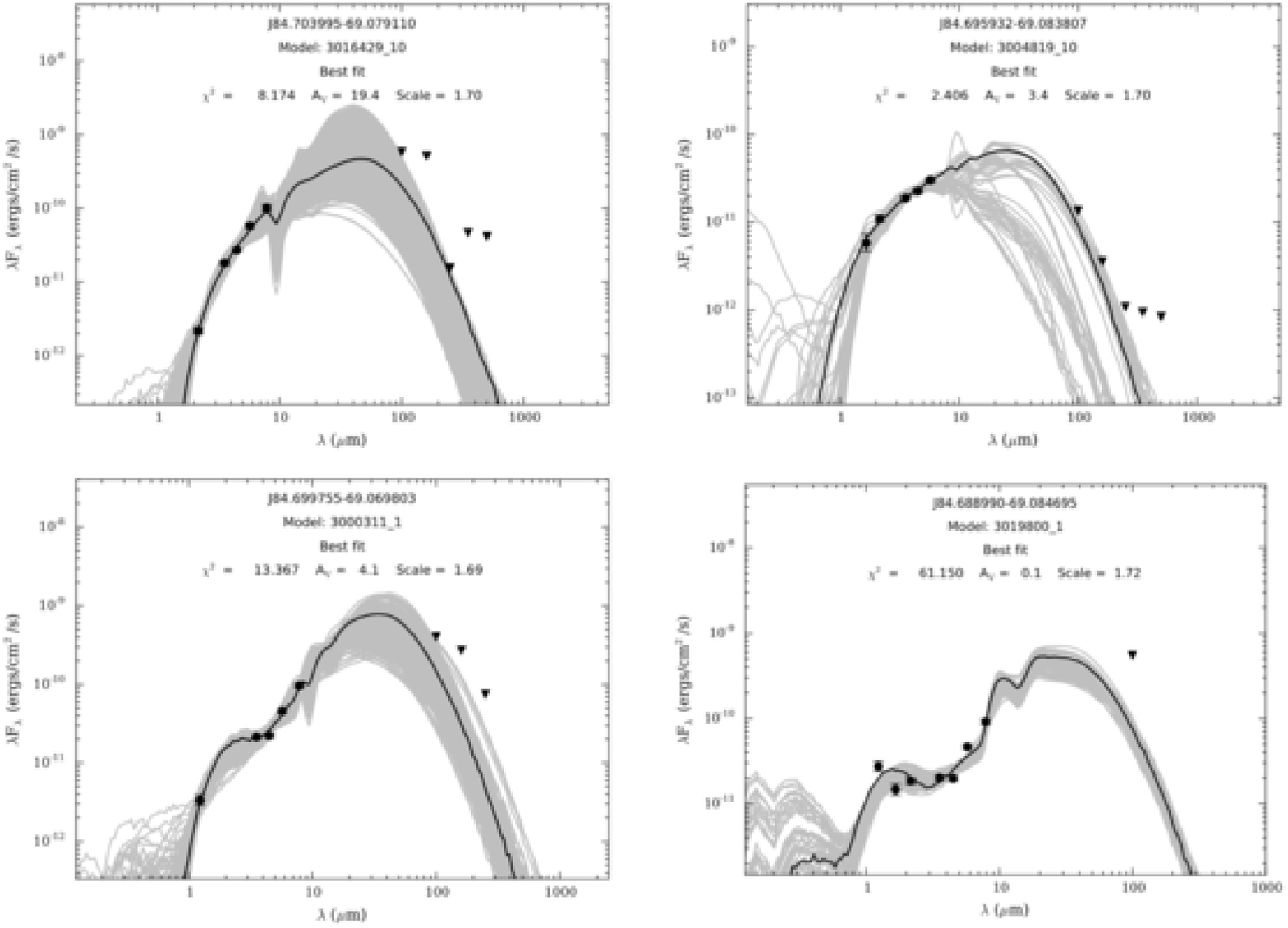}
\caption{SED fits of all point sources in the SAGE catalog that are in the ALMA footprint of this study and are YSO candidates. They are ordered from top to bottom, left to right in order of highest mass to lowest mass. Black dots are the fitted data points, black triangles are upper limits, the black line is the best fit model, and the grey lines are all models that have $\mathrm{\chi^{2}} \textless 3$ relative to the best fit model.} 
\label{fig:d5}
\end{figure}

\noindent
\begin{figure}
\setcounter{figure}{4} 
\renewcommand{\thefigure}{\arabic{figure} continued}
\centering
\includegraphics[scale=0.40,clip,trim=0cm 0cm 0cm 0cm,angle=0]{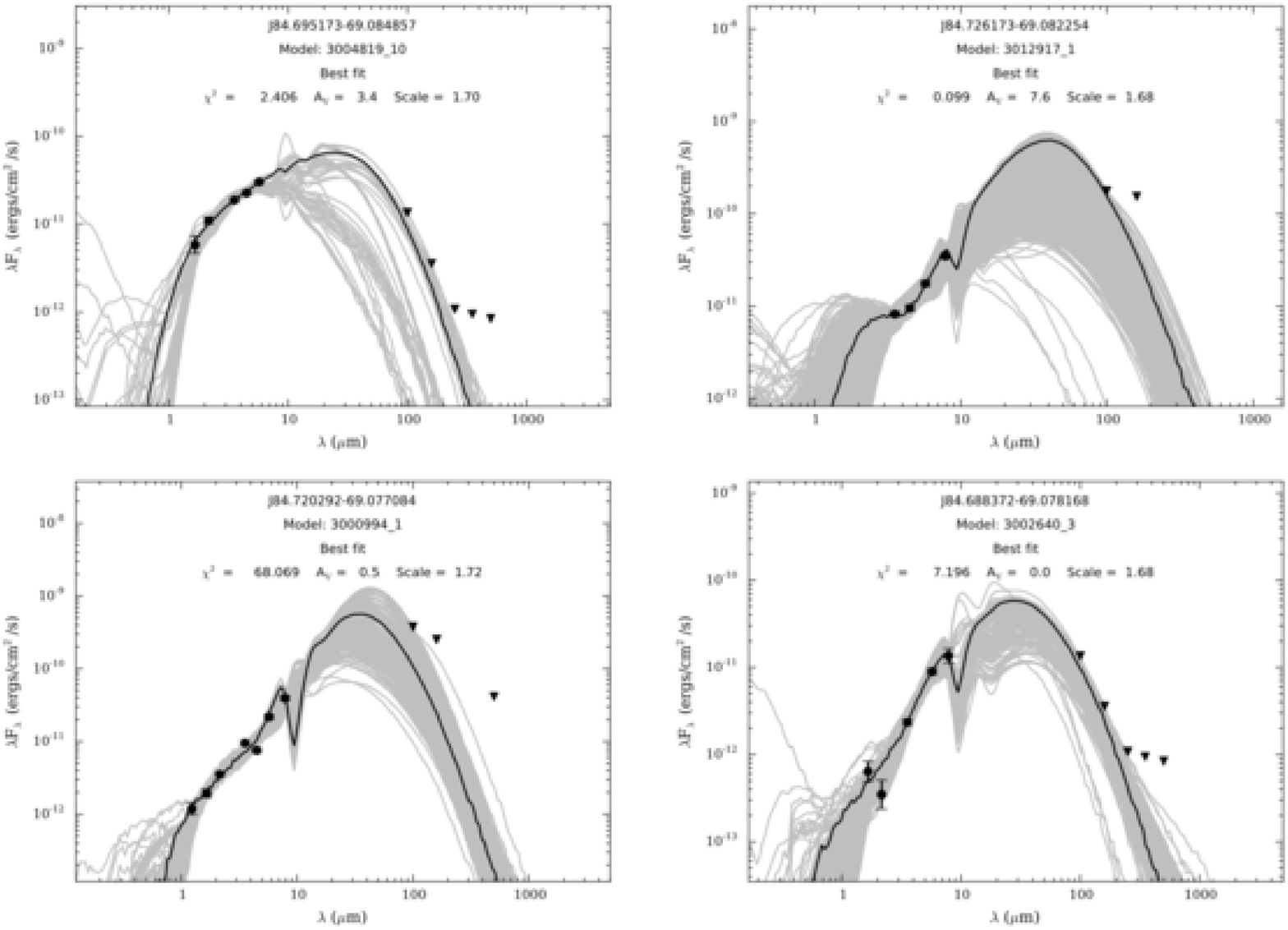}
\caption{SED fits of all point sources in the SAGE catalog that are in the ALMA footprint of this study and are YSO candidates. They are ordered from top to bottom, left to right in order of highest mass to lowest mass. Black dots are the fitted data points, black triangles are upper limits, the black line is the best fit model, and the grey lines are all models that have $\mathrm{\chi^{2}} \textless 3$ relative to the best fit model.} 
\end{figure}

\noindent
\begin{figure}
\setcounter{figure}{4} 
\renewcommand{\thefigure}{\arabic{figure} continued}
\centering
\includegraphics[scale=0.40,clip,trim=0cm 0cm 0cm 0cm,angle=0]{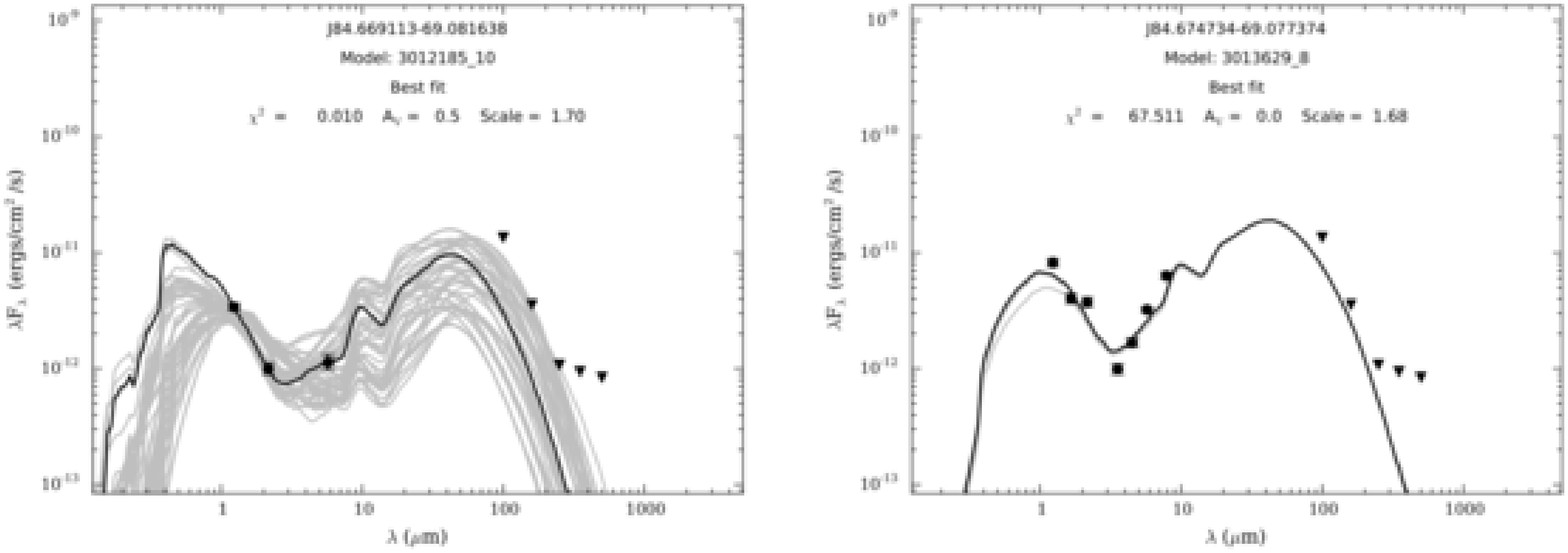}
\caption{SED fits of all point sources in the SAGE catalog that are in the ALMA footprint of this study and are YSO candidates. They are ordered from top to bottom, left to right in order of highest mass to lowest mass. Black dots are the fitted data points, black triangles are upper limits, the black line is the best fit model, and the grey lines are all models that have $\mathrm{\chi^{2}} \textless 3$ relative to the best fit model.}
\end{figure}

\noindent
\begin{figure}
\setcounter{figure}{5} 
\renewcommand{\thefigure}{\arabic{figure}}
\centering
	\vspace{-10.00\baselineskip}
	\subfloat{\includegraphics[width=3.5in]{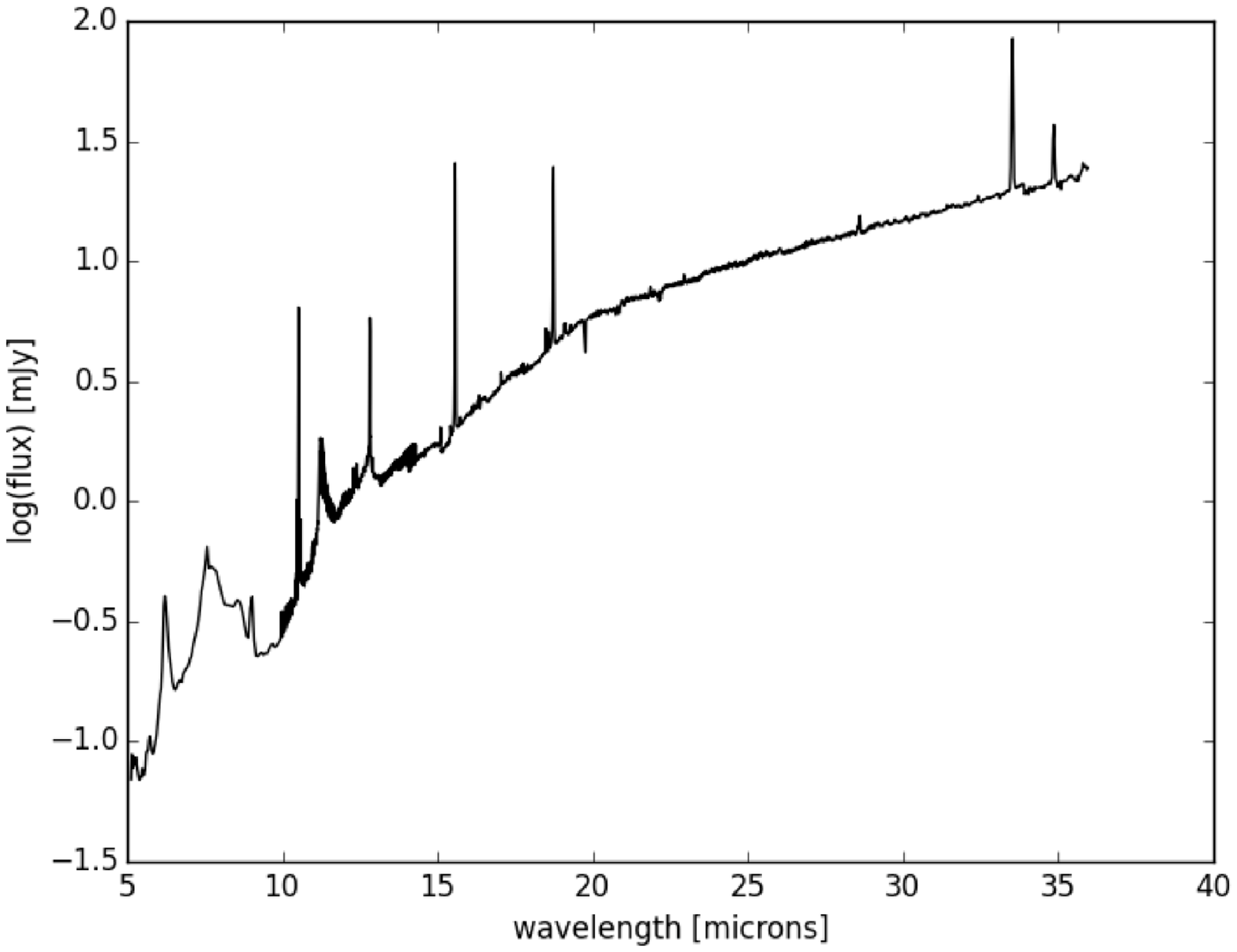}}\\
	\vspace{-10.00\baselineskip}
	\subfloat{\includegraphics[width=3.5in]{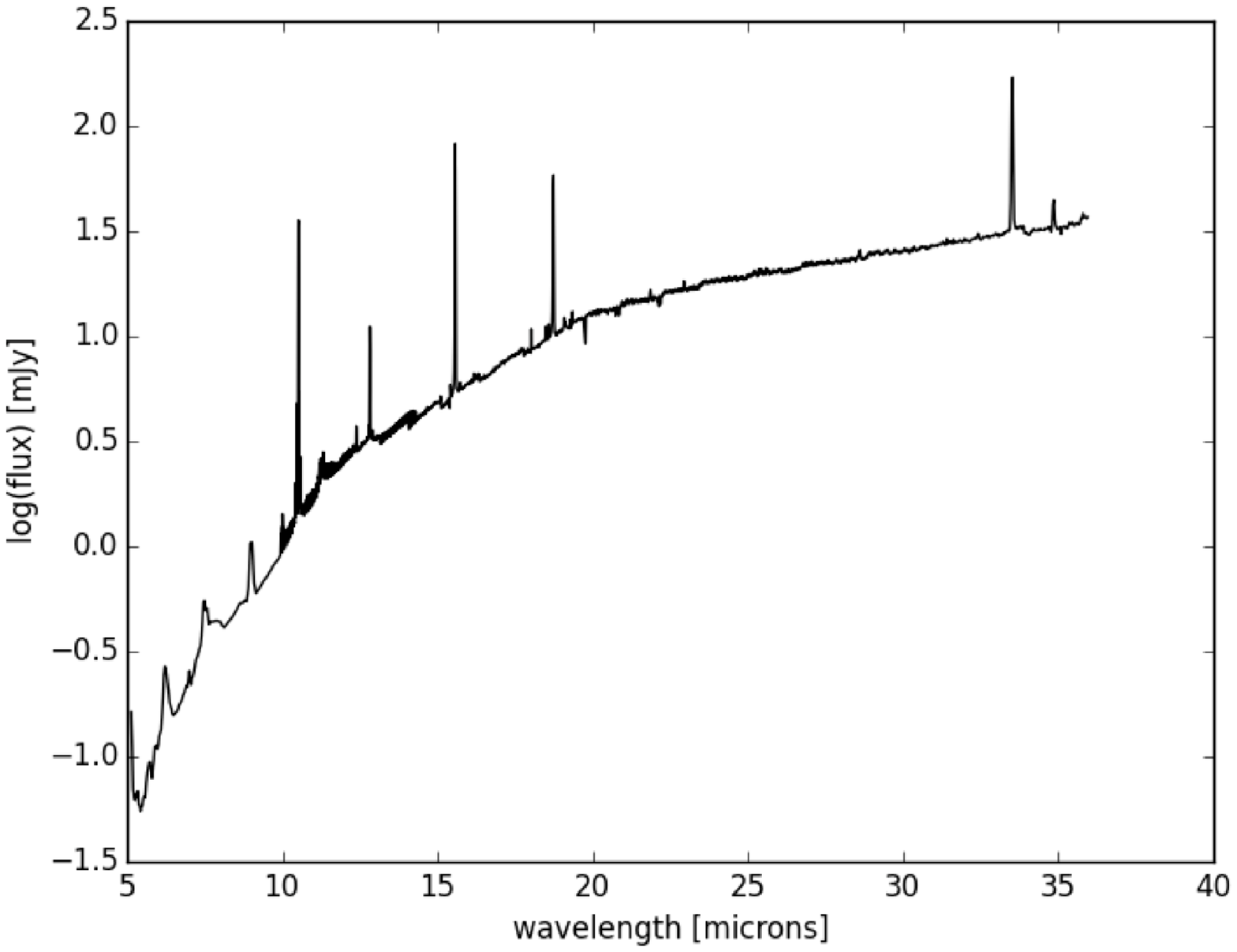}}\\	
	\vspace{-10.00\baselineskip}
	\subfloat{\includegraphics[width=3.5in]{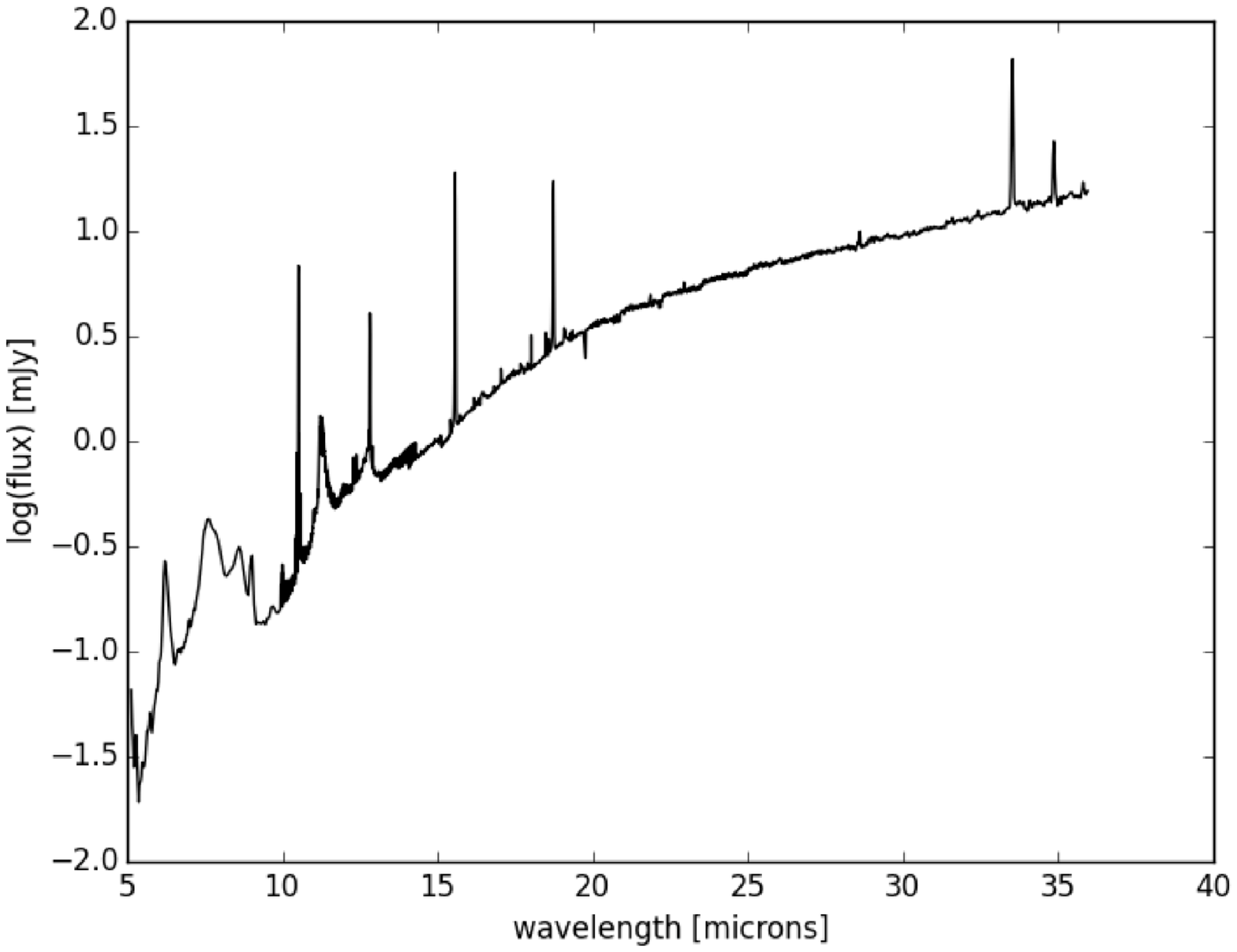}}\\
	\caption{Spectra of YSO candidates in the 30 Doradus ALMA footprint. Left: J84.703995-69.079110. Center: J84.688990-69.084695 . Right: J84.720292-69.077084. Prominent features include PAH (6.2 $\mu$m), PAH (7.7 $\mu$m), PAH (8.6 $\mu$m), [ArIII] (8.99 $\mu$m), silicon absorption (9.7 $\mu$m), $\mathrm{H_{2}}$ (9.7 $\mu$m), [SIV] (10.5 $\mu$m), PAH (11.2 $\mu$m), $\mathrm{H_{2}}$ (12.3 $\mu$m), [NeII] (12.81 $\mu$m), [NeII] (15.55 $\mu$m), $\mathrm{H_{2}}$ (17.1 $\mu$m), [SIII] (18.71 $\mu$m), [SIII] (33.48 $\mu$m), and [SiII] (34.81 $\mu$m).}
	\label{fig:d6}
\end{figure}

\noindent
\begin{figure*}
\setcounter{figure}{6} 
\renewcommand{\thefigure}{\arabic{figure}}
\centering
\includegraphics[scale=0.65,clip,trim=0cm 0cm 0cm 0cm,angle=0]{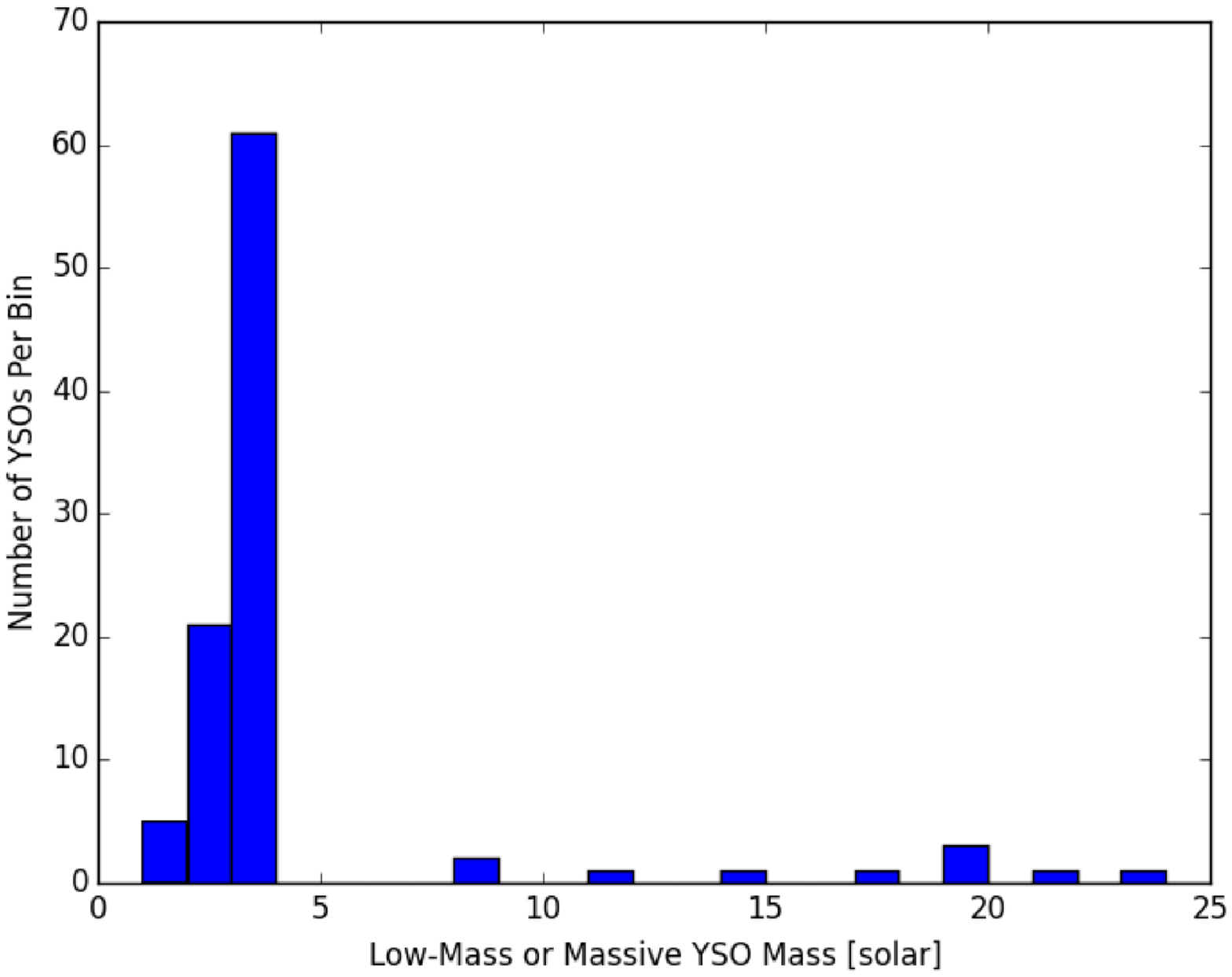}
\caption{Mass distribution of low-mass and massive YSOs.} 
\label{fig:d7}
\end{figure*}

\noindent
\begin{figure*}
\setcounter{figure}{7} 
\renewcommand{\thefigure}{\arabic{figure}}
\centering
\includegraphics[scale=0.50,clip,trim=0cm 0cm 0cm 0cm,angle=0]{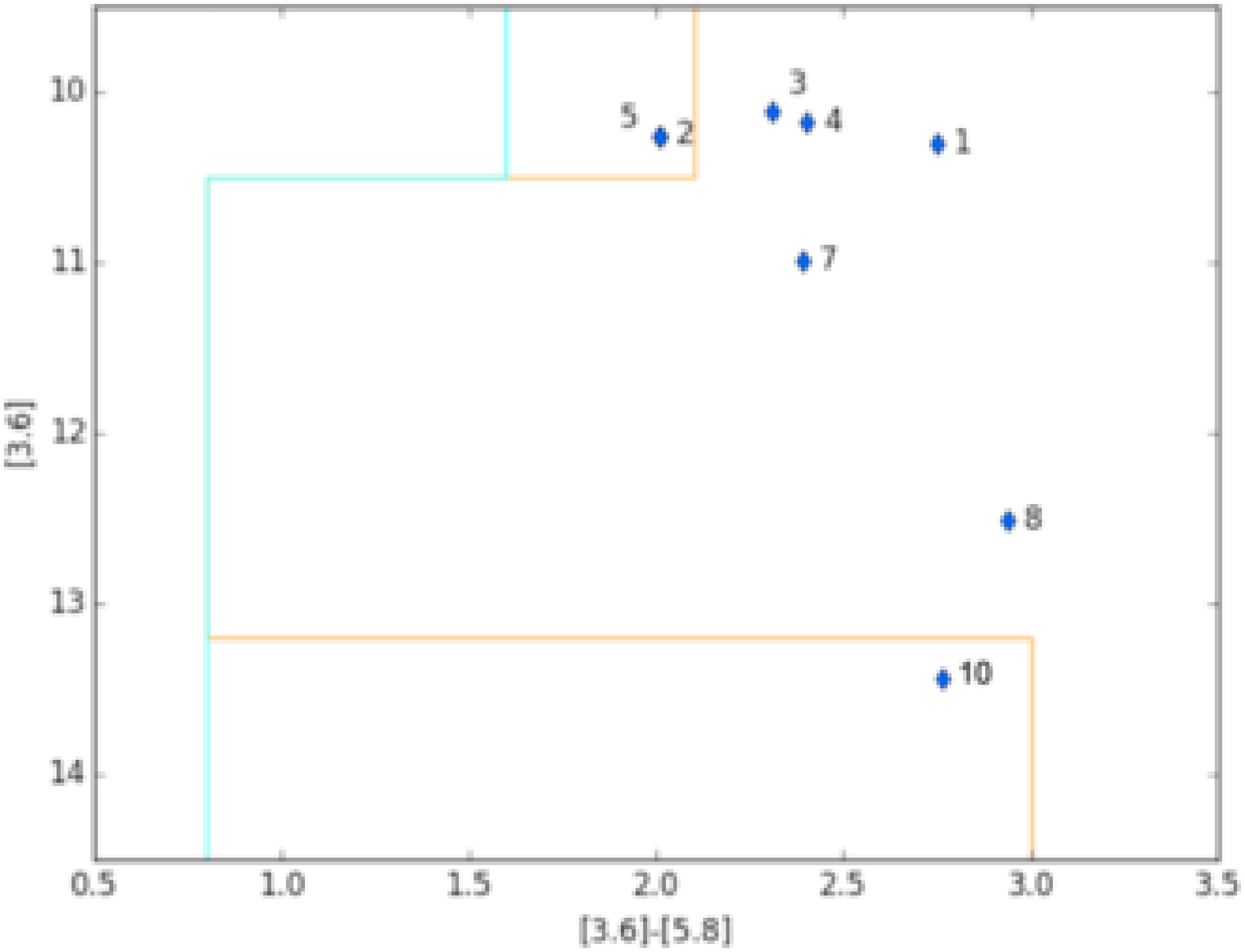}
\caption{CMD of IRAC [3.6] versus IRAC [3.6] - IRAC [5.8]. Everything to the right of the orange line fits the criteria for the \citet{carl12} $\mathrm{\alpha}$ cut and everything to the right of the cyan line fits the \citet{carl12} $\mathrm{\beta}$ cut. Numbers next to each point correspond to the rank of the YSO candidates from most massive to least massive as listed in Table 6. Even though we have a total of 11 YSO candidates, we only have 8 objects plotted because a few of them are missing [3.6] or [5.8] photometry.} 
\label{fig:d8}
\end{figure*}

\noindent
\begin{figure*}
\setcounter{figure}{8} 
\renewcommand{\thefigure}{\arabic{figure}}
\centering
\includegraphics[scale=0.65,clip,trim=0cm 0cm 0cm 0cm,angle=0]{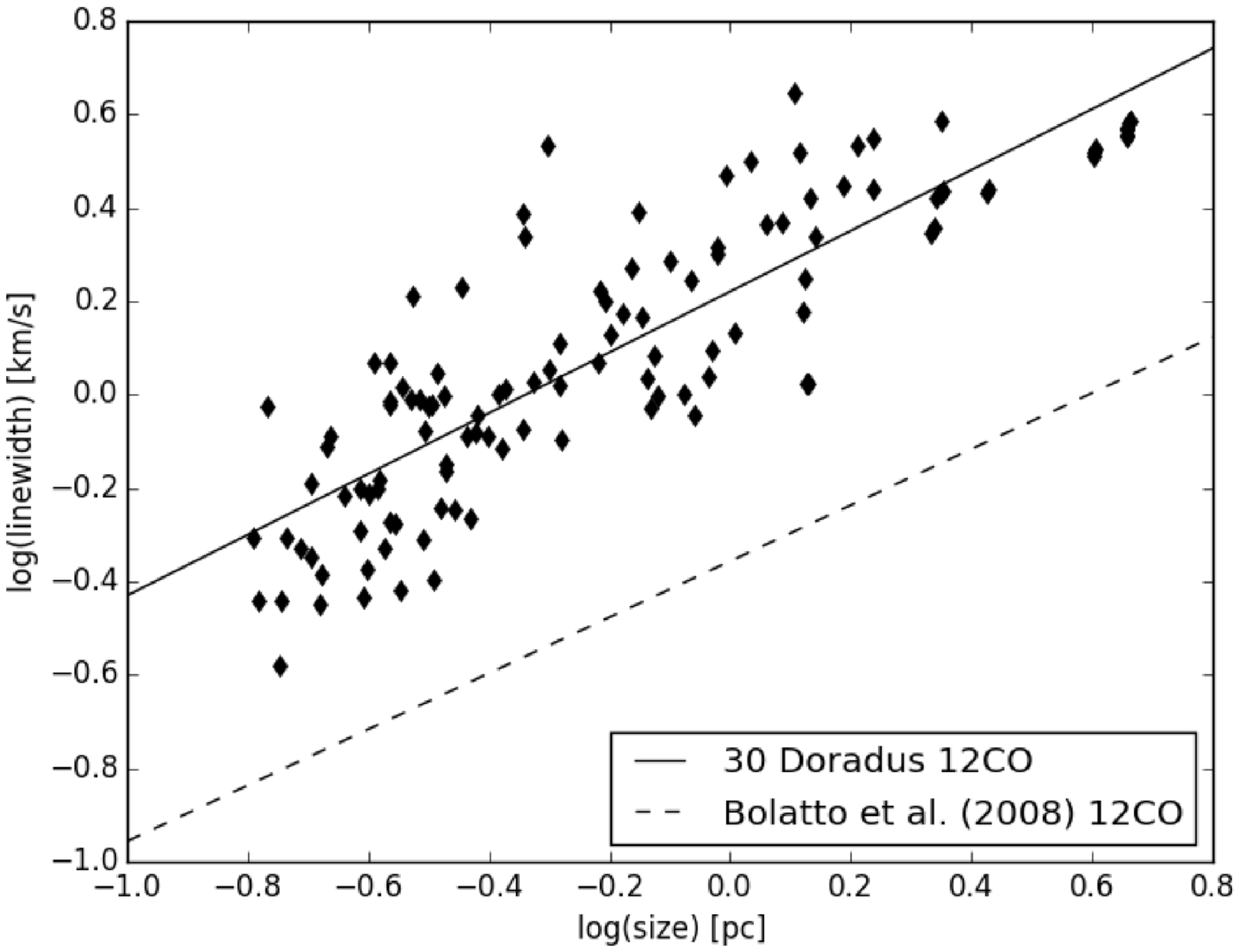}
\caption{Size-linewidth relation of $\mathrm{^{12}CO}$ (2-1) brightness temperature clumps. The best fit line is given by $\mathrm{\sigma = (1.66 \pm 0.06) r^{(0.65 \pm 0.04)}}$. The dashed line, given by the equation $\mathrm{\sigma = 0.44 r^{0.60}}$, is the best-fit line for extragalactic $\mathrm{^{12}CO}$ molecular cloud analyzed by \citet{bola08}. The $\mathrm{^{12}CO}$ (2-1) linewidths of clumps in 30 Doradus are offset from extragalactic clumps by a factor of 3.8.} 
\label{fig:d9}
\end{figure*}

\noindent
\begin{figure*}
\setcounter{figure}{9} 
\renewcommand{\thefigure}{\arabic{figure}}
\centering
\includegraphics[scale=0.65,clip,trim=0cm 0cm 0cm 0cm,angle=0]{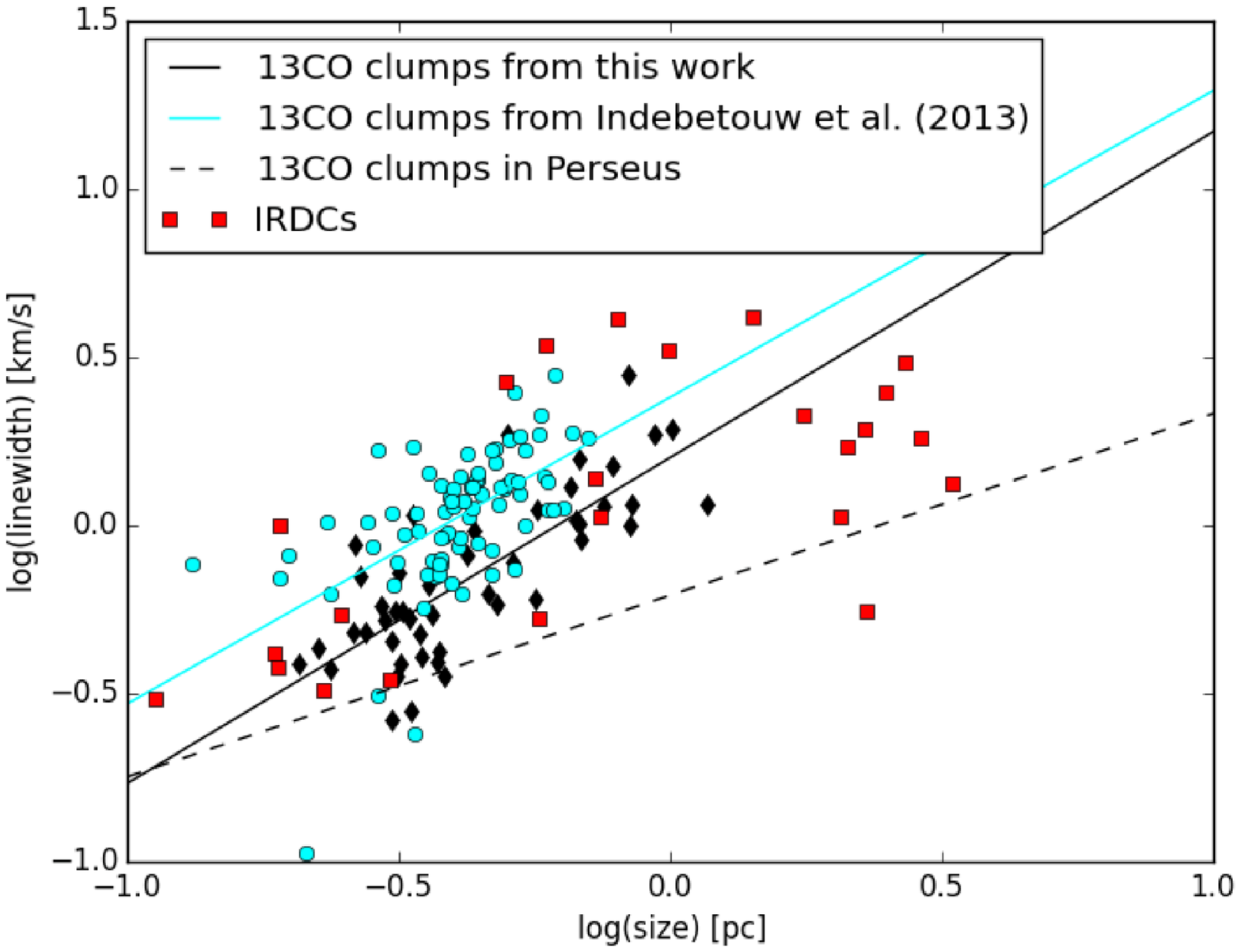}
\vspace{-8.50\baselineskip}
\caption{The size (of astrodendro clumps and cprops clumps) is calculated from the weighted second moment in two spatial directions, specifically the direction of greatest spatial extent and perpendicular to that.  The radius is then 1.91 times the geometric mean of those two spatial second moments. The linewidth is calculated using the weighted velocity second moment of the pixels assigned to the cprops clump or astrodendro structure. Black diamonds are the size-linewidth values of $\mathrm{^{13}CO}$ (2-1) clumps in this study. The best fit line of astrodendro clumps is given by the equation $\mathrm{\sigma = (1.58 \pm 0.18) r^{(0.97 \pm 0.12)}}$ and represented by the solid black line. Cyan circles are 30 Doradus $\mathrm{^{13}CO}$ (2-1) brightness temperature size-linewidth clumps from \citet{inde13}. The \citet{inde13} clumps are systematically offset by the clumps in this work by a factor of 1.5. This difference is because astrodendro systematically calculates smaller linewidths for clumps that are comparable in size to the clumps found by cprops. See Appendix B for further details. The best fit line of the cprops clumps is given by the equation $\mathrm{\sigma = (2.39 \pm 0.33) r^{(0.91 \pm 0.15)}}$ and represented by the solid cyan line. Red squares are size-linewidths from infrared dark cloud studies by \citet{gibs09, bont10, pere13}. We converted the FWHM to $\sigma$ by dividing the FWHM values given in the infrared dark cloud studies by a factor of 2.35. The dashed line ($\mathrm{\sigma = 0.62 r^{0.54}}$) is the best-fit line for molecular clumps in Perseus analysis by \citet{shet12} using dendrograms. The 30 Doradus clumps are offset from the Perseus clumps by a factor of 2.5.}
\label{fig:d10}
\end{figure*}

\clearpage

\noindent
\begin{figure*}
\setcounter{figure}{10} 
\renewcommand{\thefigure}{\arabic{figure}}
\centering
\includegraphics[scale=0.65,clip,trim=0cm 0cm 0cm 0cm,angle=0]{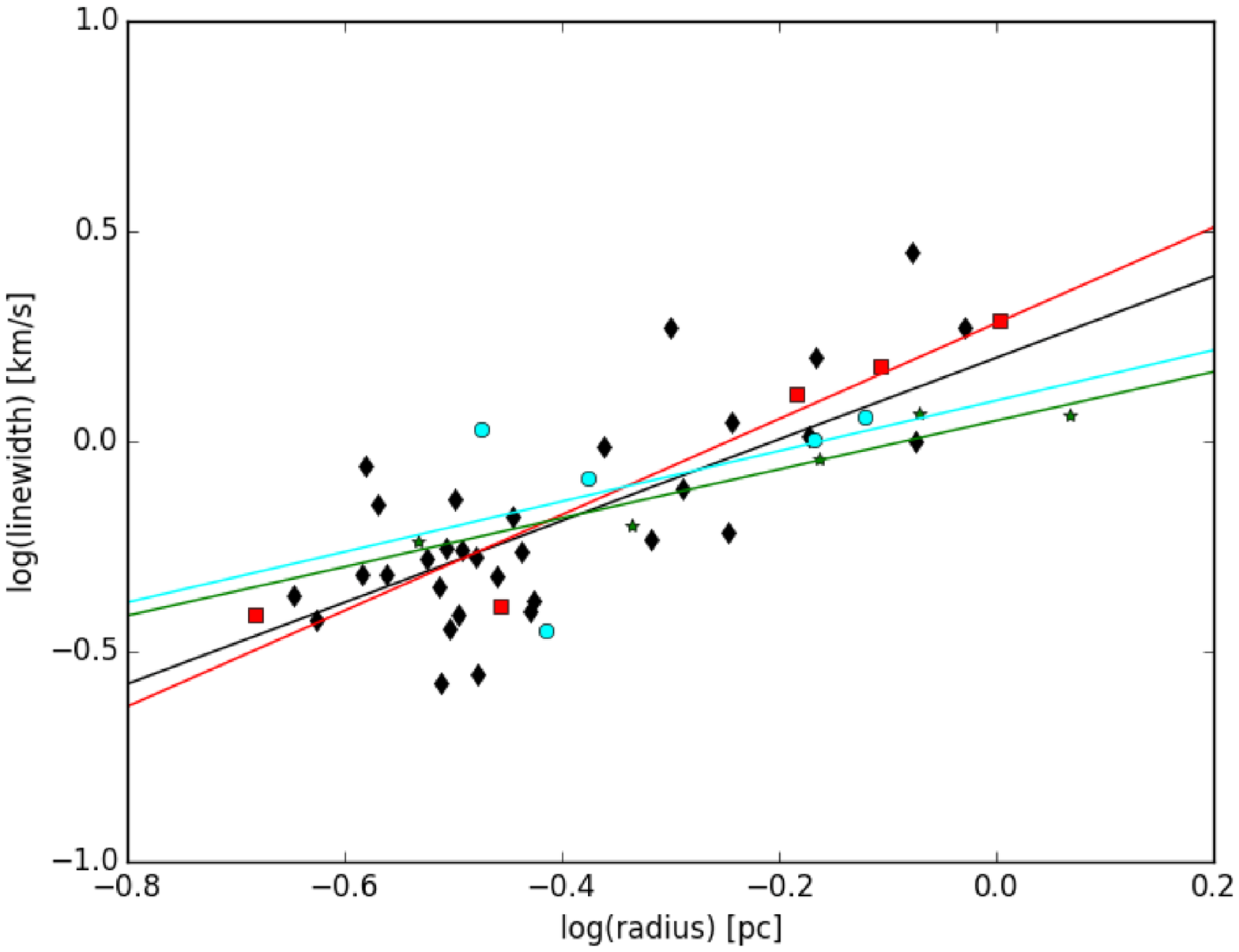}
\caption{Size-linewidth relation of $\mathrm{^{13}CO}$ (2-1) brightness temperature clumps. The best fit line of all clumps is given by $\mathrm{\sigma = (1.58 \pm 0.18) r^{(0.97 \pm 0.12)}}$. The red squares are clumps in the red dendrogram structure in Figure ~\ref{fig:d3a} ($\mathrm{\sigma = (1.91 \pm 0.31) r^{(1.14 \pm 0.19)}}$), the green stars are those that are in the green structure in Figure ~\ref{fig:d3a} ($\mathrm{\sigma = (1.12 \pm 0.11) r^{(0.58 \pm 0.11)}}$), and the cyan circles are clumps in the cyan structure in Figure ~\ref{fig:d3a} ($\mathrm{\sigma = (1.25 \pm 0.69) r^{(0.60 \pm 0.48)}}$). The red line is the best-fit line to the red squares, the green line is the best-fit line to the green stars, and the cyan line is the best-fit line to the cyan circles. There are a total of 5 each of red squares, green stars, and cyan circles plotted.} 
\label{fig:d11}
\end{figure*}

\noindent
\begin{figure*}
\setcounter{figure}{11} 
\renewcommand{\thefigure}{\arabic{figure}}
\centering
\includegraphics[scale=0.65,clip,trim=0cm 0cm 0cm 0cm,angle=0]{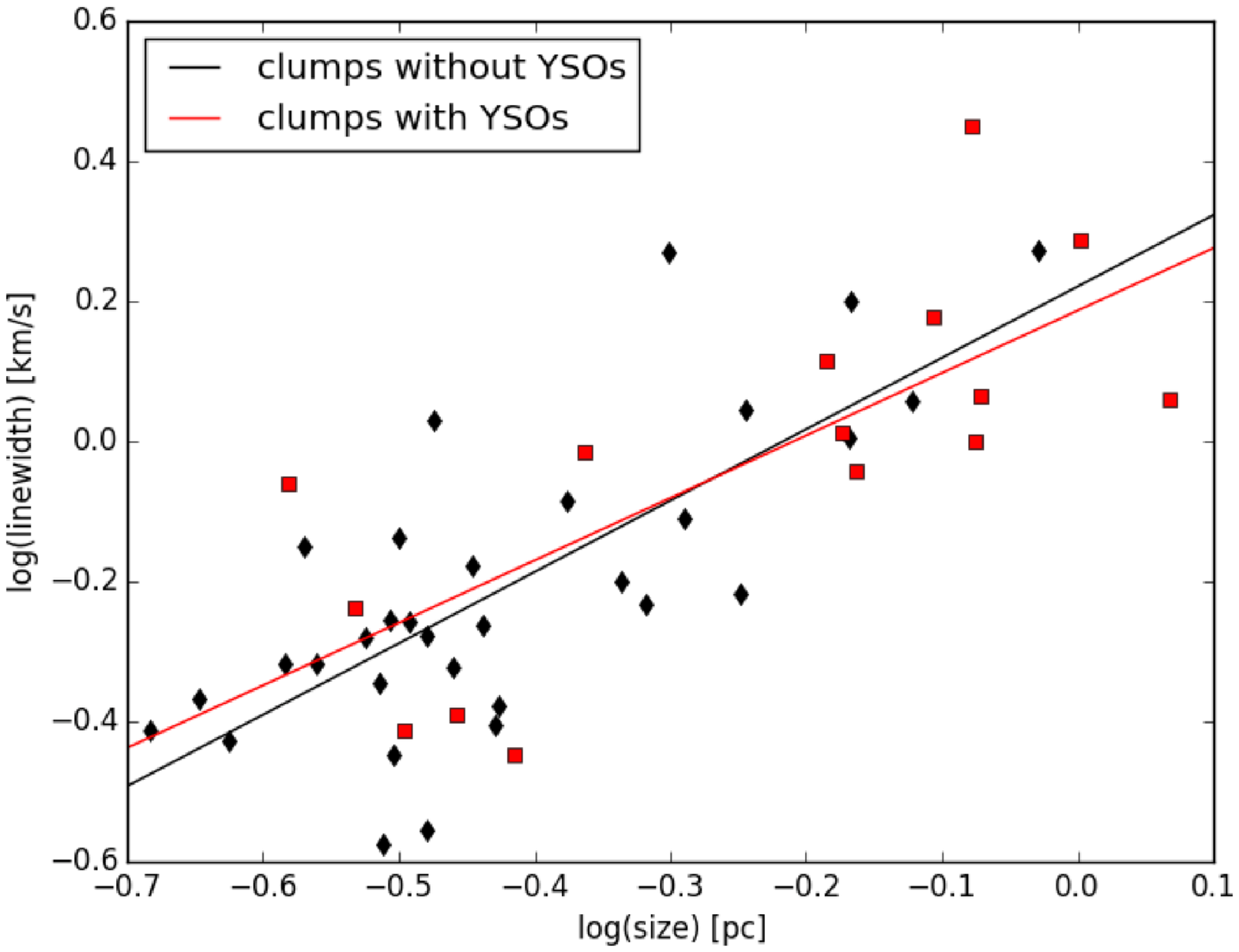}
\caption{Size-linewidth relation of $\mathrm{^{13}CO}$ (2-1) column density clumps. The black diamonds are clumps not associated with any star formation and the best fit line is given by $\mathrm{\sigma = (1.66 \pm 0.31) r^{(1.02 \pm 0.17)}}$. The red squares show the clumps in our study that are are associated with star formation ($\mathrm{\sigma = (1.53 \pm 0.25) r^{(0.89 \pm 0.22)}}$). There is no noticeable difference in the slope or the intercept of the size-linewidth relation between clumps with active star formation and those without.} 
\label{fig:d12}
\end{figure*}

\noindent
\begin{figure*}
\setcounter{figure}{12} 
\renewcommand{\thefigure}{\arabic{figure}}
\centering
\includegraphics[scale=0.65,clip,trim=0cm 0cm 0cm 0cm,angle=0]{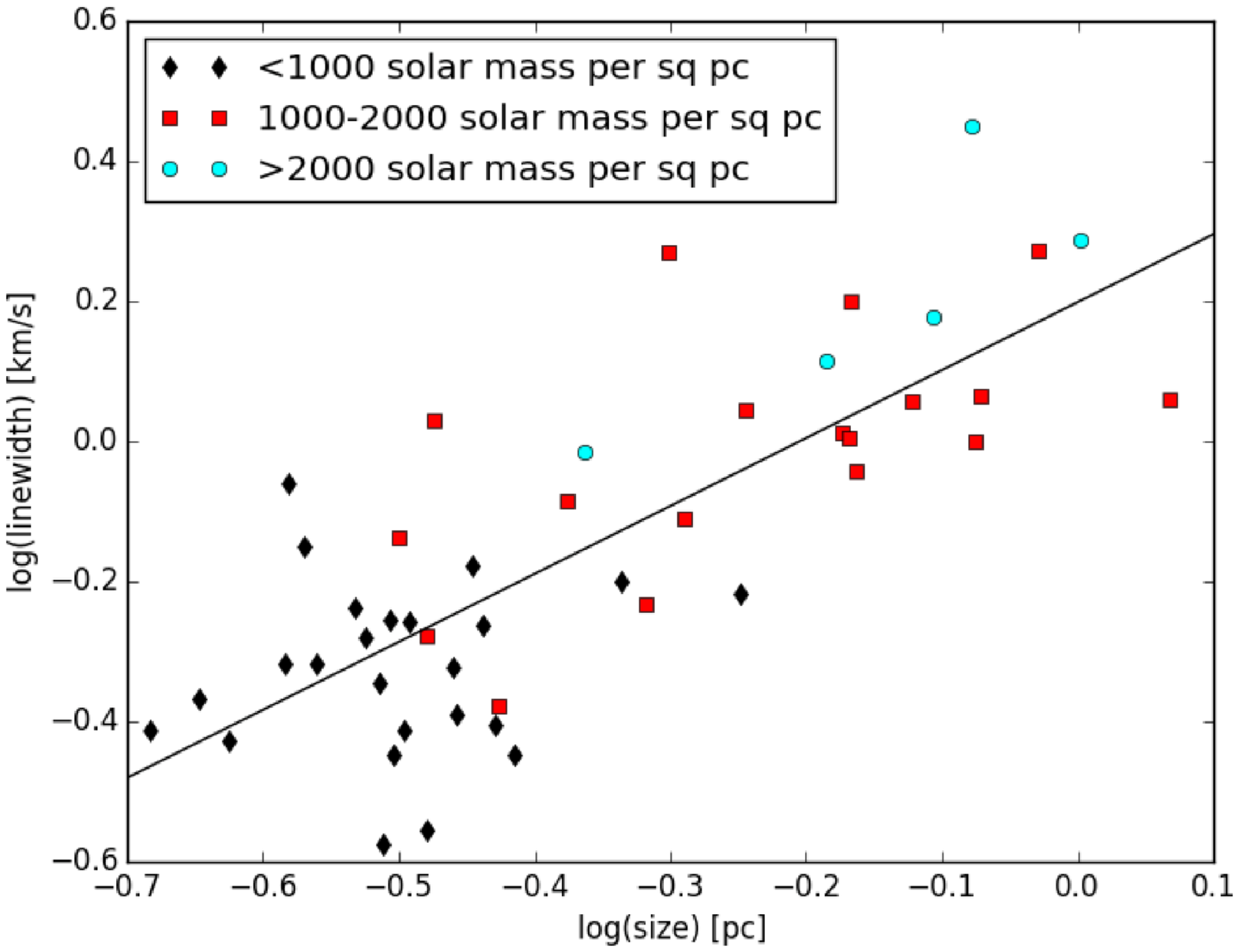}
\caption{Size-linewidth relation of $\mathrm{^{13}CO}$ (2-1) column density clumps. The best fit line is $\mathrm{\sigma = (1.58 \pm 0.18) r^{(0.97 \pm 0.12)}}$. Black diamonds are clumps with mass density less than 1000 $\mathrm{M_{\odot}\;pc^{-2}}$, red squares are clumps with mass density between 1000-2000 $\mathrm{M_{\odot}\;pc^{-2}}$, and cyan circles are clumps with mass density above 2000 $\mathrm{M_{\odot}\;pc^{-2}}$.} 
\label{fig:d13}
\end{figure*}

\noindent
\begin{figure*}
\setcounter{figure}{13} 
\renewcommand{\thefigure}{\arabic{figure}}
\centering
\includegraphics[scale=0.65,clip,trim=0cm 0cm 0cm 0cm,angle=0]{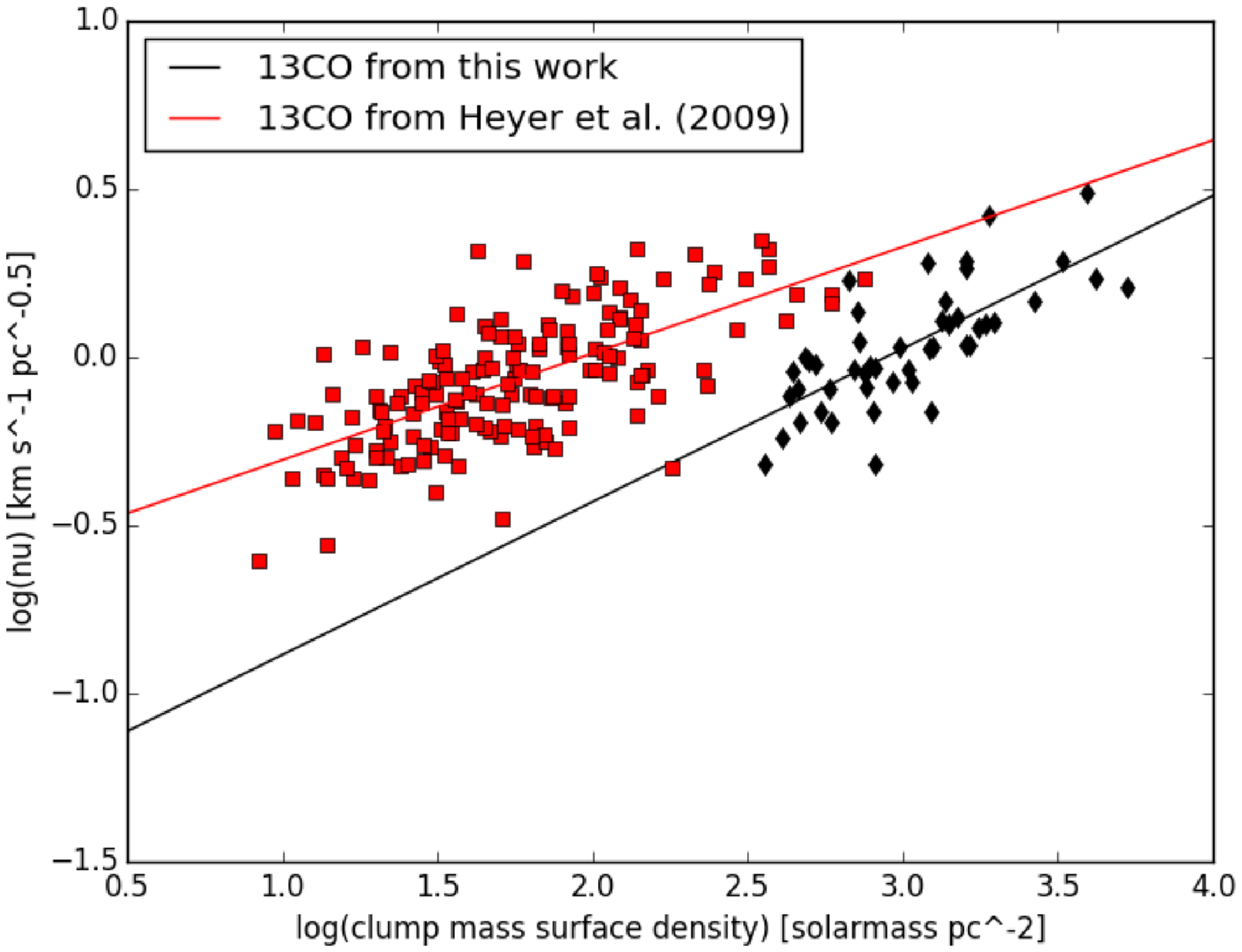}
\caption{The dependence of $\mathrm{\frac{\sigma}{r^{0.5}}}$ on $\mathrm{\Sigma}$. The black line ($\mathrm{\nu = (0.04 \pm 0.02) \Sigma^{(0.45 \pm 0.06)}}$) is the best-fit line is going through $\mathrm{\nu}$ versus $\mathrm{\Sigma}$ of 30 Doradus clumps in this study. The red line ($\mathrm{\nu = (0.24 \pm 0.02) \Sigma^{(0.31 \pm 0.04)}}$) is the best-fit line going through $\mathrm{\nu}$ versus $\mathrm{\Sigma}$ of clumps studied by \citet{heye09}. The dependence on $\Sigma$ is contradictory to Larson's scaling relation, but consistent with the slope found by \citet{heye09}.} 
\label{fig:d14}
\end{figure*}

\noindent
\begin{figure*}
\setcounter{figure}{14} 
\renewcommand{\thefigure}{\arabic{figure}}
\centering
\includegraphics[scale=0.65,clip,trim=0cm 0cm 0cm 0cm,angle=0]{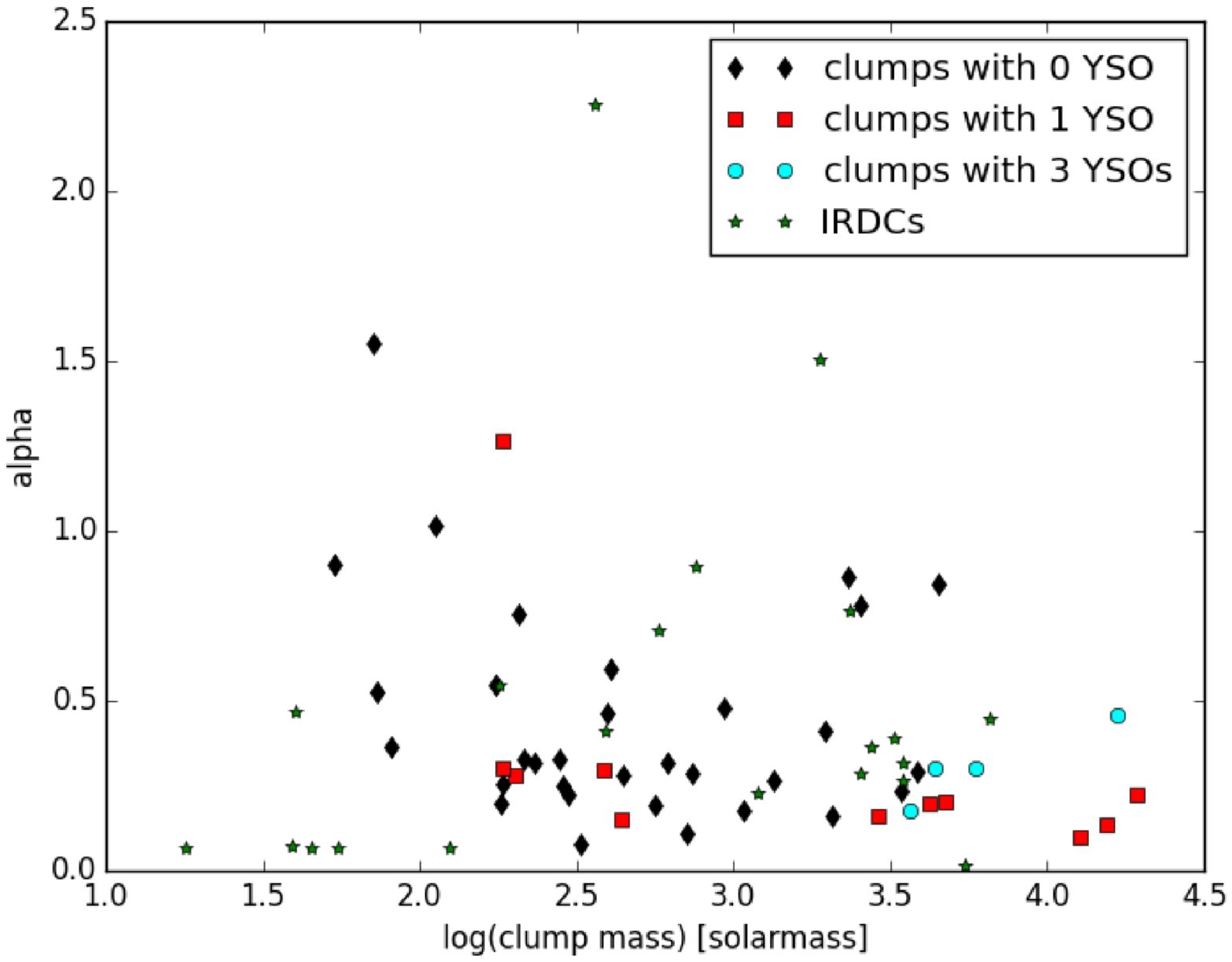}
\caption{Figure showing the relation between the virial parameter versus the mass of the clump. We define $\alpha$ to equal to $\mathrm{\frac{5 \sigma^{2}\;r}{G\;M}}$. Black diamonds are clumps with no YSO candidates, red squares are clumps with 1 YSO candidate, and cyan circles are clumps with 3 YSO candidates. Green stars are from infrared dark cloud studies by \citet{gibs09, bont10, pere13}.} 
\label{fig:d15}
\end{figure*}

\noindent
\begin{figure}
\setcounter{figure}{15} 
\renewcommand{\thefigure}{\arabic{figure}}
\centering
\includegraphics[scale=0.45,clip,trim=0cm 0cm 0cm 0cm,angle=0]{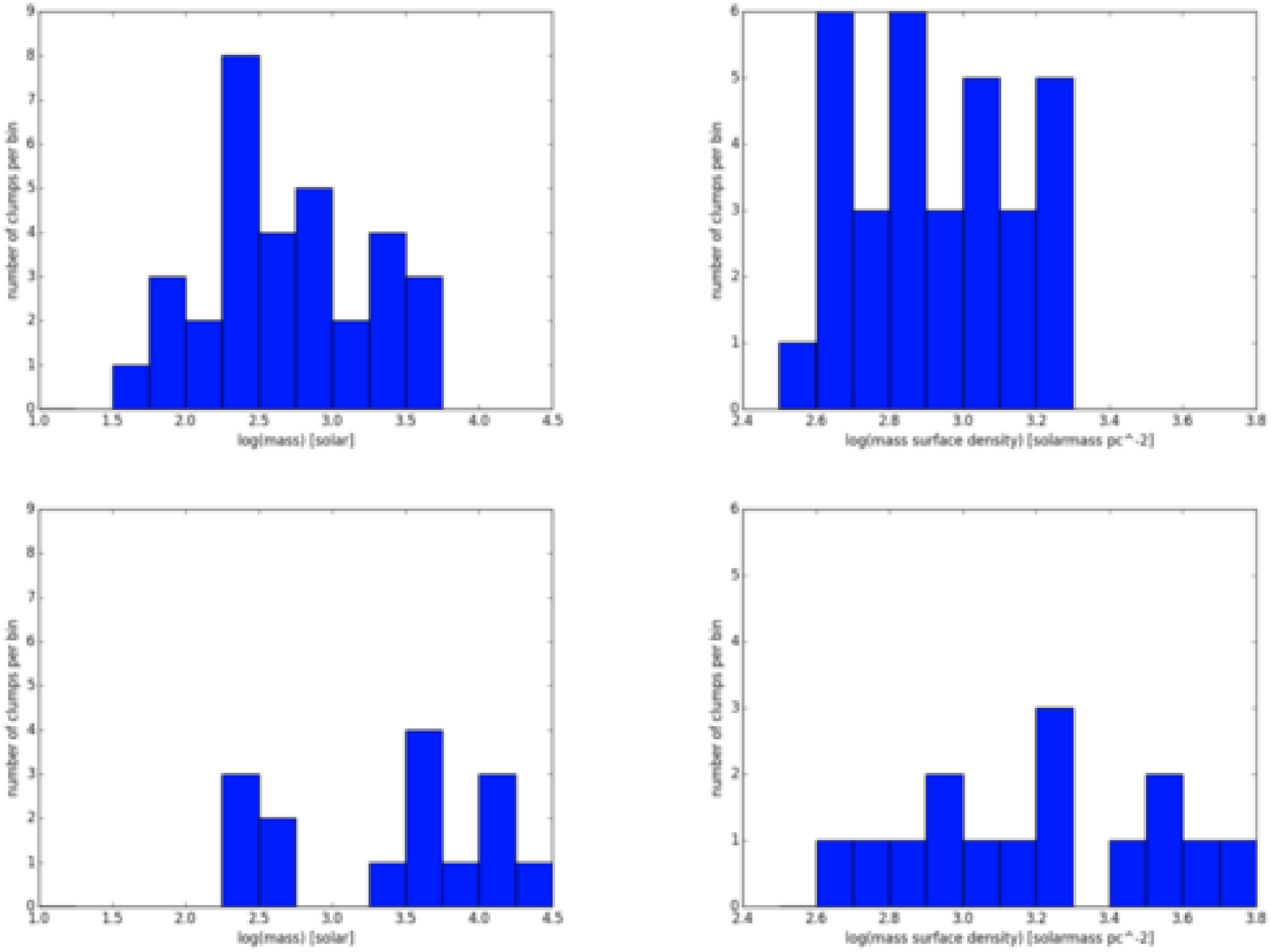}
\caption{Left Column: Mass distribution of clumps without (top) and with (bottom) YSO candidates. Right Column: Mass surface density distribution of clumps without (top) and with (bottom) YSO candidates.}
\label{fig:d16}
\end{figure}

\noindent
\setcounter{figure}{16} 
\renewcommand{\thefigure}{\arabic{figure} a}
\begin{figure*}
\centering
\includegraphics[scale=0.65,clip,trim=0cm 0cm 0cm 0cm,angle=0]{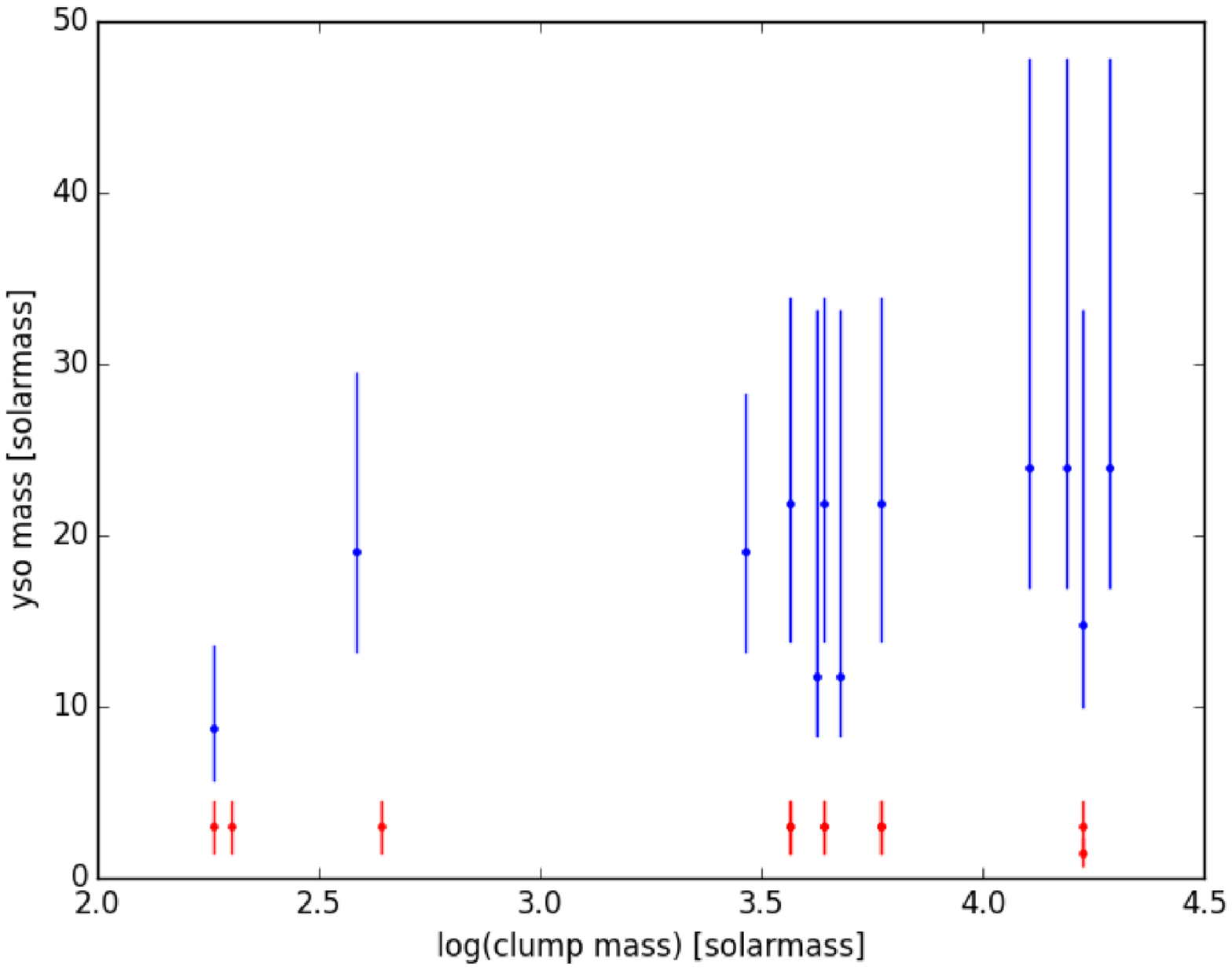}
\caption{Star mass versus clump mass for each newly forming star. Red circles represent the low-mass more evolved YSO candidates selected via H$\mathrm{\alpha}$ excess. Blue circles represent the massive YSO candidates selected by SED fitting.} 
\label{fig:d17a}
\end{figure*}

\noindent
\setcounter{figure}{16} 
\renewcommand{\thefigure}{\arabic{figure} b}
\begin{figure*}
\centering
\includegraphics[scale=0.65,clip,trim=0cm 0cm 0cm 0cm,angle=0]{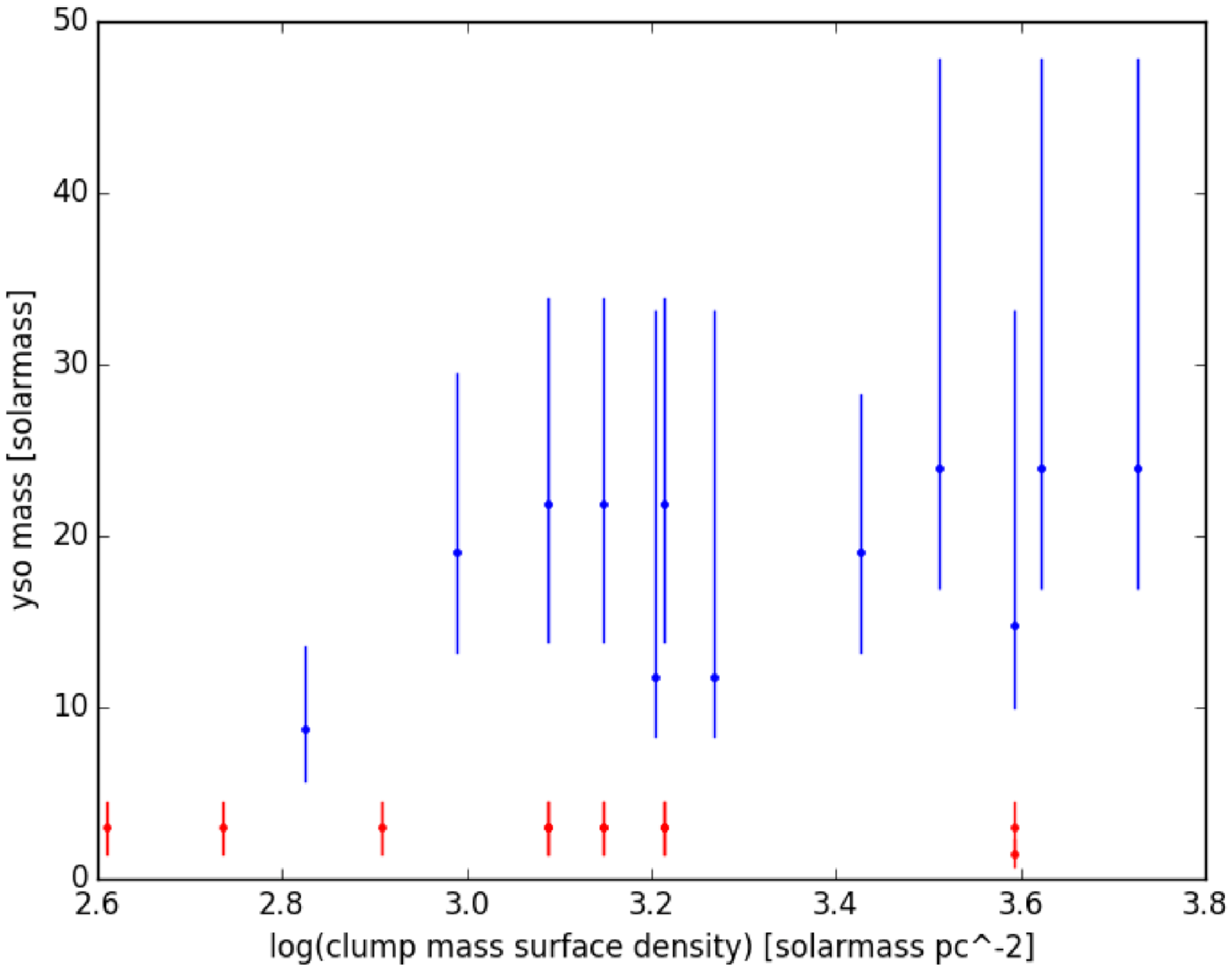}
\caption{Star mass versus clump mass surface density for each newly forming star. Red circles represent the low-mass more evolved YSO candidates selected via H$\mathrm{\alpha}$ excess. Blue circles represent the massive YSO candidates selected by SED fitting.} 
\label{fig:d17b}
\end{figure*}

\setcounter{table}{0} 
\renewcommand{\thetable}{\arabic{table}}
\tabletypesize{\tiny}
\begin{deluxetable}{ccccccc}
\tablewidth{0pt}
\tablecaption{Mass Derived from $\mathrm{^{12}CO}$ (2-1)}
\tablehead{\colhead{Clump ID} & \colhead{RA} & \colhead{Dec} & \colhead{Radius [$\mathrm{pc}$]} & \colhead{$\mathrm{F_{^{12}CO}}$ [$\mathrm{K\;km\;s^{-1}}$]} & \colhead{Total Mass [$\mathrm{M_{\odot}}$]} & \colhead{Linewidth [km/s]}}
\startdata
0	&	84.706989	&	-69.068853	&	0.522	&	21.2	$\pm$	2.55	&	321	$\pm$	38.6	&	1.29	\\
1	&	84.700807	&	-69.078029	&	4.61	&	137	$\pm$	17.5	&	98900	$\pm$	12600	&	3.86	\\
2	&	84.700775	&	-69.078057	&	4.59	&	139	$\pm$	17.2	&	97300	$\pm$	12100	&	3.83	\\
3	&	84.700755	&	-69.078118	&	4.57	&	143	$\pm$	16.7	&	93700	$\pm$	10900	&	3.74	\\
4	&	84.703518	&	-69.069242	&	0.457	&	23.2	$\pm$	2.23	&	271	$\pm$	26.1	&	2.18	\\
5	&	84.700788	&	-69.078107	&	4.57	&	144	$\pm$	16.4	&	90800	$\pm$	10300	&	3.69	\\
6	&	84.703853	&	-69.071507	&	0.454	&	47.2	$\pm$	2.42	&	647	$\pm$	33.2	&	2.44	\\
7	&	84.700814	&	-69.078139	&	4.55	&	148	$\pm$	15.5	&	83100	$\pm$	8710	&	3.58	\\
8	&	84.700823	&	-69.078138	&	4.55	&	148	$\pm$	15.5	&	83400	$\pm$	8740	&	3.59	\\
9	&	84.698556	&	-69.079379	&	4.03	&	151	$\pm$	13.5	&	63900	$\pm$	5750	&	3.35	\\
10	&	84.698567	&	-69.079381	&	4.02	&	151	$\pm$	13.2	&	61700	$\pm$	5390	&	3.3	\\
11	&	84.698536	&	-69.079390	&	4.01	&	152	$\pm$	12.8	&	58300	$\pm$	4910	&	3.24	\\
12	&	84.694103	&	-69.077254	&	2.25	&	148	$\pm$	7.42	&	19100	$\pm$	958	&	3.87	\\
13	&	84.699955	&	-69.072633	&	0.287	&	27	$\pm$	1.47	&	137	$\pm$	7.45	&	1.04	\\
14	&	84.699347	&	-69.075175	&	0.497	&	158	$\pm$	2.40	&	2130	$\pm$	32.4	&	3.41	\\
15	&	84.696479	&	-69.077376	&	1.28	&	153	$\pm$	5.07	&	9170	$\pm$	304	&	4.43	\\
16	&	84.707331	&	-69.071393	&	0.298	&	46.8	$\pm$	1.54	&	261	$\pm$	8.59	&	1.62	\\
17	&	84.699810	&	-69.069712	&	1.54	&	53.6	$\pm$	6.60	&	5440	$\pm$	669	&	2.81	\\
18	&	84.698967	&	-69.069816	&	1.36	&	61.7	$\pm$	5.43	&	4240	$\pm$	373	&	2.65	\\
19	&	84.697151	&	-69.069254	&	0.257	&	35.5	$\pm$	1.31	&	141	$\pm$	5.21	&	1.17	\\
20	&	84.693634	&	-69.080124	&	0.337	&	20.2	$\pm$	1.71	&	138	$\pm$	11.7	&	0.991	\\
21	&	84.697258	&	-69.079987	&	0.273	&	13.7	$\pm$	1.38	&	60.7	$\pm$	6.11	&	0.535	\\
22	&	84.710057	&	-69.073121	&	1.38	&	144	$\pm$	6.82	&	15600	$\pm$	738	&	2.18	\\
23	&	84.695760	&	-69.082149	&	0.471	&	35.4	$\pm$	2.23	&	411	$\pm$	25.9	&	1.07	\\
24	&	84.696537	&	-69.081386	&	0.243	&	21.7	$\pm$	1.19	&	71.1	$\pm$	3.89	&	0.511	\\
25	&	84.689353	&	-69.081503	&	0.872	&	30.1	$\pm$	3.26	&	749	$\pm$	81.1	&	0.908	\\
26	&	84.689385	&	-69.082348	&	0.338	&	34.7	$\pm$	1.79	&	259	$\pm$	13.4	&	0.683	\\
27	&	84.700869	&	-69.080518	&	2.68	&	156	$\pm$	10.0	&	36600	$\pm$	2350	&	2.76	\\
28	&	84.695538	&	-69.082384	&	0.311	&	45.4	$\pm$	1.60	&	271	$\pm$	9.55	&	0.84	\\
29	&	84.702396	&	-69.079943	&	2.26	&	161	$\pm$	9.11	&	31000	$\pm$	1770	&	2.73	\\
30	&	84.689291	&	-69.080981	&	0.526	&	29.7	$\pm$	2.39	&	396	$\pm$	31.9	&	0.803	\\
31	&	84.688452	&	-69.081129	&	0.337	&	32.9	$\pm$	1.76	&	237	$\pm$	12.7	&	0.712	\\
32	&	84.709587	&	-69.073393	&	0.864	&	159	$\pm$	3.63	&	4900	$\pm$	112	&	1.76	\\
33	&	84.700921	&	-69.080487	&	2.66	&	159	$\pm$	9.71	&	35100	$\pm$	2140	&	2.70	\\
34	&	84.702480	&	-69.079910	&	2.24	&	159	$\pm$	9.00	&	30100	$\pm$	1710	&	2.71	\\
35	&	84.701184		&	-69.082238	&	0.715	&	71.3	$\pm$	2.93	&	1440	$\pm$	59.2	&	1.47	\\
36	&	84.700356	&	-69.082942	&	0.372	&	62.3	$\pm$	1.75	&	446	$\pm$	12.5	&	0.541	\\
37	&	84.702569	&	-69.079872	&	2.20	&	162	$\pm$	8.61	&	27900	$\pm$	1480	&	2.64	\\
38	&	84.695010	&	-69.075487	&	0.358	&	24.8	$\pm$	1.58	&	144	$\pm$	9.17	&	1.70	\\
39	&	84.700337	&	-69.069687	&	1.72	&	47	$\pm$	7.23	&	5730	$\pm$	881	&	2.75	\\
40	&	84.691066	&	-69.080654	&	0.261	&	26.3	$\pm$	1.30	&	103	$\pm$	5.09	&	0.63	\\
41	&	84.699043	&	-69.069825	&	1.33	&	66.3	$\pm$	4.67	&	3370	$\pm$	237	&	1.77	\\
42	&	84.701829	&	-69.081685	&	0.381	&	68	$\pm$	1.78	&	502	$\pm$	13.1	&	0.903	\\
43	&	84.708896	&	-69.079471	&	0.229	&	26.1	$\pm$	1.13	&	77.4	$\pm$	3.35	&	0.606	\\
44	&	84.719623	&	-69.076736	&	1.73	&	146	$\pm$	8.61	&	25300	$\pm$	1490	&	3.54	\\
45	&	84.715952	&	-69.071826	&	0.38	&	58.4	$\pm$	1.56	&	332	$\pm$	8.67	&	0.829	\\
46	&	84.699158	&	-69.069782	&	1.32	&	65.3	$\pm$	4.22	&	2710	$\pm$	175	&	1.50	\\
47	&	84.692969	&	-69.085462	&	0.202	&	29.9	$\pm$	0.97	&	66.1	$\pm$	2.14	&	0.451	\\
48	&	84.688529	&	-69.085050	&	0.621	&	148	$\pm$	3.26	&	3670	$\pm$	80.9	&	1.59	\\
49	&	84.693965	&	-69.084531	&	0.181	&	33.2	$\pm$	0.84	&	54.4	$\pm$	1.38	&	0.361	\\
50	&	84.702648	&	-69.079743	&	2.18	&	174	$\pm$	8.08	&	26400	$\pm$	1230	&	2.28	\\
51	&	84.719989	&	-69.076724	&	1.62	&	174	$\pm$	7.41	&	22300	$\pm$	949	&	3.41	\\
52	&	84.720925	&	-69.076775	&	1.31	&	201	$\pm$	6.44	&	19300	$\pm$	621	&	3.30	\\
53	&	84.693749	&	-69.082392	&	0.171	&	36.5	$\pm$	0.85	&	62.1	$\pm$	1.45	&	0.941	\\
54	&	84.702724	&	-69.079712	&	2.16	&	175	$\pm$	7.79	&	24700	$\pm$	1090	&	2.21	\\
55	&	84.704743	&	-69.078626	&	1.22	&	211	$\pm$	5.87	&	16900	$\pm$	471	&	2.34	\\
56	&	84.721886	&	-69.076712	&	1.08	&	225	$\pm$	5.57	&	16300	$\pm$	404	&	3.17	\\
57	&	84.715112		&	-69.070094	&	0.295	&	36.3	$\pm$	1.53	&	201	$\pm$	8.47	&	0.973	\\
58	&	84.699190	&	-69.069651	&	1.34	&	63	$\pm$	3.74	&	2040	$\pm$	121	&	1.06	\\
59	&	84.699174	&	-69.069684	&	1.35	&	60.6	$\pm$	3.92	&	2170	$\pm$	141	&	1.06	\\
60	&	84.698488	&	-69.069178	&	0.919	&	70.0	$\pm$	2.7	&	1190	$\pm$	45.9	&	1.10	\\
61	&	84.696900	&	-69.082925	&	1.02	&	115	$\pm$	4.56	&	5590	$\pm$	222	&	1.36	\\
62	&	84.704867	&	-69.078592	&	1.15	&	221	$\pm$	5.39	&	15100	$\pm$	368	&	2.31	\\
63	&	84.695507	&	-69.078140	&	0.954	&	151	$\pm$	4.33	&	6540	$\pm$	189	&	2.07	\\
64	&	84.721953	&	-69.076703	&	0.987	&	242	$\pm$	4.84	&	13300	$\pm$	266	&	2.95	\\
65	&	84.710659	&	-69.069206	&	0.731	&	13.3	$\pm$	2.98	&	277	$\pm$	62.1	&	1.08	\\
66	&	84.696763	&	-69.069577	&	0.423	&	64.1	$\pm$	1.85	&	511	$\pm$	14.7	&	1.03	\\
67	&	84.695439	&	-69.084713	&	0.202	&	90.3	$\pm$	0.990	&	206	$\pm$	2.26	&	0.645	\\
68	&	84.705060	&	-69.078616	&	0.95	&	203	$\pm$	4.18	&	83100	$\pm$	1710	&	2.01	\\
69	&	84.700879	&	-69.068490	&	0.321	&	71.6	$\pm$	1.46	&	359	$\pm$	7.32	&	0.949	\\
70	&	84.705644	&	-69.078739	&	0.634	&	235	$\pm$	3.07	&	5170	$\pm$	67.5	&	1.34	\\
71	&	84.695601	&	-69.078200	&	0.798	&	147	$\pm$	3.60	&	4440	$\pm$	109	&	1.94	\\
72	&	84.695392	&	-69.078432	&	0.688	&	150	$\pm$	3.08	&	3310	$\pm$	67.9	&	1.87	\\
73	&	84.700724	&	-69.070683	&	0.419	&	48.6	$\pm$	2.05	&	477	$\pm$	20.1	&	0.769	\\
74	&	84.721433	&	-69.076804	&	0.704	&	248	$\pm$	3.45	&	6870	$\pm$	95.6	&	2.45	\\
75	&	84.688778	&	-69.077411	&	0.932	&	131	$\pm$	4.09	&	5110	$\pm$	159	&	1.24	\\
76	&	84.690076	&	-69.075857	&	0.328	&	83.4	$\pm$	1.59	&	490	$\pm$	9.34	&	1.11	\\
77	&	84.697577	&	-69.070949	&	0.179	&	17.5	$\pm$	0.820	&	27.7	$\pm$	1.29	&	0.263	\\
78	&	84.704109	&	-69.068918	&	0.501	&	37.7	$\pm$	2.73	&	657	$\pm$	47.6	&	1.13	\\
79	&	84.681270	&	-69.089642	&	0.273	&	36.3	$\pm$	1.39	&	163	$\pm$	6.24	&	1.18	\\
80	&	84.683899	&	-69.087016	&	0.414	&	54.4	$\pm$	2.32	&	681	$\pm$	29.1	&	1.00	\\
81	&	84.696870	&	-69.082978	&	0.758	&	109	$\pm$	3.34	&	2810	$\pm$	86.1	&	0.993	\\
82	&	84.688625	&	-69.077588	&	0.751	&	142	$\pm$	3.70	&	4530	$\pm$	118	&	1.21	\\
83	&	84.714477	&	-69.077183	&	0.61	&	103	$\pm$	3.10	&	2310	$\pm$	69.5	&	1.66	\\
84	&	84.681043	&	-69.088247	&	0.272	&	25.6	$\pm$	1.41	&	118	$\pm$	6.49	&	0.951	\\
85	&	84.692264	&	-69.083296	&	0.211	&	33.6	$\pm$	0.990	&	76.6	$\pm$	2.26	&	0.413	\\
86	&	84.701799	&	-69.079815	&	0.332	&	68.9	$\pm$	1.54	&	384	$\pm$	8.58	&	0.571	\\
87	&	84.692210	&	-69.079778	&	0.184	&	52.7	$\pm$	0.840	&	86.7	$\pm$	1.38	&	0.495	\\
88	&	84.698923	&	-69.070512	&	0.253	&	39.4	$\pm$	1.24	&	142	$\pm$	4.47	&	0.614	\\
89	&	84.696715	&	-69.069817	&	0.209	&	19.9	$\pm$	1.04	&	50.3	$\pm$	2.63	&	0.357	\\
90	&	84.692520	&	-69.083699	&	0.278	&	35.7	$\pm$	1.36	&	154	$\pm$	5.87	&	0.529	\\
91	&	84.697617	&	-69.076296	&	0.218	&	78.5	$\pm$	1.03	&	194	$\pm$	2.54	&	0.814	\\
92	&	84.703864	&	-69.079984	&	0.162	&	66.4	$\pm$	0.750	&	88.1	$\pm$	0.995	&	0.494	\\
93	&	84.711315		&	-69.080377	&	0.322	&	9.25	$\pm$	1.57	&	53.1	$\pm$	9.01	&	0.403	\\
94	&	84.694719	&	-69.079030	&	0.306	&	166	$\pm$	1.48	&	851	$\pm$	7.59	&	0.976	\\
95	&	84.710370	&	-69.079071	&	0.311	&	36.2	$\pm$	1.54	&	201	$\pm$	8.55	&	0.492	\\
96	&	84.702161	&	-69.078375	&	0.455	&	85.9	$\pm$	1.99	&	793	$\pm$	18.4	&	0.842	\\
97	&	84.703010	&	-69.078498	&	0.251	&	82.0	$\pm$	1.19	&	271	$\pm$	3.93	&	0.422	\\
98	&	84.696271	&	-69.077918	&	0.194	&	68.2	$\pm$	0.890	&	125	$\pm$	1.63	&	0.469	\\
99	&	84.724458	&	-69.075500	&	0.165	&	40.9	$\pm$	0.770	&	56.9	$\pm$	1.07	&	0.361	\\
100	&	84.701233	&	-69.078210	&	0.284	&	63.6	$\pm$	1.25	&	233	$\pm$	4.58	&	0.383	\\
101	&	84.727933	&	-69.079484	&	0.738	&	11.7	$\pm$	2.72	&	203	$\pm$	47.2	&	0.934	\\
102	&	84.728497	&	-69.079548	&	0.842	&	11.9	$\pm$	2.98	&	249	$\pm$	62.4	&	1.00	\\
103	&	84.712801	&	-69.078013	&	0.273	&	31.5	$\pm$	1.40	&	143	$\pm$	6.34	&	0.965	\\
104	&	84.709897	&	-69.076059	&	0.664	&	60.7	$\pm$	3.18	&	1440	$\pm$	75.4	&	1.49	\\
105	&	84.731833	&	-69.079910	&	0.215	&	13.8	$\pm$	1.07	&	36.6	$\pm$	2.84	&	0.774	\\
106	&	84.716524	&	-69.077616	&	0.247	&	27.3	$\pm$	1.15	&	84.6	$\pm$	3.56	&	0.368	\\
107	&	84.702366	&	-69.075318	&	0.366	&	25.7	$\pm$	1.85	&	206	$\pm$	14.8	&	0.816	\\
108	&	84.728531	&	-69.083668	&	0.605	&	22.5	$\pm$	2.89	&	439	$\pm$	56.4	&	1.17	\\
109	&	84.728261	&	-69.083084	&	0.244	&	20.5	$\pm$	1.20	&	68.7	$\pm$	4.02	&	0.631	\\
110	&	84.726557	&	-69.081757	&	0.398	&	34.1	$\pm$	2.15	&	367	$\pm$	23.1	&	0.817	\\
111	&	84.668852	&	-69.081488	&	0.315	&	22.9	$\pm$	1.59	&	134	$\pm$	9.31	&	0.943	\\
112	&	84.728558	&	-69.084019	&	0.35	0	&	24.9	$\pm$	1.73	&	173	$\pm$	12.01	&	0.569	\\
113	&	84.723553	&	-69.078390	&	0.524	&	20.8	$\pm$	2.40	&	179	$\pm$	20.7	&	1.04	\\
114	&	84.669455	&	-69.075850	&	0.262	&	15.8	$\pm$	1.30	&	61.7	$\pm$	5.08	&	0.655	\\
115	&	84.729553	&	-69.081604	&	0.267	&	14.3	$\pm$	1.30	&	55.9	$\pm$	5.08	&	0.469	\\
\enddata
\tablecomments{Column 1: The ID we give each clump. Column 4: The geometric mean radius calculated using the major and minor axis of the best-fit ellipse: $\mathrm{\sqrt{major\;axis\;x\;minor\;axis}}$. Column 6: The total $\mathrm{H_{2}}$ mass. We use a $\mathrm{X_{CO}}$ conversion factor of $8.4 \pm 3\;\mathrm{M_{\odot}\;pc^{-2}\;(K\;km\;s^{-1})^{-1}}$ \citep{inde13}. Column 7: The measured linewidth.}
\end{deluxetable}

\setcounter{table}{1} 
\renewcommand{\thetable}{\arabic{table}}
\tabletypesize{\tiny}
\begin{deluxetable}{cccccccc}
\tablewidth{0pt}
\tablecaption{Column Densities and Mass Derived from $\mathrm{^{13}CO}$ (2-1)}
\tablehead{\colhead{Clump ID} & \colhead{RA} & \colhead{Dec} & \colhead{Radius [$\mathrm{pc}$]} & \colhead{$\mathrm{N_{^{13}CO}}$ [$\mathrm{cm^{-2}}$]} & \colhead{$\mathrm{N_{H_{2}}}$ [$\mathrm{cm^{-2}}$]} & \colhead{Total $\mathrm{H_{2}}$ Mass [$\mathrm{M_{\odot}}$]} & \colhead{Linewidth [km/s]}}
\startdata
0	&	84.689449	&	-69.082378	&	0.373	&	7.44	$\pm$	0.13	$\times	10^{15}	$	&	3.72	$\pm$	0.07	$\times	10^{22}	$	&	297	$\pm$	5.59	&	0.394	\\
1	&	84.687939	&	-69.081210	&	0.320	&	6.87	$\pm$	0.11	$\times	10^{15}	$	&	3.43	$\pm$	0.05	$\times	10^{22}	$	&	183	$\pm$	2.67	&	0.388	\\
2	&	84.709746	&	-69.073602	&	0.936	&	2.02	$\pm$	0.32	$\times	10^{16}	$	&	1.01	$\pm$	0.02	$\times	10^{23}	$	&	4510	$\pm$	89.3	&	1.87	\\
3	&	84.711099	&	-69.073085	&	0.501	&	2.39	$\pm$	0.21	$\times	10^{16}	$	&	1.19	$\pm$	0.01	$\times	10^{23}	$	&	2310	$\pm$	19.4	&	1.86	\\
4	&	84.695409	&	-69.082384	&	0.347	&	7.39	$\pm$	0.13	$\times	10^{15}	$	&	3.69	$\pm$	0.07	$\times	10^{22}	$	&	281	$\pm$	5.33	&	0.477	\\
5	&	84.708707	&	-69.079363	&	0.312	&	6.13	$\pm$	0.07	$\times	10^{15}	$	&	3.06	$\pm$	0.04	$\times	10^{22}	$	&	71.5	$\pm$	0.935	&	0.556	\\
6	&	84.701895	&	-69.081726	&	0.359	&	9.14	$\pm$	0.14	$\times	10^{15}	$	&	4.57	$\pm$	0.07	$\times	10^{22}	$	&	396	$\pm$	6.07	&	0.665	\\
7	&	84.691777	&	-69.086031	&	0.226	&	5.56	$\pm$	0.07	$\times	10^{15}	$	&	2.78	$\pm$	0.01	$\times	10^{22}	$	&	53.3	$\pm$	0.192	&	0.429	\\
8	&	84.687966	&	-69.085250	&	0.434	&	3.38	$\pm$	0.20	$\times	10^{16}	$	&	1.69	$\pm$	0.10	$\times	10^{23}	$	&	2910	$\pm$	172	&	0.969	\\
9	&	84.721610	&	-69.076879	&	0.837	&	4.96	$\pm$	0.39	$\times	10^{16}	$	&	2.48	$\pm$	0.19	$\times	10^{23}	$	&	16700	$\pm$	1280	&	2.81	\\
10	&	84.697005	&	-69.069689	&	0.571	&	1.74	$\pm$	0.22	$\times	10^{16}	$	&	8.68	$\pm$	0.01	$\times	10^{22}	$	&	1960	$\pm$	2.26	&	1.11	\\
11	&	84.705166	&	-69.078640	&	1.010	&	4.11	$\pm$	0.46	$\times	10^{16}	$	&	2.06	$\pm$	0.02	$\times	10^{23}	$	&	19300	$\pm$	187	&	1.94	\\
12	&	84.707623	&	-69.074421	&	0.515	&	1.57	$\pm$	0.20	$\times	10^{16}	$	&	7.86	$\pm$	0.98	$\times	10^{22}	$	&	1350	$\pm$	168	&	0.774	\\
13	&	84.697312	&	-69.082689	&	1.170	&	1.55	$\pm$	0.41	$\times	10^{16}	$	&	7.73	$\pm$	0.21	$\times	10^{22}	$	&	5920	$\pm$	161	&	1.15	\\
14	&	84.696562	&	-69.083241	&	0.849	&	1.78	$\pm$	0.33	$\times	10^{16}	$	&	8.90	$\pm$	0.02	$\times	10^{22}	$	&	4380	$\pm$	9.84	&	1.16	\\
15	&	84.695307	&	-69.084753	&	0.294	&	1.23	$\pm$	0.12	$\times	10^{16}	$	&	6.15	$\pm$	0.06	$\times	10^{22}	$	&	385	$\pm$	3.76	&	0.579	\\
16	&	84.693610	&	-69.082348	&	0.323	&	6.25	$\pm$	0.09	$\times	10^{15}	$	&	3.12	$\pm$	0.04	$\times	10^{22}	$	&	112	$\pm$	1.44	&	0.553	\\
17	&	84.705443	&	-69.078644	&	0.784	&	5.30	$\pm$	0.36	$\times	10^{16}	$	&	2.65	$\pm$	0.02	$\times	10^{23}	$	&	15400	$\pm$	116	&	1.51	\\
18	&	84.705570	&	-69.078625	&	0.654	&	6.73	$\pm$	0.29	$\times	10^{16}	$	&	3.37	$\pm$	0.01	$\times	10^{23}	$	&	12700	$\pm$	37.7	&	1.30	\\
19	&	84.715115	&	-69.070060	&	0.261	&	9.88	$\pm$	0.10	$\times	10^{15}	$	&	4.94	$\pm$	0.05	$\times	10^{22}	$	&	214	$\pm$	2.17	&	0.482	\\
20	&	84.696728	&	-69.069400	&	0.375	&	1.56	$\pm$	0.14	$\times	10^{16}	$	&	7.82	$\pm$	0.07	$\times	10^{22}	$	&	711	$\pm$	6.36	&	0.419	\\
21	&	84.710215	&	-69.069043	&	0.333	&	4.54	$\pm$	0.09	$\times	10^{15}	$	&	2.27	$\pm$	0.04	$\times	10^{22}	$	&	81.4	$\pm$	1.43	&	0.279	\\
22	&	84.700917	&	-69.068526	&	0.317	&	1.69	$\pm$	0.13	$\times	10^{16}	$	&	8.44	$\pm$	0.06	$\times	10^{22}	$	&	612	$\pm$	4.35	&	0.727	\\
23	&	84.700929	&	-69.070926	&	0.566	&	5.82	$\pm$	0.18	$\times	10^{15}	$	&	2.91	$\pm$	0.88	$\times	10^{22}	$	&	405	$\pm$	122	&	0.607	\\
24	&	84.714332	&	-69.077009	&	0.682	&	1.52	$\pm$	0.27	$\times	10^{16}	$	&	7.60	$\pm$	0.01	$\times	10^{22}	$	&	2520	$\pm$	3.32	&	1.58	\\
25	&	84.696657	&	-69.083134	&	0.688	&	2.06	$\pm$	0.28	$\times	10^{16}	$	&	1.03	$\pm$	0.01	$\times	10^{23}	$	&	3660	$\pm$	35.5	&	0.906	\\
26	&	84.697115	&	-69.070046	&	0.308	&	1.03	$\pm$	0.12	$\times	10^{16}	$	&	5.13	$\pm$	0.06	$\times	10^{22}	$	&	325	$\pm$	3.80	&	0.266	\\
27	&	84.683912	&	-69.087133	&	0.275	&	8.84	$\pm$	0.11	$\times	10^{15}	$	&	4.42	$\pm$	0.05	$\times	10^{22}	$	&	232	$\pm$	2.62	&	0.482	\\
28	&	84.701028	&	-69.080010	&	0.463	&	1.03	$\pm$	0.18	$\times	10^{16}	$	&	5.16	$\pm$	0.09	$\times	10^{22}	$	&	735	$\pm$	12.8	&	0.629	\\
29	&	84.688237	&	-69.077655	&	0.673	&	2.34	$\pm$	0.28	$\times	10^{16}	$	&	1.17	$\pm$	0.01	$\times	10^{23}	$	&	4210	$\pm$	36.0	&	1.03	\\
30	&	84.688363	&	-69.077495	&	0.843	&	2.03	$\pm$	0.32	$\times	10^{16}	$	&	1.01	$\pm$	0.02	$\times	10^{23}	$	&	4760	$\pm$	94.3	&	1.00	\\
31	&	84.714727	&	-69.077451	&	0.332	&	1.31	$\pm$	0.14	$\times	10^{16}	$	&	6.55	$\pm$	0.07	$\times	10^{22}	$	&	563	$\pm$	6.02	&	0.529	\\
32	&	84.681295	&	-69.089657	&	0.237	&	5.46	$\pm$	0.08	$\times	10^{15}	$	&	2.73	$\pm$	0.04	$\times	10^{22}	$	&	72.7	$\pm$	1.07	&	0.374	\\
33	&	84.690069	&	-69.085793	&	0.299	&	6.63	$\pm$	0.11	$\times	10^{15}	$	&	3.31	$\pm$	0.54	$\times	10^{22}	$	&	174	$\pm$	28.4	&	0.523	\\
34	&	84.694844	&	-69.078758	&	0.756	&	1.89	$\pm$	0.30	$\times	10^{16}	$	&	9.47	$\pm$	0.02	$\times	10^{22}	$	&	3870	$\pm$	8.17	&	1.14	\\
35	&	84.689538	&	-69.076042	&	0.365	&	9.50	$\pm$	0.14	$\times	10^{15}	$	&	4.75	$\pm$	0.07	$\times	10^{22}	$	&	443	$\pm$	6.53	&	0.545	\\
36	&	84.695008	&	-69.078698	&	0.680	&	2.22	$\pm$	0.26	$\times	10^{16}	$	&	1.11	$\pm$	0.01	$\times	10^{23}	$	&	3410	$\pm$	30.7	&	1.01	\\
37	&	84.714061	&	-69.076682	&	0.481	&	1.36	$\pm$	0.19	$\times	10^{16}	$	&	6.80	$\pm$	0.09	$\times	10^{22}	$	&	1070	$\pm$	14.2	&	0.585	\\
38	&	84.692053	&	-69.079906	&	0.386	&	5.14	$\pm$	0.13	$\times	10^{15}	$	&	2.57	$\pm$	0.01	$\times	10^{22}	$	&	201	$\pm$	0.782	&	0.356	\\
39	&	84.696185	&	-69.077785	&	0.336	&	2.02	$\pm$	0.14	$\times	10^{16}	$	&	1.01	$\pm$	0.07	$\times	10^{23}	$	&	934	$\pm$	64.7	&	1.07	\\
40	&	84.703806	&	-69.080023	&	0.208	&	1.18	$\pm$	0.08	$\times	10^{16}	$	&	5.90	$\pm$	0.01	$\times	10^{22}	$	&	182	$\pm$	0.308	&	0.387	\\
41	&	84.694412	&	-69.079160	&	0.422	&	2.50	$\pm$	0.19	$\times	10^{16}	$	&	1.25	$\pm$	0.10	$\times	10^{23}	$	&	2050	$\pm$	164	&	0.820	\\
42	&	84.692162	&	-69.083837	&	0.314	&	5.88	$\pm$	0.12	$\times	10^{15}	$	&	2.94	$\pm$	0.06	$\times	10^{22}	$	&	184	$\pm$	3.76	&	0.358	\\
43	&	84.701180 	&	-69.078147	&	0.349	&	1.02	$\pm$	0.14	$\times	10^{16}	$	&	5.10	$\pm$	0.07	$\times	10^{22}	$	&	438	$\pm$	6.01	&	0.407	\\
44	&	84.728132	&	-69.083090	&	0.270	&	9.00	$\pm$	0.10	$\times	10^{15}	$	&	4.50	$\pm$	0.05	$\times	10^{22}	$	&	206	$\pm$	2.29	&	0.706	\\
45	&	84.726124	&	-69.081820	&	0.306	&	9.63	$\pm$	0.12	$\times	10^{15}	$	&	4.81	$\pm$	0.06	$\times	10^{22}	$	&	285	$\pm$	3.56	&	0.451	\\
46	&	84.668881	&	-69.081515	&	0.263	&	8.44	$\pm$	0.10	$\times	10^{15}	$	&	4.22	$\pm$	0.05	$\times	10^{22}	$	&	183	$\pm$	2.17	&	0.873	\\
\enddata
\tablecomments{Column 1: The ID we give each clump. Column 4: The geometric mean radius calculated using the major and minor axis of the best-fit ellipse: $\mathrm{\sqrt{major\;axis\;x\;minor\;axis}}$. Column 6: The $\mathrm{H_{2}}$ column density. We use a $\mathrm{\frac{H_{2}}{^{13}CO (2-1)}}$ conversion factor of $5 \times 10^{6}$ \citep{inde13}. Column 7: The measured linewidth.}
\end{deluxetable}

\setcounter{table}{2} 
\renewcommand{\thetable}{\arabic{table}}
\tabletypesize{\tiny}
\begin{deluxetable}{cccccccccccccccccc}
\rotate
\tablewidth{0pt}
\tablecaption{Properties of More Evolved Low-Mass YSO Candidates}
\tablehead{\colhead{ID} & \colhead{RA} & \colhead{Dec} & \colhead{mag275} & \colhead{err275} &\colhead{mag336} &\colhead{err336} &\colhead{mag555} &\colhead{err555} &\colhead{mag658} &\colhead{err658} &\colhead{mag775} &\colhead{err775} &\colhead{mag110} &\colhead{err110} &\colhead{mag160} &\colhead{err160} &\colhead{Mass [$\mathrm{M_{\odot}}$]}}
\startdata
1	&	84.701393	&	-69.091209	&	23.952	&	11.111	&	23.267	&	0.01	&	22.361	&	0.038	&	21.595	&	0.046	&	21.684	&	0.08	&	21.096	&	0.014	&	20.713	&	0.037	&	2.5	\\
2	&	84.700844	&	-69.091339	&	25.854	&	1.585	&	24.366	&	0.394	&	22.735	&	0.016	&	21.325	&	0.039	&	21.561	&	0.03	&	20.605	&	0.032	&	19.721	&	0.077	&	2.5	\\
3	&	84.698189	&	-69.092331	&	27.005	&	3.316	&	25.313	&	0.537	&	23.901	&	0.063	&	22.201	&	0.079	&	22.474	&	0.1	&	20.977	&	0.332	&	19.319	&	0.278	&	1.5	\\
4	&	84.694267	&	-69.089218	&	20.805	&	0.01	&	20.652	&	0.063	&	20.515	&	0.03	&	19.76	&	0.043	&	20.274	&	0.0	&	20.184	&	0.061	&	20.006	&	0.05	&	3.0	\\
5	&	84.690453	&	-69.090652	&	19.027	&	0.049	&	18.941	&	0.005	&	19.438	&	0.008	&	19.063	&	0.015	&	19.232	&	0.00	&	19.077	&	0.004	&	18.987	&	0.002	&	3.0	\\
6	&	84.685699	&	-69.091995	&	23.538	&	0.735	&	22.406	&	0.243	&	21.454	&	0.02	&	20.393	&	0.074	&	20.615	&	0.029	&	19.928	&	0.049	&	19.359	&	0.006	&	3.0	\\
7	&	84.728310	&	-69.079735	&	24.373	&	0.354	&	23.11	&	0.293	&	21.554	&	0.027	&	20.095	&	0.061	&	20.297	&	0.02	&	19.147	&	0.041	&	18.088	&	0.1	&	3.0	\\
8	&	84.688332	&	-69.090294	&	24.766	&	0.465	&	23.007	&	0.053	&	21.771	&	0.031	&	20.535	&	0.047	&	20.521	&	0.005	&	19.52	&	0.009	&	18.598	&	0.005	&	3.0	\\
9	&	84.696457	&	-69.088234	&	21.887	&	0.088	&	20.39	&	0.01	&	19.361	&	0.014	&	18.385	&	0.032	&	18.446	&	0	&	17.832	&	0.002	&	17.322	&	0.001	&	3.0	\\
10	&	84.718658	&	-69.081444	&	25.214	&	0.731	&	24.044	&	0.171	&	23.416	&	0.044	&	21.513	&	0.014	&	22.818	&	0.04	&	22.106	&	0.001	&	21.737	&	0.065	&	1.5	\\
11	&	84.700165	&	-69.087601	&	19.105	&	0.024	&	19.102	&	0.004	&	19.518	&	0.008	&	18.823	&	0.098	&	19.204	&	0.01	&	18.895	&	0.001	&	18.736	&	0.036	&	3.0	\\
12	&	84.689651	&	-69.089500	&	23.131	&	0.061	&	22.369	&	0.027	&	21.569	&	0.023	&	20.592	&	0.091	&	21.036	&	0	&	20.586	&	0.16	&	20.271	&	0.046	&	3.0	\\
13	&	84.685539	&	-69.090813	&	24.617	&	11.111	&	24.09	&	0.18	&	22.892	&	0.035	&	21.34	&	0.083	&	21.679	&	0	&	20.784	&	0.054	&	19.943	&	0.001	&	2.5	\\
14	&	84.718697	&	-69.080963	&	25.410	&	11.111	&	24.707	&	0.062	&	23.868	&	0.084	&	21.977	&	0.099	&	23.175	&	0	&	22.14	&	0.082	&	21.6	&	0.135	&	1.5	\\
15	&	84.716599	&	-69.080612	&	23.791	&	0.458	&	22.922	&	0.098	&	22.384	&	0.008	&	21.402	&	0.082	&	21.76	&	0.03	&	21.261	&	0.076	&	20.918	&	0.011	&	2.5	\\
16	&	84.694611	&	-69.086166	&	22.047	&	0.065	&	21.775	&	11.111	&	21.237	&	0.051	&	20.233	&	0.021	&	21.03	&	0.05	&	20.36	&	0.097	&	20.093	&	0.035	&	3.0	\\
17	&	84.687714	&	-69.089203	&	22.63	&	0.011	&	22.081	&	0.019	&	21.387	&	0.023	&	19.703	&	0.069	&	21.167	&	0.01	&	20.579	&	0.006	&	20.337	&	0.022	&	3.0	\\
18	&	84.685844	&	-69.089600	&	18.848	&	0.00	&	18.821	&	0.006	&	19.425	&	0.017	&	18.702	&	0.02	&	19.221	&	0	&	18.996	&	0.001	&	18.802	&	0.001	&	3.0	\\
19	&	84.681122	&	-69.090309	&	19.398	&	0.003	&	19.305	&	0.003	&	19.657	&	0.017	&	19.152	&	0.02	&	19.279	&	0.0	&	19.19	&	0.223	&	19.462	&	0.097	&	3.0	\\
20	&	84.680565	&	-69.089874	&	24.373	&	0.302	&	24.17	&	0.099	&	23.191	&	0.02	&	20.741	&	0.039	&	22.118	&	0.04	&	20.806	&	0.043	&	20.375	&	0.007	&	3.0	\\
21	&	84.715660	&	-69.079575	&	25.377	&	2.033	&	24.136	&	0.134	&	23.716	&	0.009	&	22.409	&	0.048	&	22.919	&	0.01	&	22.427	&	0.011	&	22.003	&	0.099	&	1.5	\\
22	&	84.709999	&	-69.080421	&	19.673	&	0.011	&	19.516	&	0.003	&	19.604	&	0.008	&	18.975	&	0.012	&	19.046	&	0.01	&	18.619	&	0.074	&	18.288	&	0.002	&	3.0	\\
23	&	84.711174	&	-69.080360	&	24.552	&	0.313	&	23.915	&	0.021	&	23.306	&	0.032	&	21.728	&	0.099	&	21.856	&	0.00	&	20.13	&	0.042	&	19.034	&	0.01	&	3.0	\\
24	&	84.707649	&	-69.081543	&	22.929	&	0.054	&	21.278	&	0.015	&	19.564	&	0.007	&	18.4	&	0.031	&	18.459	&	0	&	17.686	&	0.022	&	17.075	&	0.001	&	3.0	\\
25	&	84.692261	&	-69.085777	&	25.221	&	0.943	&	24.795	&	0.516	&	23.432	&	0.055	&	21.361	&	0.045	&	22.348	&	0.09	&	20.886	&	0.094	&	20.211	&	0.077	&	2.5	\\
26	&	84.694832	&	-69.085495	&	21.403	&	0.007	&	21.625	&	0.069	&	21.233	&	0.006	&	19.054	&	0.028	&	20.674	&	0.0	&	19.643	&	0.009	&	18.66	&	0.01	&	3.0	\\
27	&	84.693153	&	-69.085175	&	25.468	&	1.708	&	23.337	&	11.111	&	22.243	&	0.091	&	20.213	&	0.045	&	21.772	&	0.01	&	20.779	&	0.452	&	20.256	&	0.063	&	2.5	\\
28	&	84.694252	&	-69.085899	&	25.125	&	3.617	&	23.721	&	0.402	&	22.436	&	0.058	&	21.091	&	0.012	&	21.279	&	0.05	&	20.438	&	0.036	&	19.671	&	0.087	&	2.5	\\
29	&	84.697914	&	-69.083656	&	18.271	&	0.003	&	18.413	&	0.029	&	19.125	&	0.008	&	18.453	&	0.044	&	18.538	&	0.0	&	17.826	&	0.004	&	17.313	&	0.006	&	3.0	\\
30	&	84.698174	&	-69.084839	&	24.63	&	0.767	&	23.609	&	0.121	&	23.154	&	0.074	&	21.342	&	0.071	&	22.72	&	0.0	&	21.912	&	0.143	&	21.493	&	0.042	&	2.0	\\
31	&	84.688438	&	-69.086906	&	25.489	&	0.289	&	25.063	&	11.111	&	23.423	&	0.069	&	20.814	&	0.039	&	22.788	&	0.01	&	21.08	&	0.294	&	20.035	&	0.066	&	2.5	\\
32	&	84.686890	&	-69.086388	&	21.415	&	0.015	&	21.337	&	0.042	&	21.163	&	0.023	&	20.128	&	0.093	&	20.792	&	0.00	&	20.146	&	0.262	&	19.701	&	0.042	&	3.0	\\
33	&	84.686668	&	-69.087700	&	20.576	&	0.117	&	20.288	&	11.111	&	20.394	&	0.027	&	19.853	&	0.011	&	20.023	&	0.01	&	19.66	&	0.158	&	19.494	&	0.046	&	3.0	\\
34	&	84.681587	&	-69.088242	&	24.7	&	0.209	&	23.442	&	11.111	&	22.992	&	0.074	&	21.031	&	0.096	&	22.229	&	0.0	&	21.314	&	0.364	&	20.741	&	0.03	&	2.0	\\
35	&	84.680145	&	-69.089180	&	18.576	&	0.03	&	18.55	&	0.004	&	19.181	&	0.017	&	18.656	&	0.06	&	18.954	&	0.01	&	18.824	&	0.013	&	18.727	&	0.001	&	3.0	\\
36	&	84.680260	&	-69.089043	&	19.101	&	0.00	&	19.088	&	0.003	&	19.562	&	0.011	&	18.34	&	0.043	&	19.323	&	0.01	&	19.164	&	0.015	&	18.643	&	0.004	&	3.0	\\
37	&	84.681496	&	-69.089035	&	20.331	&	0.203	&	20.078	&	0.028	&	20.255	&	0.015	&	19.576	&	0.095	&	20.048	&	0.0	&	19.63	&	0.156	&	19.473	&	0.073	&	3.0	\\
38	&	84.680847	&	-69.088844	&	25.694	&	4.887	&	23.776	&	0.315	&	22.364	&	0.018	&	20.934	&	0.085	&	21.213	&	0.0	&	20.189	&	0.047	&	19.587	&	0.024	&	3.0	\\
39	&	84.679916	&	-69.088379	&	24.738	&	0.994	&	22.854	&	11.111	&	21.546	&	0.004	&	20.299	&	0.019	&	20.532	&	0.026	&	19.619	&	0.045	&	18.926	&	0.016	&	3.0	\\
40	&	84.680016	&	-69.088806	&	23.422	&	0.1	&	21.996	&	11.111	&	21.067	&	0.027	&	19.927	&	0.079	&	20.257	&	0.0	&	19.523	&	0.058	&	19.065	&	0.023	&	3.0	\\
41	&	84.719650	&	-69.077087	&	24.578	&	0.162	&	23.923	&	11.111	&	22.9	&	0.025	&	20.202	&	0.078	&	20.897	&	0.026	&	17.669	&	0.013	&	15.892	&	0.039	&	3.0	\\
42	&	84.719276	&	-69.077477	&	24.742	&	1.02	&	23.502	&	0.342	&	23.176	&	0.038	&	22.133	&	0.078	&	22.558	&	0.069	&	22.158	&	0.02	&	21.849	&	0.003	&	1.5	\\
43	&	84.725426	&	-69.076439	&	25.037	&	0.87	&	24.234	&	0.302	&	22.865	&	0.024	&	21.62	&	0.07	&	21.627	&	0.05	&	20.665	&	0.02	&	19.904	&	0.005	&	2.5	\\
44	&	84.694695	&	-69.083130	&	23.626	&	0.11	&	22.407	&	0.133	&	21.854	&	0.085	&	20.873	&	0.073	&	21.394	&	0.02	&	20.799	&	0.315	&	20.552	&	0.145	&	3.0	\\
45	&	84.692535	&	-69.081985	&	23.418	&	0.21	&	23.36	&	0.266	&	22.533	&	0.036	&	21.333	&	0.074	&	21.395	&	0.02	&	19.786	&	0.115	&	18.943	&	0.133	&	3.0	\\
46	&	84.681580	&	-69.084602	&	20.725	&	0.289	&	20.002	&	0.009	&	20.064	&	0.012	&	18.832	&	0.005	&	19.385	&	0.02	&	18.452	&	0.019	&	17.501	&	0.009	&	3.0	\\
47	&	84.679291	&	-69.084869	&	23.978	&	0.619	&	23.60	&	0.086	&	22.44	&	0.028	&	21.321	&	0.095	&	21.499	&	0.04	&	20.656	&	0.055	&	20.084	&	0.058	&	2.5	\\
48	&	84.680473	&	-69.085472	&	24.218	&	0.904	&	23.013	&	0.244	&	21.825	&	0.04	&	20.47	&	0.084	&	20.816	&	0.02	&	19.806	&	0.199	&	19.308	&	0.032	&	3.0	\\
49	&	84.679359	&	-69.086403	&	22.02	&	11.111	&	21.624	&	0.008	&	21.111	&	0.027	&	19.982	&	0.017	&	20.838	&	0.032	&	20.168	&	0.058	&	19.949	&	0.162	&	3.0	\\
50	&	84.701096	&	-69.077980	&	19.052	&	0.014	&	19.115	&	0.006	&	19.62	&	0.019	&	18.665	&	0.043	&	19.086	&	0.01	&	18.484	&	0.01	&	18.091	&	0.004	&	3.0	\\
51	&	84.692154	&	-69.080086	&	19.636	&	0.017	&	19.721	&	0.022	&	20.027	&	0.012	&	18.825	&	0.02	&	19.17	&	0.004	&	18.189	&	0.003	&	17.624	&	0.01	&	3.0	\\
52	&	84.688080	&	-69.081558	&	22.126	&	0.247	&	21.674	&	0.034	&	21.073	&	0.042	&	20.211	&	0.028	&	20.356	&	0.003	&	19.527	&	0.121	&	18.864	&	0.013	&	3.0	\\
53	&	84.688301	&	-69.081680	&	22.141	&	11.111	&	21.955	&	0.09	&	21.773	&	0.042	&	20.386	&	0.078	&	21.3	&	0.041	&	20.195	&	0.208	&	19.896	&	0.042	&	3.0	\\
54	&	84.684372	&	-69.083496	&	26.444	&	0.638	&	24.486	&	11.111	&	23.044	&	0.068	&	21.17	&	0.095	&	22.455	&	0.091	&	21.227	&	0.022	&	21.037	&	0.027	&	2.5	\\
55	&	84.677231	&	-69.085037	&	24.325	&	2.261	&	24.993	&	0.17	&	23.349	&	0.04	&	22.139	&	0.093	&	22.324	&	0.04	&	21.53	&	0.191	&	20.809	&	0.091	&	2.0	\\
56	&	84.691353	&	-69.079071	&	23.208	&	0.027	&	22.895	&	0.127	&	22.093	&	0.015	&	21.138	&	0.055	&	21.538	&	0.01	&	20.857	&	0.025	&	20.525	&	0.094	&	3.0	\\
57	&	84.684090	&	-69.081886	&	22.047	&	0.199	&	21.94	&	0.02	&	22.037	&	0.04	&	20.575	&	0.084	&	21.282	&	0.025	&	20.26	&	0.088	&	19.667	&	0.021	&	3.0	\\
58	&	84.686569	&	-69.080055	&	18.204	&	0.011	&	18.377	&	0.003	&	19.192	&	0.014	&	18.656	&	0.047	&	18.857	&	0.023	&	18.526	&	0.003	&	18.239	&	0.001	&	3.0	\\
59	&	84.681427	&	-69.082382	&	20.446	&	0.122	&	20.422	&	0.007	&	20.577	&	0.005	&	19.773	&	0.078	&	20.252	&	0.036	&	19.747	&	0.011	&	19.595	&	0.011	&	3.0	\\
60	&	84.681511	&	-69.082527	&	23.645	&	1	&	22.558	&	0.309	&	21.584	&	0.066	&	19.362	&	0.072	&	20.929	&	0.036	&	19.701	&	0.117	&	19.125	&	0.071	&	3.0	\\
61	&	84.674179	&	-69.084106	&	24.864	&	0.449	&	24.045	&	0.15	&	22.919	&	0.032	&	21.645	&	0.052	&	21.811	&	0.016	&	20.685	&	0.044	&	19.936	&	0.029	&	2.5	\\
62	&	84.682388	&	-69.080818	&	23.704	&	0.238	&	23	&	0.148	&	21.976	&	0.048	&	20.734	&	0.017	&	21.059	&	0.012	&	20.261	&	0.019	&	19.688	&	0.008	&	3.0	\\
63	&	84.700630	&	-69.075279	&	21.251	&	0.004	&	21.143	&	0.048	&	20.988	&	0.008	&	20.515	&	0.042	&	20.692	&	0.017	&	20.309	&	0.002	&	20.14	&	0.027	&	3.0	\\
64	&	84.684479	&	-69.077736	&	20.953	&	11.111	&	20.351	&	0.06	&	19.988	&	0.02	&	19.186	&	0.05	&	19.377	&	0.002	&	18.941	&	0.006	&	18.576	&	0.005	&	3.0	\\
65	&	84.683121	&	-69.078804	&	23.388	&	0.285	&	21.243	&	0.007	&	19.748	&	0.013	&	18.758	&	0.054	&	18.81	&	0.011	&	18.01	&	11.111	&	17.451	&	0.013	&	3.0	\\
66	&	84.681145	&	-69.078850	&	24.586	&	0.455	&	23.169	&	0.163	&	22.572	&	0.069	&	21.311	&	0.058	&	22.08	&	0.066	&	21.552	&	0.103	&	21.341	&	0.064	&	2.0	\\
67	&	84.706413	&	-69.071678	&	23.375	&	0.203	&	22.644	&	0.132	&	22.222	&	0.035	&	21.53	&	0.022	&	21.647	&	0.014	&	21.185	&	0.017	&	20.892	&	0.094	&	2.5	\\
68	&	84.671455	&	-69.081024	&	23.496	&	0.461	&	22.51	&	0.015	&	21.653	&	0.055	&	20.501	&	0.095	&	21.156	&	0.005	&	20.521	&	0.016	&	20.177	&	0.016	&	3.0	\\
69	&	84.680443	&	-69.077477	&	24.546	&	11.111	&	22.594	&	0.046	&	21.444	&	0.076	&	19.966	&	0.077	&	20.34	&	0.042	&	19.366	&	0.119	&	18.574	&	0.009	&	3.0	\\
70	&	84.697449	&	-69.073441	&	20.468	&	0.005	&	20.087	&	0.032	&	20.184	&	0.012	&	18.785	&	0.037	&	19.018	&	0.011	&	17.973	&	0.002	&	17.257	&	0.004	&	3.0	\\
71	&	84.702751	&	-69.072067	&	24.741	&	0.111	&	23.873	&	0.151	&	22.527	&	0.039	&	21.219	&	0.067	&	21.429	&	0.026	&	20.474	&	0.004	&	19.892	&	0.013	&	3.0	\\
72	&	84.700760	&	-69.071678	&	23.645	&	11.111	&	22.721	&	0.105	&	22.061	&	0.03	&	21.197	&	0.026	&	21.289	&	0.018	&	20.693	&	0.123	&	20.385	&	0.001	&	3.0	\\
73	&	84.704803	&	-69.071213	&	26.26	&	4.198	&	24.966	&	0.087	&	23.708	&	0.061	&	21.821	&	0.084	&	22.775	&	0.01	&	21.529	&	0.021	&	20.87	&	0.031	&	2.0	\\
74	&	84.673096	&	-69.078514	&	22.993	&	0.209	&	22.404	&	0.108	&	21.963	&	0.006	&	21.113	&	0.087	&	21.562	&	0.047	&	21.009	&	0.06	&	20.876	&	0.032	&	2.5	\\
75	&	84.671036	&	-69.080124	&	23.411	&	0.078	&	21.635	&	0.037	&	19.874	&	0.014	&	18.717	&	0.032	&	18.764	&	0.011	&	17.94	&	0.006	&	17.319	&	0.001	&	3.0	\\
76	&	84.672043	&	-69.080139	&	20.922	&	0.097	&	20.636	&	0.036	&	20.311	&	0.028	&	19.598	&	0.048	&	19.832	&	0.024	&	19.196	&	0.075	&	18.827	&	0.037	&	3.0	\\
77	&	84.692680	&	-69.072685	&	20.098	&	0.004	&	20.085	&	0.079	&	20.235	&	0.018	&	19.695	&	0.018	&	19.973	&	0.016	&	19.707	&	0.009	&	19.594	&	0.023	&	3.0	\\
78	&	84.691689	&	-69.073006	&	22.974	&	11.111	&	22.228	&	0.016	&	21.377	&	0.024	&	19.984	&	0.098	&	20.593	&	0.032	&	19.795	&	0.026	&	19.204	&	0.005	&	3.0	\\
79	&	84.704056	&	-69.069389	&	23.665	&	11.111	&	23.258	&	0.265	&	22.4	&	0.034	&	20.781	&	0.055	&	21.333	&	0.008	&	19.861	&	0.037	&	18.97	&	0.054	&	3.0	\\
80	&	84.700874	&	-69.069824	&	23.756	&	0.265	&	23.379	&	0.112	&	22.702	&	0.043	&	19.331	&	0.051	&	20.07	&	0.016	&	17.726	&	0.033	&	15.823	&	0.017	&	3.0	\\
81	&	84.700478	&	-69.069885	&	23.218	&	0.389	&	22.995	&	0.009	&	22.792	&	0.031	&	20.95	&	0.098	&	21.142	&	0.006	&	19.505	&	0.059	&	18.11	&	0.057	&	3.0	\\
82	&	84.681694	&	-69.073654	&	22.861	&	0.048	&	22.266	&	0.005	&	21.321	&	0.018	&	20.124	&	0.067	&	20.659	&	0.038	&	20.012	&	0.001	&	19.634	&	0.001	&	3.0	\\
83	&	84.669319	&	-69.076431	&	25.196	&	0.191	&	24.124	&	0.17	&	22.645	&	0.016	&	21.469	&	0.04	&	21.797	&	0.003	&	20.747	&	0.003	&	20.297	&	0.032	&	3.0	\\
84	&	84.669365	&	-69.077171	&	23.675	&	0.15	&	22.898	&	0.298	&	21.919	&	0.022	&	20.701	&	0.074	&	21.015	&	0.034	&	20.237	&	0.002	&	19.627	&	0.068	&	3.0	\\
85	&	84.667603	&	-69.076630	&	25.749	&	0.349	&	23.583	&	0.323	&	22.688	&	0.013	&	21.151	&	0.094	&	21.846	&	0.023	&	20.934	&	0.101	&	20.499	&	0.034	&	3.0	\\
86	&	84.667641	&	-69.077843	&	24.436	&	0.675	&	23.754	&	0.157	&	23.107	&	0.023	&	21.28	&	0.091	&	22.691	&	0.051	&	21.438	&	0.064	&	21.384	&	0.09	&	2.0	\\
87	&	84.676857	&	-69.073395	&	25.452	&	11.111	&	23.088	&	11.111	&	22.445	&	0.026	&	20.899	&	0.08	&	21.784	&	0.017	&	20.781	&	0.147	&	20.184	&	0.064	&	2.5	\\
\enddata
\tablecomments{Column 1: The ID we give each more evolved low-mass YSOs (discussed in section 4.1). Columns 4-17: The magnitude and error of the star in F275W, F336W, F555W, F658N, F775W, F110W, and F160W respectively. Column 18: Mass of star determined from isochrone fitting. Note: The magnitude and errors vary in significant figures. These numbers are the same as \citet{sabb16}, therefore we do not change them in order to be consistent.}
\end{deluxetable}

\setcounter{table}{3} 
\renewcommand{\thetable}{\arabic{table}}
\tabletypesize{\tiny}
\begin{deluxetable}{ccccccccccccccc}
\tablewidth{0pt}
\rotate
\tablecaption{Near Infrared Photometry For SAGE Sources in the ALMA Footprint}
\tablehead{\colhead{Name} & \colhead{J} & \colhead{eJ} & \colhead{H} & \colhead{eH} & \colhead{K} & \colhead{eK} & \colhead{I1} & \colhead{eI1} & \colhead{I2} & \colhead{eI2} & \colhead{I3} & \colhead{eI3} & \colhead{I4} & \colhead{eI4}}
\startdata
J84.703995-69.079110 & & & & & 1.57011 & 0.14461 & 21.38444 & 1.37386 & 40.415511 & 2.23344 & 109.8241473 & 8.92142 & 264.88736 & 26.83673 \\
J84.695932-69.083807 & & & 3.30100 & 0.74500 & 7.93499 & 0.75269 & 22.27629 & 1.13999 & 33.91525 & 1.57547 & 57.79999 & 2.56299 & & \\
J84.699755-69.069803 & 1.38831 & 0.21737 & & & & & 25.38351 & 1.42744 & 33.37963 & 1.84666 & 87.23642 & 4.82862 & 252.96546 & 16.39286 \\
J84.688990-69.084695 & 11.33674 & 1.46181 & 8.20918 & 1.28535 & 13.30241 & 1.22519 & 23.79861 & 1.53435 & 29.54935 & 1.95117 & 88.85826 & 5.72894 & 241.58146 & 17.82829 \\
J84.695173-69.084857 & & & 3.30100 & 0.74500 & 7.93499 & 0.75269 & 22.27629 & 1.13999 & 33.91525 & 1.57547 & 57.79999 & 2.56299 & & \\
J84.726173-69.082254 & & & & & & & 9.7133293 & 0.66765012 & 14.2948082 & 1.03271423 & 33.45063937 & 2.21671575 & 91.7597332 & 5.56257797 \\
J84.720292-69.077084 & 0.49486 & 0.09115 & 1.08218 & 0.13954 & 2.53472 & 0.14007 & 11.28633 & 0.72765 & 11.33833 & 0.73174 & 41.75397 & 2.69197 & 103.52881 & 7.62828 \\
J84.688372-69.078168 & & & 0.36669 & 0.09798 & 0.26890 & 0.10230 & 2.78999 & 0.21799 &  &  & 17.13815 & 1.15499 & 36.29999 & 6.95368 \\
J84.669113-69.081638 & 1.40499 & 0.09706 & & & 0.72229 & 0.08202 &  &  &  &  & 2.19572 & 0.33989 &  & \\
J84.674734-69.077374 & 3.38299 & 0.25519 & 2.24200 & 0.21950 & 2.70099 & 0.24979 & 1.17999 & 0.12334 & 2.5999 & 0.22548 & 6.15999 & 0.33131 & 16.63815 & 1.11835 \\
J84.709403-69.075682 & 1.03600 & 0.08567 & 0.63669 & 0.07874 & 0.87559 & 0.10589 &  &  & 0.67424 & 0.11927 &  &  &  & \\
J84.688168-69.071013 & 1.89400 & 0.17440 & 1.30099 & 0.17139 & 1.16499 & 0.15129 & 0.95843 & 0.81513 &  &  &  &  &  & \\
J84.694286-69.074499 & 6.28200 & 0.16609 & 4.59499 & 0.12399 & 2.94799 & 0.09258 &  &  & 1.74770 & 0.38999 &  &  &  & \\
J84.676469-69.082774 & 39.75000 & 0.87860 & 27.96999 & 0.54100 & 19.31999 & 0.33799 & 9.89999 & 0.33199 & 7.49540 & 0.83999 &  &  &  &  \\
J84.671132-69.077168 & 2.13800 & 0.12620 & 2.94199 & 0.19650 & 3.41899 & 0.21259 & 0.77319 & 0.71533 &  &  &  &  &  & \\
\enddata
\tablecomments{Columns 2-7: JHK photometry and error in mJy. Columns 8-15: IRAC photometry and error in mJy. Note: The flux and errors vary in significant figures. These numbers are the same as those listed in the SAGE catalog as well studies done by \citet{grue09, chen10, seal14}. Therefore we do not change them in order to be consistent.}
\end{deluxetable}

\setcounter{table}{4} 
\renewcommand{\thetable}{\arabic{table}}
\tabletypesize{\tiny}
\begin{deluxetable}{ccccccccccc}
\tablewidth{0pt}
\rotate
\tablecaption{Far Infrared Photometry For SAGE Sources in the ALMA Footprint}
\tablehead{\colhead{Name} & \colhead{PACS} & \colhead{error} & \colhead{PACS} & \colhead{error} & \colhead{SPIRE} & \colhead{error} & \colhead{SPIRE} & \colhead{error} & \colhead{SPIRE} & \colhead{error}\\
& \colhead{100 $\mu$m} & \colhead{PACS 100 $\mu$m} & \colhead{160 $\mu$m} & \colhead{PACS 160 $\mu$m} & \colhead{250 $\mu$m} & \colhead{SPIRE 250 $\mu$m} & \colhead{350 $\mu$m} & \colhead{SPIRE 350 $\mu$m} & \colhead{500 $\mu$m} & \colhead{SPIRE 500 $\mu$m}}
\startdata
J84.703995-69.079110 & 19450 & 5076 & 27370 & 2511 & 1295 & 667 & 5397 & 698 & 6875 & 593\\
J84.695932-69.083807 &  &  &  &  &  &  &  &  &  & \\
J84.699755-69.069803 & 13460 & 2594 & 14490 & 1408 & 6252 & 597 &  &  &  & \\
J84.688990-69.084695 & 18340 & 3100 &  &  &  &  &  &  &  & \\
J84.695173-69.084857 &  &  &  &  &  &  &  &  &  &  \\
J84.726173-69.082254 & 5890 & 2529 & 8198 & 1126 &  &  &  &  &  & \\
J84.720292-69.077084 & 12520 & 2699 & 13480 & 1567 &  &  &  &  & 6875 & 593\\
J84.688372-69.078168 &  &  &  &  &  &  &  &  &  & \\
J84.669113-69.081638 &  &  &  &  &  &  &  &  &  & \\
J84.674734-69.077374 &  &  &  &  &  &  &  &  &  & \\
J84.709403-69.075682 &  &  &  &  &  &  &  &  &  & \\
J84.688168-69.071013 &  &  &  &  &  &  &  &  &  & \\
J84.694286-69.074499 &  &  &  & &  &  &  &  &  &  \\
J84.676469-69.082774 &  &  &  & &  &  &  &  &  &  \\
J84.671132-69.077168 &  &  &  &  &  & &  &  &  &  \\
\enddata
\tablecomments{Columns 2-5: PACS 100 $\mu$m and 160 $\mu$m photometry and error in mJy. Columns 6-11: SPIRE 250 $\mu$m, 350 $\mu$m, and 500 $\mu$m photometry and error in mJy. Note: The flux and errors vary in significant figures. These numbers are the same as those listed in the SAGE catalog as well studies done by \citet{grue09, chen10, seal14}. Therefore we do not change them in order to be consistent.}
\end{deluxetable}

\setcounter{table}{5} 
\renewcommand{\thetable}{\arabic{table}}
\tabletypesize{\tiny}
\tablewidth{0pt}
\rotate
\begin{deluxetable}{cccccccccc}
\tablecaption{Properties of SAGE Sources in the ALMA Footprint}
\tablehead{\colhead{ID Number} & \colhead{Name} & \colhead{RA} & \colhead{Dec} & \colhead{Mass [$\mathrm{M_{\odot}}$]} & \colhead{Luminosity[$\mathrm{L_{\odot}}$]} & \colhead{Reduced $\mathrm{\chi^{2}_{stellar\;photosphere}}$} & \colhead{Reduced $\mathrm{\chi^{2}_{YSO\;SED}}$} & \colhead{Is it a YSO Candidate?} & \colhead{References}}
\startdata
1 & J84.703995-69.079110 & 84.703995 & -69.079110 & $23.9_{-7.01}^{+23.8}$ & $6.81_{-4.20}^{+23.8} \times 10^{4}$ & 209 & 1.63 & yes & a, b, c, d, e\\
2 & J84.695932-69.083807 & 84.695932 & -69.083807 & $21.8_{-8.07}^{+12.0}$ & $5.62_{-4.16}^{+6.49} \times 10^{4}$ & 43.3 & 0.482 & yes & e\\
3 & J84.699755-69.069803 & 84.699755 & -69.069803 & $19.0_{-5.25}^{+10.5}$ & $5.62_{-4.16}^{+6.49} \times 10^{4}$ & 333 & 2.67 & yes & b, c, d, e\\
4 & J84.688990-69.084695 & 84.688990 & -69.084695 & $19.0_{-5.87}^{+9.13}$ & $4.64_{-3.17}^{+5.36} \times 10^{4}$ & 241 & 8.73 & yes & a, b, c, e\\
5 & J84.695173-69.084857 & 84.695173 & -69.084857 & $19.0_{-5.87}^{+10.5}$ & $5.62_{-4.16}^{+6.49} \times 10^{4}$ & 43.3 & 0.482 & yes & d, e\\
6 & J84.726173-69.082254 & 84.726173 & -69.082254 & $17.3_{-7.37}^{+15.7}$ & $3.16_{-2.34}^{+5.09} \times 10^{4}$ & 292 & 0.025 & yes & b, e\\
7 & J84.720292-69.077084 & 84.720292 & -69.077084 & $14.8_{-4.79}^{+18.3}$ & $3.16_{-2.48}^{+6.84} \times 10^{4}$ & 234 & 9.72 & yes & a, b, c, d, e\\
8 & J84.688372-69.078168 & 84.688372 & -69.078168 & $11.8_{-3.43}^{+21.4}$ & $8.25_{-5.09}^{+207} \times 10^{3}$ & 82.1 & 1.44 & yes & e\\
9 & J84.669113-69.081638 & 84.669113 & -69.081638 & $8.71_{-2.96}^{+4.78}$ & $3.16_{-2.34}^{+11.5} \times 10^{3}$ & 36.5 & 0.003 & yes & e\\
10 & J84.674734-69.077374 & 84.674734 & -69.077374 & $8.51_{-1.75}^{+4.98}$ & $3.83_{-2.62}^{+13.9} \times 10^{3}$ & 209 & 9.64 & yes & e\\
11 & J84.709403-69.075682 & 84.709403 & -69.075682 & $8.71_{-2.25}^{+9.07}$ & $8.25_{-7.04}^{+23.4} \times 10^{3}$ & 6.67 & 2.28 & no & e\\
12 & J84.688168-69.071013 & 84.688168 & -69.071013 & $8.71_{-2.68}^{+5.74}$ & $3.16_{-2.34}^{+14.6} \times 10^{3}$ & 2.17 & 1.13 & no & e\\
13 & J84.694286-69.074499 & 84.694286 & -69.074499 & $11.2_{-3.09}^{+7.83}$ & $8.25_{-5.64}^{+30.0} \times 10^{3}$ & 3.21 & 64.6 & no & e\\
14 & J84.676469-69.082774 & 84.676469 & -69.082774 & $11.2_{-1.67}^{+4.63}$ & $8.25_{-3.61}^{+13.3} \times 10^{3}$ & 4.37 & 935 & no & e\\
15 & J84.671132-69.077168 & 84.671132 & -69.077168 & $8.51_{-2.20}^{+4.67}$ & $3.16_{-2.16}^{+11.5} \times 10^{3}$ & 11.3 & 25.9 & no & e\\
\enddata
\tablecomments{Column 1: ID Number. Columns 5: Mass of the YSO candidate as given by the peak of the liklihood distribution and 1$\mathrm{sigma}$ error. Columns 6: Luminosity of the YSO candidate as given by the peak of the liklihood distribution and 1$\mathrm{sigma}$ error. Column 7: Reduced $\mathrm{\chi^{2}}$ of the best-fit stellar photosphere ($\mathrm{\chi^{2}}$ divided by the number of fitted points). Column 8: Reduced $\mathrm{\chi^{2}}$ of the best-fit SED model ($\mathrm{\chi^{2}}$ divided by the number of fitted points). Column 10: References: a for \citet{seal09}, b for \citet{seal14}, c for \citet{grue09}, d for \citet{walb13}, and e for this work.}
\end{deluxetable}

\end{document}